\documentclass[11pt]{article}
\usepackage{etex}

\usepackage[table]{xcolor}
\usepackage{amsfonts}
\usepackage{natbib}											
\usepackage{graphicx, adjustbox} 									
\usepackage[small,bf,hang]{caption}		
\usepackage{subcaption}									
\usepackage{wrapfig}										
\usepackage{multirow}										
\usepackage[latin1]{inputenc}						
\usepackage{amsmath}										
\allowdisplaybreaks[1]									
\usepackage{amssymb}										
\usepackage{amsthm}											
\usepackage{booktabs}										
\usepackage{url}                        
\usepackage{fancyhdr}                   
\usepackage{listings} 									
\usepackage{pdfpages}										
\usepackage{float}                      
\usepackage[title,titletoc,toc]{appendix}				
\usepackage{enumerate}									
\usepackage{enumitem}										
\usepackage{mathtools}										
\usepackage{algorithm}										
\usepackage{algpseudocode}								
\usepackage{authblk}											
\usepackage{bm}
\usepackage{pdflscape}
\usepackage{mathtools}
\usepackage{setspace}
\usepackage[labelformat=simple]{subcaption}

\makeatother
\usepackage[plainpages=false,hyperfootnotes=false]{hyperref}
\hypersetup{
  colorlinks   = true, 
  urlcolor     = blue, 
  linkcolor    = red, 
  citecolor   = blue 
}

\usepackage{everypage}
\newlength{\hfoot}
\newlength{\vfoot}
\AddEverypageHook{\ifdim\textwidth=\linewidth\relax
\else\setlength{\hfoot}{-\topmargin}%
\addtolength{\hfoot}{-\headheight}%
\addtolength{\hfoot}{-\headsep}%
\addtolength{\hfoot}{-.5\linewidth}%
\ifodd\value{page}\setlength{\vfoot}{\oddsidemargin}%
\else\setlength{\vfoot}{\evensidemargin}\fi%
\addtolength{\vfoot}{\textheight}%
\addtolength{\vfoot}{\footskip}%
\raisebox{\hfoot}[0pt][0pt]{\rlap{\hspace{\vfoot}\rotatebox[origin=cB]{90}{\thepage}}}\fi}

\DeclareCaptionLabelFormat{andtable}{#1~#2  \&  \tablename~\thetable}

\author[1,*]{Sander Devriendt}
\author[1,2]{Katrien Antonio}
\author[1]{Tom Reynkens}
\author[1]{Roel Verbelen}

\affil[1]{Faculty of Economics and Business, KU Leuven, Belgium.}
\affil[2]{Faculty of Economics and Business, University of Amsterdam, The Netherlands.}
\affil[*]{Corresponding author. Email address: \href{mailto:sander.devriendt@kuleuven.be}{sander.devriendt@kuleuven.be}}

\title{\textbf{Sparse regression with Multi-type Regularized Feature modeling}}

\begin{document}

\maketitle

\begin{abstract}
Within the statistical and machine learning literature, regularization techniques are often used to construct sparse (predictive) models. Most regularization strategies only work for data where all predictors are treated identically, such as Lasso regression for (continuous) predictors treated as linear effects. However, many predictive problems involve different types of predictors and require a tailored regularization term. We propose a multi-type Lasso penalty that acts on the objective function as a sum of subpenalties, one for each type of predictor. As such, we allow for predictor selection and level fusion within a predictor in a data-driven way, simultaneous with the parameter estimation process. We develop a new estimation strategy for convex predictive models with this multi-type penalty. Using the theory of proximal operators, our estimation procedure is computationally efficient, partitioning the overall optimization problem into easier to solve subproblems, specific for each predictor type and its associated penalty. Earlier research applies approximations to non-differentiable penalties to solve the optimization problem. The proposed SMuRF algorithm removes the need for approximations and achieves a higher accuracy and computational efficiency. This is demonstrated with an extensive simulation study and the analysis of a case-study on insurance pricing analytics.
\end{abstract}

\noindent%
\vfill

\section{Introduction}\label{sec:intro}

With the arrival of big data, many companies and institutions struggle to infer meaningful information from their data sets. We propose a novel estimation framework for sparse regression models that can simultaneously handle: (1) the selection of the most relevant predictors (or: features), and (2) the binning or level fusion of different predictor types, taking into account their structural properties. 

Recently, the use of regularization techniques became very popular as a strategy to identify the predictors with the most predictive power, enabling the construction of sparse regression models. When performing the estimation, the regularization terms, or penalties, effectively put a budget constraint (\cite{bookhastie}) on the parameter space, in order to reduce the dimensionality to avoid overfitting and multicollinearity issues, and to improve the interpretability of the fitted model. The statistical and machine learning community developed numerous regularization methods to obtain sparse and hence more interpretable predictive models, such as the penalized smoothing splines of \cite{eilers} or the least absolute shrinkage and selection operator (Lasso) in \cite{lasso}, followed by its many extensions in \cite{elnet}, \cite{fusedlasso}, \cite{grouplasso}, \cite{gfl} and \cite{genlasso}.

Most of these methods, however, are developed for linear regression and specific data sets where all predictors are of the same type and thus the same type of penalty is applied to all parameters or coefficients. For example, the Lasso is originally developed for linear regression with continuous predictors, using one parameter (or: coefficient) per predictor.
However, many predictive problems require more general loss functions and rely on various types of predictors requiring different kinds of regularization applied to the coefficients. For example, the levels of a discretized continuous or ordinal predictor (e.g.\ age) have a different structure compared to a spatial predictor (e.g.\ postal code), where a two-dimensional layout determines the relationship between the levels. This also applies to nominal predictors (e.g.\ type of industry) where the underlying structure is often predictor-specific. This level structure within a predictor needs to be taken into account when assigning regularization terms to coefficients, leading to different penalties tailored to the specific structure of the corresponding predictor. The use of such penalties enables the analyst not only to select the relevant predictors, but also to fuse levels within a predictor (e.g.\ clusters of postal codes or industry types). This fusion of levels is often challenging in large data sets where many predictors are present which may consist of many levels. An automatic selection and fusion strategy is then very helpful. To the best of our knowledge, the first attempt at regularized regression with multiple predictor types is \cite{gertheiss2010} who provide a regularization method which can simultaneously deal with binary, ordered and nominal predictors for linear models. This was later extended to generalized linear models (GLMs) in \cite{gertheiss2017}. We inherit their formulation where the multi-type penalty acts on the objective function as a sum of subpenalties, one for each predictor type.

The design of an effective and accurate estimation strategy is one of the main challenges for the aforementioned regularization techniques. On the one hand, the machine learning community often employs specialized optimization techniques such as the algorithm for Least Angle Regression (LARS, as in \cite{lars}), parametric flow (e.g.\ in \cite{xin}) or subgradient finding algorithms (see \cite{liu}). These techniques are well suited for data sets containing a single type of predictor and its corresponding penalty, but are very difficult to extend when different predictor types come into play simultaneously. On the other hand, \cite{gertheiss2017}, within the statistical community, propose local quadratic approximations of the penalties such that the penalized iteratively reweighted least squares (PIRLS) algorithm is applicable in the context of regularized GLMs. This approach can solve the more general, multi-type Lasso regularization setting, but the quadratic approximations lead to non-exact predictor selection and level fusion of the predictor coefficients. In addition, the PIRLS algorithm requires the calculation of large matrix inverses and therefore does not scale well to big data, both tall and wide.

Combining insights from both the machine learning and statistics literature, our main contribution is the design of an efficient calibration strategy, suitable for regularization with different predictor types and more general loss functions. Our solution builds on the properties of proximal operators (see \cite{parikh} for a comprehensive overview) which have been studied for Lasso-type penalties in for example \cite{fista} and \cite{xin}. Using proximal operators, our algorithm decomposes the optimization procedure with the multi-type penalty into a set of smaller optimization subproblems for each predictor type and its associated penalty. As such, our approach nullifies the need for approximations as in \cite{gertheiss2017}. Next to this, our proposed estimation procedure applies specialized machine learning algorithms to each subproblem. Furthermore, this partition enables the use of distributed or parallel computing for the optimization process. Additionally, we provide an open-source implementation of our algorithm in the \texttt{R} package \texttt{smurf}, available on CRAN (\url{https://cran.r-project.org/web/packages/smurf/}).

The paper is structured as follows. Section \ref{sec:lasso} explains the different Lasso penalties. Section \ref{sec:opt} presents our optimization algorithm for a multi-type penalty. Our approach is then applied to a simulation study in Section \ref{sec:simulation} and to a motor insurance case-study in Section \ref{sec:MTPLcase}. Section \ref{sec:conclusion} concludes.

\section{Multi-type Lasso regularization}\label{sec:lasso}

Consider a response $\bm{y}$ and the corresponding model matrix $\bm{X}$. We assume that continuous and binary predictors are coded with one parameter, and thus one column in $\bm{X}$, while the ordinal, nominal and spatial predictors are represented through dummy or one-hot encoding. The regularized objective function for a multi-type predictive model is
\begin{equation}
\mathcal{O}(\bm\beta; \bm{X}, \bm{y}) = f(\bm\beta; \bm{X},\bm{y}) + \lambda \cdot \sum_{j=0}^J g_j(\bm\beta_j), \label{eq:penmultireg}
\end{equation}
where $f$ is a convex, differentiable function, $g_j$ a convex function for all $j \in \{1,\ldots, J\}$ and $\bm\beta_j$ represents a subset of the full parameter vector $\bm\beta$ such that $(\beta_0, \bm\beta_1,\ldots, \bm\beta_J) = \bm\beta$, with $\beta_0$ the intercept if present. For simplicity, we set $g_0(\beta_0) = 0$ since the intercept is typically not regularized, though our approach also works for regularized intercepts. Here, $f$ is the loss function, measuring the distance between the observed and the fitted data, e.g.\ the least squares criterion in \cite{lasso} or minus the log-likelihood for GLMs in \cite{restrglm}. The penalty functions $g_j$ serve as a measure to avoid overfitting the data, while the tuning parameter $\lambda$ controls the strength of the penalty. A high value of $\lambda$ increases its importance in the objective function $\mathcal{O}$ and will increase the regularization. The partition of $\bm\beta$ in subvectors $\bm\beta_j$ and the choice of $g_j$ is based on the predictor types and level structures such that for each $j$, the penalty $g_j$ reflects this structure on the individual coefficients in $\bm\beta_j$. For simplicity, we assume that $\bm\beta$ is partitioned per predictor with the coefficients in $\bm\beta_j$ corresponding to the levels used to code predictor $j$. For a continuous predictor, $\bm\beta_j$ contains a single coefficient while (e.g.) for an ordinal predictor $\bm\beta_j$ comprises as many coefficients as there are levels.

To enhance sparsity, we design multi-type Lasso penalties for \eqref{eq:penmultireg} which can remove or fuse coefficients due to the use of the non-differentiable $L_1$-norm. Since we apply only one overall tuning parameter $\lambda$, it is important to incorporate penalty weights to improve predictive performance, to obtain asymptotic consistency and to correct for structural differences such as the number of levels within a predictor. Section~\ref{sec:pentypes} gives an overview of the implemented penalties and motivates their use to enhance sparsity in multi-type predictive models, while Section~\ref{sec:weights} explains the purpose and our implementation of the penalty weights.

\subsection{Lasso and generalizations}\label{sec:pentypes}

\paragraph{Lasso}\hspace*{-0.2cm}[\cite{lasso}].
The Lasso penalty applies the $L_1$-norm to predictor coefficients:
\begin{equation}
g_{\text{Lasso}}(\bm\beta_j) = \sum_{i=1}^{p_j} w_{j,i}|\beta_{j,i}| = ||\bm{w}_j * \bm\beta_j||_1,
\label{eq:lasso}
\end{equation}
where $p_j$ is the number of individual coefficients $\beta_{j,i}$ within the vector $\bm\beta_j$, $\bm{w}_j$ is a vector of penalty weights and `$*$' denotes the componentwise multiplication. Depending on the tuning parameter $\lambda$ and the penalty weight vector $\bm{w}_j$, this penalty will encourage some coefficients to become zero, effectively removing them from the model. The other coefficients will have estimates closer to 0 compared to the unregularized setting, reducing estimation variance but increasing bias. To have a fair regularization over all coefficients $\beta_{j,i}$, the respective columns of $\bm{X}$ should be centered and standardized, also for one-hot encoded predictors, as explained in \cite{tibshirani1997}. For continuous or binary predictors, represented by one coefficient, the Lasso penalty serves as a predictor selection tool where the most important predictors receive non-zero coefficients. With ordinal or nominal predictors, Lasso selects the relevant coefficients (or: levels) within each predictor. In this case, no reference category should be chosen, as this would change the interpretation of the coefficients and subsequently of the Lasso penalty. Instead of being a level selection tool, the penalty would then result in a method to fuse levels with the reference category.

\paragraph{Group Lasso}\hspace{-0.2cm}[\cite{grouplasso}].
The Group Lasso penalty uses an $L_2$-norm to encourage the coefficients in $\bm\beta_j$ to be removed from the model in an all-in or all-out approach:
\begin{equation*}
g_{\text{grpLasso}}(\bm\beta_j) = w_j \sqrt{\sum_{i=1}^{p_j} \beta_{j,i}^2}= ||w_{j}\bm\beta_{j}||_2,
\end{equation*}
where $w_{j}$ is a penalty weighting factor. In contrast to the $L_1$-norm, the $L_2$-norm is not separable for each coefficient in $\bm\beta_j$ and is only non-differentiable when all $\beta_{j,i}$ are 0. This penalty is appropriate to determine if $\bm\beta_j$ has adequate predictive power as a whole, because the estimates for $\beta_{j,i}$ will be either all zero or all non-zero. When $\bm\beta_j$ consists of only one coefficient, the $L_2$-norm reduces to the $L_1$-norm and the standard Lasso penalty is retrieved. This Group Lasso penalty is particularly useful for selecting ordinal or nominal predictors. When applied to an ordinal or nominal predictor, the Group Lasso requires no reference category, similar to the case of the standard Lasso penalty.

\paragraph{Fused Lasso}\hspace{-0.2cm}[\cite{fusedlasso}].
To group consecutive levels within a predictor, the Fused Lasso penalty puts an $L_1$-penalty on the differences between subsequent coefficients:\\
\begin{equation}
g_{\text{fLasso}}(\bm\beta_j) = \sum_{i=2}^{p_j} w_{j,i-1}|\beta_{j,i} - \beta_{j,i-1}| = ||\bm{D}(\bm{w}_j) \bm\beta_j||_1, \label{eq:flasso}
\end{equation}
with $\bm{D}(\bm{w_j})$ the first order difference matrix with dimensions $(p_j-1)\times p_j$ where the rows are weighted by the elements in $\bm{w}_j$:
\begin{align}
\bm{D}(\bm{w}_j) = 
\begin{bmatrix*}[r]
    -w_{j,1} & w_{j,1} & 0 & &  0 & 0\\
    0& -w_{j,2} & w_{j,2}  & \cdots & 0& 0\\
    0&  0   & -w_{j,3}   & & 0  &0\\
    &  \vdots & & \ddots   &  w_{j,p_j-2} & 0\\
    0& 0&0&& -w_{j,p_j-1} & w_{j,p_j-1}
    \end{bmatrix*}.
\label{eq:Dmat}
\end{align}
This penalty is suitable for ordinal predictors and continuous predictors coded as ordinal predictors to capture non-linear effects. Because \eqref{eq:flasso} only regularizes differences, a reference level needs to be chosen to get a unique minimizer $\bm\beta$ in optimization problem \eqref{eq:penmultireg}. The coefficient of $\bm\beta_j$ corresponding to this reference level is then set to 0 or, equivalently, the relevant column in \eqref{eq:Dmat} is omitted. For high values of $\lambda$ in \eqref{eq:penmultireg}, all differences between subsequent coefficients from $\bm\beta_j$ will become 0, such that they are fused with the reference category, and the predictor is then effectively removed from the model. The Fused Lasso is ideal to bin ordinal predictors. It also applies to continuous predictors for which non-linear effects are expected by starting from a very crude binning, for example by rounding to the integer and assigning each integer its own coefficient.

\paragraph{Generalized Fused Lasso}\hspace*{-0.2cm}[\cite{gfl}].
The Generalized Fused Lasso (GFL) allows the user to set a graph $\mathcal{G}$ that indicates which coefficient differences should be regularized:
\begin{align}
g_{\text{gfLasso}}(\bm\beta_j) &= \sum_{(i,l)\in \mathcal{G}} w_{j,il}|\beta_{j,i} - \beta_{j,l}| = ||\bm{G}(\bm{w}_j)\bm\beta_j||_1,\label{eq:gflasso} 
\end{align}
where $\bm{G}(\bm{w}_j)$ is the matrix with dimensions $r_{\bm{G}}\times p_j$ of the linear map projecting $\bm\beta_j$ onto all differences of coefficients connected by the $r_{\bm{G}}$ edges in the graph $\mathcal{G}$, with the rows weighted by the elements in $\bm{w}_j$. The matrix $\bm{G}(\bm{w}_j)$ thus generalizes $\bm{D}(\bm{w}_j)$ in \eqref{eq:Dmat}. Similar to the Fused Lasso, a reference category is needed to obtain a unique minimizer $\bm\beta$ of \eqref{eq:penmultireg}. This penalty is useful to bin predictors whenever a straightforward graph is available. Section~\ref{sec:MTPLcase} shows an example with a spatial predictor, for which the logical penalty regularizes the coefficient differences for municipalities sharing a physical border. For nominal predictors without any underlying structure, we follow \cite{gertheiss2010} to use the graph leading to a regularization of all possible coefficient differences. Another special case of the Generalized Fused Lasso is the 2D-Fused Lasso (\cite{fusedlasso}), known from image recognition, which can be used for modeling interaction effects as we illustrate in the simulation study of Section~\ref{sec:simulation}.

\subsection{Penalty weights}\label{sec:weights}

\cite{adlasso} shows that the standard Lasso penalty might lead to inconsistent selection of coefficients. We therefore investigate the use of penalty weights $\bm{w}_j$ to improve the performance of the different penalties. \cite{adlasso} suggests to incorporate adaptive penalty weights $\bm{w}_j^{(\text{ad})}$ into the Lasso penalty to obtain the oracle properties of consistency and asymptotic normality for the resulting estimates. The adaptive penalty weights are based on initial consistent parameter estimates $\hat{\bm\beta}$ as obtained with GLM or ridge estimation, where the latter uses a small tuning parameter. Using the notation of \eqref{eq:lasso}, the adaptive penalty weights are defined by
\begin{equation*}
w_{j,i}^{(\text{ad})} = |\hat{\beta}_{j,i}|^{-\gamma},
\end{equation*}
with $\gamma>0$ a tuning parameter. Intuitively, these weights `adapt' the penalty to data-driven prior information in the form of an initial estimator. As a consequence, coefficients initially estimated as small will be regularized more than large ones. In this paper, we set $\gamma=1$ as in \cite{gertheiss2010} and use the penalty-specific adaptive weights listed in Table~\ref{tab:weights}, adopted from \cite{adgrouplasso}, \cite{adfl} and \cite{adgfl}, respectively.
\begin{table}[ht!]
    \begin{center}
    \begin{adjustbox}{max width = \textwidth}
    \begin{tabular}{p{4.5cm} p{4.5cm} p{4.5cm}}
    \toprule
    Penalty name & $\bm{w}_j^{(\text{ad})}$ & $\bm{w}_j^{(\text{st})}$\\
    \midrule
    Lasso & $w_{j,i}^{(\text{ad})} = |\hat{\beta}_{j,i}|^{-1}$ & $w_{j,i}^{(\text{st})} = 1$ \\
    \addlinespace[1em]
    Group Lasso & $w_{j}^{(\text{ad})} = ||\hat{\bm\beta}_{j}||_2^{-1}$ & $w_{j}^{(\text{st})} = 1$ \\
    \addlinespace[1em]
    Fused Lasso & $w_{j,i-1}^{(\text{ad})} = |\hat{\beta}_{j,i} - \hat{\beta}_{j,i-1}|^{-1}$ & $w^{\text{(st)}}_{j,i-1} = \sqrt{\frac{n_{j,i}+n_{j,i-1}}{n}}$ \\
    \addlinespace[1em]
    Generalized Fused Lasso & $w_{j,il}^{(\text{ad})} = |\hat{\beta}_{j,i} - \hat{\beta}_{j,l}|^{-1}$ & $w^{\text{(st)}}_{j,il} = \frac{p_j-1}{r_{\bm{G}}}\sqrt{\frac{n_{j,i}+n_{j,l}}{n}}$ \\
      \bottomrule
    \end{tabular}
    \end{adjustbox}
    \caption{\label{tab:weights}Overview of the proposed adaptive and standardization penalty weights for each penalty type.}
    \end{center}
\end{table}

As mentioned in Section \ref{sec:pentypes}, the columns of $\bm{X}$ linked to subvector $\bm\beta_j$ should be centered and standardized when applying the Lasso and Group Lasso penalties, also for dummy-coded predictors, as argued in \cite{tibshirani1997}. This is necessary to counteract the influence of different measuring scales for different predictors. Post-estimation, one can transform $\bm{X}$ and $\bm{\beta}$ back to the original scale for easier interpretation and prediction. However, when applying the (Generalized) Fused Lasso to predictors, standardization is not possible because the levels would lose their initial interpretation, making subsequent level fusion meaningless. Therefore, \cite{weights} and \cite{gertheiss2010} propose an alternative penalty weighting scheme for the (Generalized) Fused Lasso penalty, when used with ordered or nominal predictors. With $n$ the number of observations in the data set and $n_{j,i}$ the number of observations in level $i$ of predictor $j$, they define the standardization penalty weights $\bm{w}_j^{(\text{st})}$ for the Fused Lasso as follows:
\begin{equation}
w^{\text{(st)}}_{j,i-1} = \sqrt{\frac{n_{j,i}+n_{j,i-1}}{n}}.\label{eq:standweights}
\end{equation}
The standardization penalty weights $w^{\text{(st)}}_{j,i-1}$ adjust for possible level imbalances, where some levels may contain more observations than others. Extending \eqref{eq:standweights} to the Generalized Fused Lasso is possible by adding an extra factor, taking into account the number of regularized differences, relative to the Fused Lasso. For a predictor with $p_j$ levels, the Fused Lasso penalty contains $p_j-1$ terms. However, for the Generalized Fused Lasso this number is determined by the number of edges $r_{\bm{G}}$ of the graph $\mathcal{G}$. Similar to \cite{gertheiss2010}, we construct a penalty of the same order as the one used in the Fused Lasso by multiplying the standardization penalty weights in \eqref{eq:standweights} by a factor $\frac{p_j-1}{r_{\bm{G}}}$, see Table~\ref{tab:weights}. Without this factor, applying a Generalized Fused Lasso with large $r_{\bm{G}}$ would make the penalty artificially larger compared to the Fused Lasso, only because there are more regularized coefficient differences. The extra factor reduces to 1 for the Fused Lasso and to $\frac{2}{p_j}$ when all pairwise differences are regularized.

One can choose to use the adaptive or the standardization penalty weights, or combine the objectives of both the adaptive and the standardization weights by multiplying them: $\bm{w}_j = \bm{w}_j^{(\text{ad})}\cdot \bm{w}_j^{(\text{st})}$. We evaluate the performance of the adaptive and standardization penalty weights in the simulation study in Section~\ref{sec:simulation}.

\section{Optimization}\label{sec:opt}

We consider the predictive model \eqref{eq:penmultireg}, where $g_j$ can be any convex penalty, such as the Lasso-type examples in Section~\ref{sec:pentypes}. \cite{gertheiss2010} solve this regularization problem for linear models with Lasso-type penalties using linear programming. \cite{gertheiss2017} propose the PIRLS algorithm to minimize \eqref{eq:penmultireg}. The main disadvantage of the PIRLS approach is the use of quadratic approximations on the penalties. This decreases the estimation accuracy and leads to inexact selection and fusion of the coefficients in the parameter vector $\bm\beta$. Our strategy creates a leap forward by using proximal operators to solve the subproblems per penalty type exactly. 

\subsection{Proximal operators}\label{sec:po}

Using a standard procedure from the gradient descent algorithm in \cite{nesterov}, we rewrite the objective function in \eqref{eq:penmultireg} and replace $f$ by a local linearization around a point $\bm{\beta}^{(*)}$ including a Tikhonov regularization term:
\begin{align}
\tilde{\mathcal{O}}(\bm\beta) &= f(\bm{\beta}^{(*)}) + \left(\bm\beta - \bm{\beta}^{(*)}\right)^{T}\nabla f\left(\bm{\beta}^{(*)}\right) + \frac{1}{2s}||\bm\beta - \bm{\beta}^{(*)}||_2^2 + \lambda \sum_{j=0}^J g_j(\bm\beta_j), \nonumber \\
&= \frac{1}{2s}\left|\left|\bm\beta^{(*)} - s\nabla f\left(\bm{\beta}^{(*)}\right) - \bm{\beta}\right|\right|_2^2 + \lambda \sum_{j=0}^J g_j(\bm\beta_j) + C, \label{eq:penmultireg.linear}
\end{align}
where we omit $\bm{X}$ and $\bm{y}$ to ease notation and $C$ is a rest term independent of $\bm\beta$. Standard results from convex optimization (see e.g.\ \cite{convexopt}) show that minimizing \eqref{eq:penmultireg} is equivalent to iteratively minimizing \eqref{eq:penmultireg.linear} with a well-chosen step size $s$.  This procedure allows us to reformulate the problem using proximal operators (\cite{parikh}). Let $h: \mathbb{R}^n \rightarrow \mathbb{R}$ be a closed proper convex function. Then, the proximal operator (PO) $\text{prox}_h: \mathbb{R}^n \rightarrow \mathbb{R}^n$ of $h$ is defined by
\begin{equation*}
\text{prox}_h(\bm{x}) = \underset{\bm{z}}{\mathrm{argmin}} \;\; \frac12 ||\bm{x}-\bm{z}||_2^2 + h(\bm{z}).
\label{eq:proxop}
\end{equation*}
By setting $\tilde{\bm\beta}^{(k)} = \bm\beta^{(k)} - s\nabla f\left(\bm{\beta}^{(k)}\right)$, it is straightforward to see that finding the minimizer of \eqref{eq:penmultireg.linear} is equivalent to calculating the PO
\begin{equation}
\text{prox}_{s\lambda\sum_j g_j}\left(\tilde{\bm\beta}^{(k)}\right) = \underset{\bm{z}}{\mathrm{argmin}} \;\; \frac12 \left|\left|\tilde{\bm\beta}^{(k)}-\bm{z}\right|\right|_2^2 + s\lambda \sum_{j=0}^J g_j(\bm{z}_j).\label{eq:proxop.applied}
\end{equation}
The first term in \eqref{eq:proxop.applied} can be partitioned into a sum of squared $L_2$-norms relative to the partition  $\left(\tilde{\beta}_0^{(k)}, \tilde{\bm\beta}_1^{(k)}, \ldots, \tilde{\bm\beta}_J^{(k)}\right)$. Together with the second penalty term, this shows us that the proximal operator in \eqref{eq:proxop.applied} is separable and solving it is equivalent to solving the subproblems
\begin{equation}
\text{prox}_{s\lambda g_j}\left(\tilde{\bm\beta}_j^{(k)}\right) = \underset{\bm{z}_j}{\mathrm{argmin}} \;\; \frac12 \left|\left|\tilde{\bm\beta}_j^{(k)}-\bm{z}_j\right|\right|_2^2 + s\lambda g_j(\bm{z}_j) \;\; \text{ for } j \in \{0,\ldots, J\}. \label{eq:proxop.applied.separated}
\end{equation}
For each $j$, \eqref{eq:proxop.applied.separated} is now a classical regularized linear model that only involves one type of penalty. We can then use the available statistical and machine learning literature to solve the different POs efficiently, as explained in Section~\ref{sec:individualPO}. An appropriate value for the step size $s$ is determined using a backtracking approach as explained in Appendix~A.2.

\subsection{The SMuRF algorithm}\label{sec:individualPO}

We use the insights of Section \ref{sec:po} to build SMuRF, an algorithm that enables Sparse Multi-type Regularized Feature modeling. The critical point is to find the solution of the POs in \eqref{eq:proxop.applied.separated} for the different penalties discussed in Section \ref{sec:pentypes}. We briefly sketch our solvers for each penalty type below and provide an overview in Table~\ref{tab:pensolvers}. Appendix~A of the supplementary material provides all further details on the implementation of SMuRF.

\paragraph{PO Intercept.}\hspace*{-0.2cm}No penalty is applied to the intercept $\beta_0$. Therefore, the PO in \eqref{eq:proxop.applied.separated} is calculated with $g_0(\cdot) = 0$. The resulting PO then reduces to the identity operator:
\begin{equation*}
\text{prox}_{g_0}\left(\tilde{\beta}_0^{(k)}\right) = \tilde{\beta}_0^{(k)}.
\end{equation*}

\paragraph{PO Lasso and Group Lasso.}\hspace*{-0.2cm}\cite{parikh} show that the PO in \eqref{eq:proxop.applied.separated} has an analytic solution for the Lasso and Group Lasso penalties. The PO is partitioned per coefficient $\tilde{\beta}_{j,i}$ (Lasso) or per group of coefficients $\tilde{\bm\beta}_{j}$ (Group Lasso) and then the (group) soft thresholding operator gives the solution:
\begin{align}
\text{prox}_{s \lambda  g_{\text{Lasso}}}\left(\tilde{\beta}_{j,i}\right) &= \tilde{\beta}_{j,i}\left(1-\frac{w_{j,i}  s  \lambda}{|\tilde{\beta}_{j,i}|}\right)_{+} := \mathcal{S}\left(\tilde{\beta}_{j,i};w_{j,i}  s  \lambda\right),\label{eq:soft.thr}\\
\text{prox}_{s\lambda g_{\text{GroupLasso}}}\left(\tilde{\bm\beta}_{j}\right) &= \tilde{\bm\beta}_{j}\left(1-\frac{w_{j} s\lambda}{||\tilde{\bm\beta}_{j}||_2}\right)_{+} := \mathcal{S}_{\text{grp}}\left(\tilde{\bm\beta}_{j}; w_j s\lambda\right),\label{eq:grpsoft.thr}
\end{align}
where $(x)_+$ returns the maximum of $x$ and $0$. 

\paragraph{PO (Generalized) Fused Lasso.}\hspace*{-0.2cm}No analytic solution exists for the PO of the (Generalized) Fused Lasso. To solve \eqref{eq:proxop.applied.separated} for these penalties, we implement the Alternating Direction Methods of Multipliers (ADMM) algorithm of \cite{admmoriginal} and \cite{admmoriginal2} incorporating some minor adjustments suggested in \cite{zhu2017}. The ADMM algorithm has previously been used to solve similar Fused Lasso (\cite{admmfused}) as well as Trend Filtering (\cite{admmtrend}) problems. This algorithm can handle any penalty of the type $||M\bm\beta_j||_1$ with $M$ an arbitrary linear transformation. We refer to Appendix~B of the supplementary material for more details on our implementation.

\begin{table}[ht!]
    \begin{center}
    \begin{adjustbox}{max width=\textwidth}
    \begin{tabular}{p{4.5cm} p{2.7cm} p{4.5cm} p{6cm}}
    \toprule
    Penalty name & Formula $g(\bm\beta_j)$ & PO solver & Typical use\\
    \midrule
    (Adaptive) Lasso & $||\bm{w}_j * \bm\beta_j||_1$ & soft-thresholding per coefficient \newline $\mathcal{S}\left(\bm\beta_j; \bm{w}_j s\lambda\right)$ & selection of continuous and binary predictors \\
    \addlinespace[1em]
    (Adaptive) Group Lasso & $ ||w_{j} \bm\beta_{j}||_2$ & group soft-thresholding \newline $\mathcal{S}_{\text{grp}}\left(\bm\beta_j; w_j s\lambda\right)$ & selection of a group of parameters: all-in or all-out\\
    \addlinespace[1em]
    (Generalized) Fused Lasso & $||\bm{G}(\bm{w}_j)\bm\beta_j||_1$ & ADMM algorithm & binning of predictors incorporating the underlying structure specified by the graph\\
    \bottomrule
    \end{tabular}
    \end{adjustbox}
    \caption{\label{tab:pensolvers}Overview of the penalties implemented for sparse modeling with SMuRF. The implemented solver for each corresponding PO as well as the typical use of these penalties are given. $w_j$ and $\bm{w}_j$ represent the relevant penalty weights, as listed in Section~\ref{sec:weights}.}
    \end{center}
\end{table}

Having these efficient solvers available for all POs, we combine them into the SMuRF algorithm, of which the naive form is given in Algorithm~\ref{alg:mtpPGA}. We improve the computational efficiency of this naive version using techniques from optimization theory. Appendix~A of the supplementary material provides the full implementation details for these improvements and the convergence properties of the algorithm. The implementation of SMuRF is modular, allowing for straightforward extension to new penalties by implementing the solver of the accompanying PO. SMuRF has the same asymptotic properties as the base proximal gradient algorithm (\cite{parikh}) which converges to the optimal solution when the number of iterations $k$ goes to infinity. 

\begin{algorithm}
\caption{Naive SMuRF algorithm}\label{alg:mtpPGA}
{\fontsize{10}{20}\selectfont
\begin{algorithmic}[1]
\State \textbf{Input} $\bm{\beta}^{(0)}, \bm{X}, \bm{y}, s, \lambda$
\For{$k=1$ to $m$}
	\State $\tilde{\bm\beta} \gets \bm{\beta}^{(k-1)} - s\nabla f(\bm{\beta}^{(k-1)})$\label{alg1:gradupdate} \Comment gradient update
	\State $\left(\tilde{\beta}_0, \tilde{\bm\beta}_1, \ldots, \tilde{\bm\beta}_J\right) \gets \tilde{\bm\beta}$ \Comment partition full vector in components per predictor
	\State $\bm{\beta}^{(k)}_j \gets \text{prox}_{s\lambda g_j}\left(\tilde{\bm\beta}_j\right)$\label{alg1:proxupdate} \Comment calculate PO for all $j$ in $\{0,\ldots,J\}$
	\State $\bm\beta^{(k)} \gets \left(\beta_0^{(k)}, \bm{\beta}^{(k)}_1, \ldots, \bm{\beta}^{(k)}_J\right)$ \Comment recombine to full vector
\EndFor
\State \textbf{return} $\bm{\beta}^{(m)}$
\end{algorithmic}
}
\end{algorithm}

\subsection{Tuning \texorpdfstring{$\lambda$}{l} and re-estimation}\label{sec:tuning}

Algorithm \ref{alg:mtpPGA} works for a single input of the regularization parameter $\lambda$. However, it is difficult to find the correct value for $\lambda$ a priori. Therefore, we run the algorithm over a grid of values for $\lambda$ and evaluate the performance of the resulting predictive models using several criteria. Additionally, we use re-estimation when possible as in \cite{gertheiss2010} to reduce the bias of the regularized estimates.

\paragraph{Tuning $\lambda$}
We evaluate a performance criterion to determine an optimal value for $\lambda$. This criterion is then evaluated over a grid of $\lambda$ values and the preferred $\hat{\lambda}$ is chosen such that the criterion is optimal. We focus on typical criteria used in a GLM context, such as the Akaike (AIC - \cite{AIC}) and Bayesian (BIC - \cite{BIC}) information criteria used for in-sample training, or the mean squared prediction error (MSPE) and the Dawid-Sebastiani scoring rule (DSS - \cite{DSS}) for out-of-sample training. The DSS is a proper scoring rule, developed for comparing predictive models, measuring both the accuracy and the sharpness of the predictions. See Table~\ref{tab:criteria} for a list of these criteria.
\begin{table}[ht!]
    \begin{center}
    \begin{adjustbox}{max width=\textwidth}
    \begin{tabular}{p{1.5cm} p{5cm} p{1.5cm} p{5cm}}
    \toprule
    Name & Formula & Name & Formula\\
    \midrule
    AIC & $-2\log \mathcal{L} + 2d$ & deviance & $-2\log \mathcal{L}$   \\
      \addlinespace[0.5em]
    BIC & $-2\log \mathcal{L} + \log(n)d$ & MSPE & $\frac1n \sqrt{\sum_{i=1}^n \left(y_i-\mu_i\right)^2}$ \\
        \addlinespace[0.5em]

     & & DSS & $\sum_{i=1}^n \left(\frac{y_i-\mu_i}{\sigma_i}\right)^2 + 2\log(\sigma_i)$\\
%
    \bottomrule
    \end{tabular}
    \end{adjustbox}
    \caption{\label{tab:criteria}Overview of performance criteria with $\mathcal{L}$ the likelihood of the model, $n$ the number of observations, $d$ the degrees of freedom of the model and $y_i$, $\mu_i$ and $\sigma_i$ the observed response, the predicted value and its standard deviation for observation $i$. The degrees of freedom $d$ are estimated as the number of unique, non-zero coefficients.}
    \end{center}
\end{table}

In Section~\ref{sec:simulation}, we compare the in-sample tuning of $\lambda$ using the AIC or BIC with the out-of-sample tuning, based on evaluating the deviance, MSPE or DSS score on a test data set. Additionally, we evaluate the latter using stratified $K$-fold cross-validation (as in \cite{stratcv}). This strategy partitions the data into $K$ disjoint sets (or: folds) such that each level of the response is equally represented in each set. For every fold, the model performance (e.g.~using one of the criteria from Table~\ref{tab:criteria}) is then evaluated on that fold after training on the $k-1$ other folds. The optimal $\lambda$ then minimizes this performance criterion (e.g.\ the average deviance or mean squared error over all folds) or is the highest $\lambda$ for which the performance criterion is within one standard deviation of this minimum. This last strategy is refered to as the `one standard error rule' in \cite{bookhastie}.

\paragraph{Re-estimation}
As with most regularization methods, the finite sample coefficient estimates and predictions obtained with the fitted model will be biased. To reduce this bias, we propose to re-estimate the model without penalties, but with a reduced model matrix $\bm{\tilde{X}}$, based on the parameter estimates obtained with SMuRF. Hereto we remove the columns of $\bm{X}$ for which the coefficients are estimated to be 0, and collapse the columns for which the coefficient estimates are fused. The re-estimated coefficients will thus have the same non-zero and fused coefficients as the regularized estimates, but will not be biased. This approach is closely related to the idea of the LARS-OLS hybrid of \cite{lars} which can be interpreted as a special case of the Relaxed Lasso from \cite{relaxedLasso}.

\section{Simulation study}\label{sec:simulation}

\subsection{Set-up}

We carefully evaluate the performance of the SMuRF algorithm with a simulation study. We model credit worthiness of customers in the presence of 7 assumed predictors and an interaction effect, based on the case study of \cite{creditscore}. Table \ref{tab:variablesSIM} lists the predictors and their levels.

\begin{table}[ht!]
    \begin{center}
    \begin{adjustbox}{max width=\textwidth}
    \begin{tabular}{p{2cm} p{1.85cm} p{9cm}}
    \arrayrulecolor{black}\toprule
    Type & Name & Description\\
	\arrayrulecolor{black}\midrule
	Response & \texttt{paid} & credit worthiness, used as response: 1 if all payments were made on time and 0 otherwise.\\
	\arrayrulecolor{lightgray}\midrule
    Ordinal & \texttt{age} & Age of the customer: 20-70.\\
\addlinespace[1em]
    & \texttt{stability} & Consecutive time, in years, spent with current job/employer: 0-20.\\
\addlinespace[1em]
    & \texttt{salary} & Monthly net income of customer in EUR, rounded to the nearest 100: 1000-5000.\\
\addlinespace[1em]
    & \texttt{loan} & Monthly loan payment, in EUR, rounded to the nearest 100: 100-3000.\\
    \arrayrulecolor{lightgray}\midrule
    Binary & \texttt{sex} & Sex of the customer: \texttt{female} or \texttt{male}.\\
    \arrayrulecolor{lightgray}\midrule
    Nominal & \texttt{prof} & Profession of the customer, coded in 10 levels.\\
            & \texttt{drink} & Type of drink customer had during acceptance interview, coded in 5 levels.\\
    \arrayrulecolor{lightgray}\midrule
    Interaction & \texttt{salxloan} & Interaction effect between the \texttt{salary} and \texttt{loan} predictors.\\
    \arrayrulecolor{black}\bottomrule
    \end{tabular}
    \end{adjustbox}
    \caption{\label{tab:variablesSIM}Overview of the response, the predictors and their levels used in the simulated data sets.}
    \end{center}
\end{table}

\begin{figure}[!ht]
\centering
\includegraphics[width = 1\textwidth]{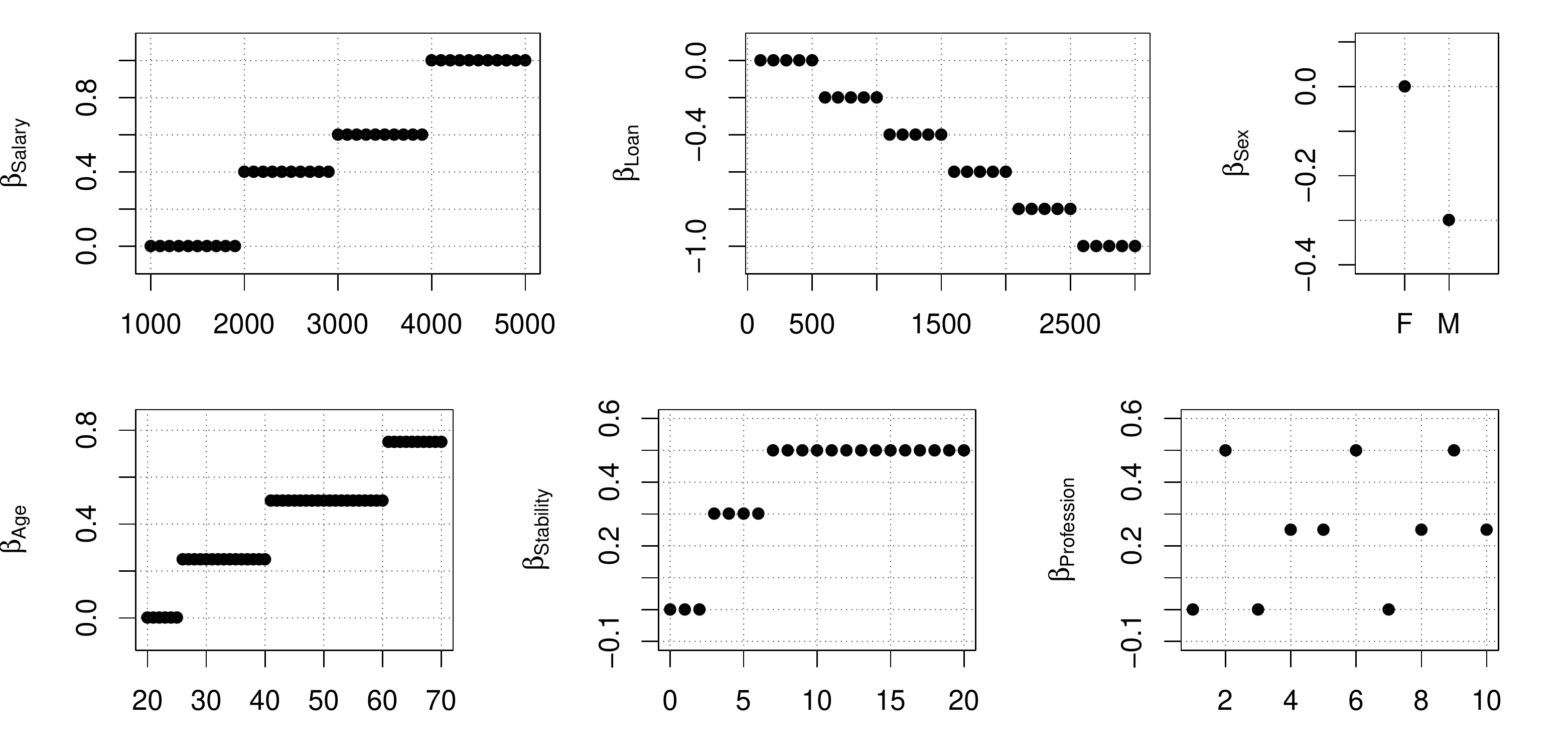}
\caption{True parameter values of the main effects used in the simulated data sets.}
\label{fig:simbeta1}
\end{figure}

\begin{figure}[!ht]	
	\centering
	\begin{subfigure}[t]{0.55\textwidth}
		\centering
		\includegraphics[width=\textwidth]{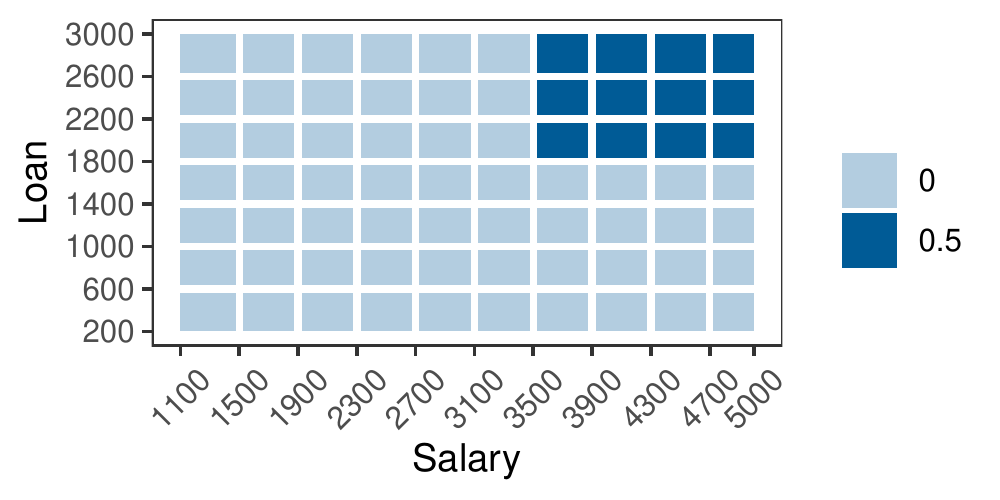}
		\caption{}\label{fig:simbeta2a}		
	\end{subfigure}
	\hspace{-0.5cm}
	\begin{subfigure}[t]{0.44\textwidth}
		\centering
		\includegraphics[width=\textwidth]{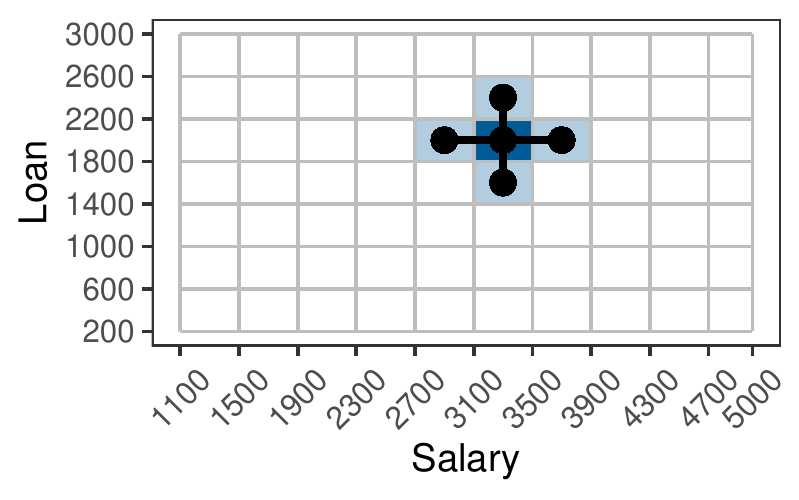}
		\caption{}\label{fig:simbeta2b}
	\end{subfigure}
	\caption{(a) True parameter values of the \texttt{salxloan} interaction used in the simulated data sets, and (b) 2-D layout of the regularized differences for one coefficient of the interaction effect.}\label{fig:simbeta2}
\end{figure}

For each observation $i$, the response $\texttt{paid}_i$ is simulated from a binomial distribution using the credit worthiness score $p_i = 1/(1+\exp(-\bm{x}_i\bm\beta))$ as its mean, with $\bm{x}_i$ the row vector with the predictor information for observation $i$ and $p_i$ denoting the probability of observation $i$ paying on time. Figure~\ref{fig:simbeta1} and Figure~\ref{fig:simbeta2a} show the values of the true coefficients $\bm\beta_j$ for most predictors $j$, used for simulating the response. The specification of the levels as well as the coefficient values for all $\bm\beta_j$ are based on the findings of \cite{creditscore}. For the predictor \texttt{drink}, all 5 coefficients are set to 0, indicating that this predictor has no predictive value. Appendix~C of the supplementary material lists the individual true coefficient values used. This parameter setup implies that the default risk declines for older, more stable and higher earning customers while it increases for higher loan sizes. The 10 professions are effectively fused into 3 categories (for example: blue-collar, white collar and others). The interaction effect in Figure~\ref{fig:simbeta2a} has to be interpreted on top of the main \texttt{salary} and \texttt{loan} effects. The credit worthiness of customers with a high income ($\geq$3,500 EUR) decreases less with high loan sizes ($\geq$2,000 EUR) compared to lower income clients.

The goal of the simulation study is to evaluate whether the SMuRF algorithm is capable of effectively fusing the coefficients into groups as displayed in Figures~\ref{fig:simbeta1} and \ref{fig:simbeta2a}. Therefore, we start from a highly over-parameterized setting where each level within a predictor, indicated by a dot in Figure~\ref{fig:simbeta1}, gets its own parameter. The interaction effect in Figure~\ref{fig:simbeta2a} is divided into a $7\times10$ grid amounting to an extra 70 parameters, one for each cell in the grid. We fix the true intercept at $\beta_0=0$ to ensure that around 30\% of the observed customers have late payments. The simulation is performed with balanced design for each predictor except for \texttt{stability} which should not be higher than $\texttt{age}-18$, since a customer can only start working from age 18 onwards. This results in a slightly higher prevalence of observations with lower stability. The negative scaled binomial log-likelihood $f$ for this setup results to
\begin{equation}
f(\bm\beta; \bm{X}, \texttt{paid}) = -\frac1n \sum_{i=1}^n \Big( \texttt{paid}_i \ \bm{x}_i\bm\beta - \log\left( 1+\exp(\bm{x}_i\bm\beta)\right) \Big).
\end{equation}
Using this setup, we simulate 100 times a data set of 80,000 observations and a single hold-out data set of 20,000 observations to be used for evaluating the performance of the models after the training and tuning process.

\subsection{Model settings}\label{sec:simsettings}

We apply a standard Lasso penalty to \texttt{sex}, a Fused Lasso penalty to \texttt{age}, \texttt{stability}, \texttt{salary} and \texttt{loan}, a Generalized Fused Lasso, regularizing all differences, to \texttt{profession} and a 2D Fused Lasso penalty to the interaction effect. This 2D Fused Lasso regularizes differences of parameters corresponding to cells which are directly left to, right to, under or above each other, as illustrated in Figure~\ref{fig:simbeta2b}. We assume a priori that drink has no predictive power and apply a Group Lasso penalty to this predictor. This encourages all coefficients of \texttt{drink} to be removed from the model simultaneously. We code the first level, as given in Table~\ref{tab:variablesSIM}, of \texttt{age}, \texttt{stability}, \texttt{salary}, \texttt{loan}, \texttt{sex} and \texttt{prof} as reference category. Despite applying a Lasso penalty, we choose a reference level for \texttt{sex}, as it is a binary predictor (see Section~\ref{sec:pentypes}). We adopt no reference category for \texttt{drink} because of the Group Lasso penalty.

The resulting full parameter vector $\bm\beta$ (including the intercept $\beta_0$) has length 225. In comparison, the true coefficient vector has 17 unique, non-zero groups of coefficients. For each generated data set, we apply the SMuRF algorithm with different settings \texttt{w|t}, where \texttt{w} denotes the setting for the penalty weights and \texttt{t} denotes the $\lambda$ tuning criterion. Table~\ref{tab:simsettings} lists all investigated combinations. For the adaptive penalty weights $\bm{w}_j^{\text{ad}}$, we use an initial estimate $\hat{\bm{\beta}}$ from a binomial GLM including a very small ridge penalty to make the model identifiable. The GLM with ridge penalty is treated as the baseline setting for performance comparison and is further denoted by \texttt{GLM.ridge}.

We compare the different techniques for tuning $\lambda$ explained in Section~\ref{sec:tuning}. A first approach fits the model on all observations and minimizes the in-sample AIC or BIC. We call this the `in-sample approach'. A second strategy splits each simulated data set into a training set of size $60,000$, used to fit the model for different values of $\lambda$, and a validation set of size 20,000. The deviance, MSPE or DSS statistic is then calculated on the validation set and minimized to tune $\lambda$. We refer to this as the `out-of-sample' approach. We also perform 10-fold stratified cross-validation with the deviance as measure of fit, with and without the one standard error rule. After the tuning of $\lambda$, we re-estimate the coefficients using the strategy outlined in Section~\ref{sec:tuning}.

\begin{table}[ht!]
    \begin{center}
    \begin{adjustbox}{width=\textwidth}
    \begin{tabular}{p{5cm} p{2cm} p{6cm} p{3cm}}
    \arrayrulecolor{black}\toprule
    Penalty weight settings & \texttt{ w|t} & Tuning settings & \texttt{w|t}\\
	\arrayrulecolor{black}\midrule
	equal weights \newline $\bm{w}_j = \bm{1}$ & \texttt{eq|t} & fit on whole data,\newline minimize AIC/BIC & \texttt{w|in.AIC}\newline \texttt{w|in.BIC}\\
\addlinespace[1em]
	GLM adaptive weights \newline $\bm{w}_j = \bm{w}^{\text{ad}}_j$& \texttt{ad|t} & fit on training sample,\newline minimize the deviance/\newline MSPE/DSS on validation sample & \texttt{w|out.dev}\newline \texttt{w|out.MSPE}\newline \texttt{w|out.DSS}\\
\addlinespace[1em]
	standardization weights \newline $\bm{w}_j = \bm{w}^{\text{st}}_j$& \texttt{st|t} & 10-fold stratified CV with deviance as measure of fit & \texttt{w|cv}\\
\addlinespace[1em]
	combined weights \newline $\bm{w}_j = \bm{w}^{\text{ad}}_j\cdot\bm{w}^{\text{st}}_j$& \texttt{ad.st|t} & 10-fold stratified CV\newline with one standard error rule and deviance as measure of fit & \texttt{w|cv.1se}\\
    \arrayrulecolor{black}\bottomrule
    \end{tabular}
    \end{adjustbox}
    \caption{\label{tab:simsettings}List of settings tested in the simulation study.}
    \end{center}
\end{table}

\subsection{Results and discussion}

\begin{figure}[!ht]
\centering
\begin{subfigure}[t]{\textwidth}
		\centering
\includegraphics[width = \textwidth]{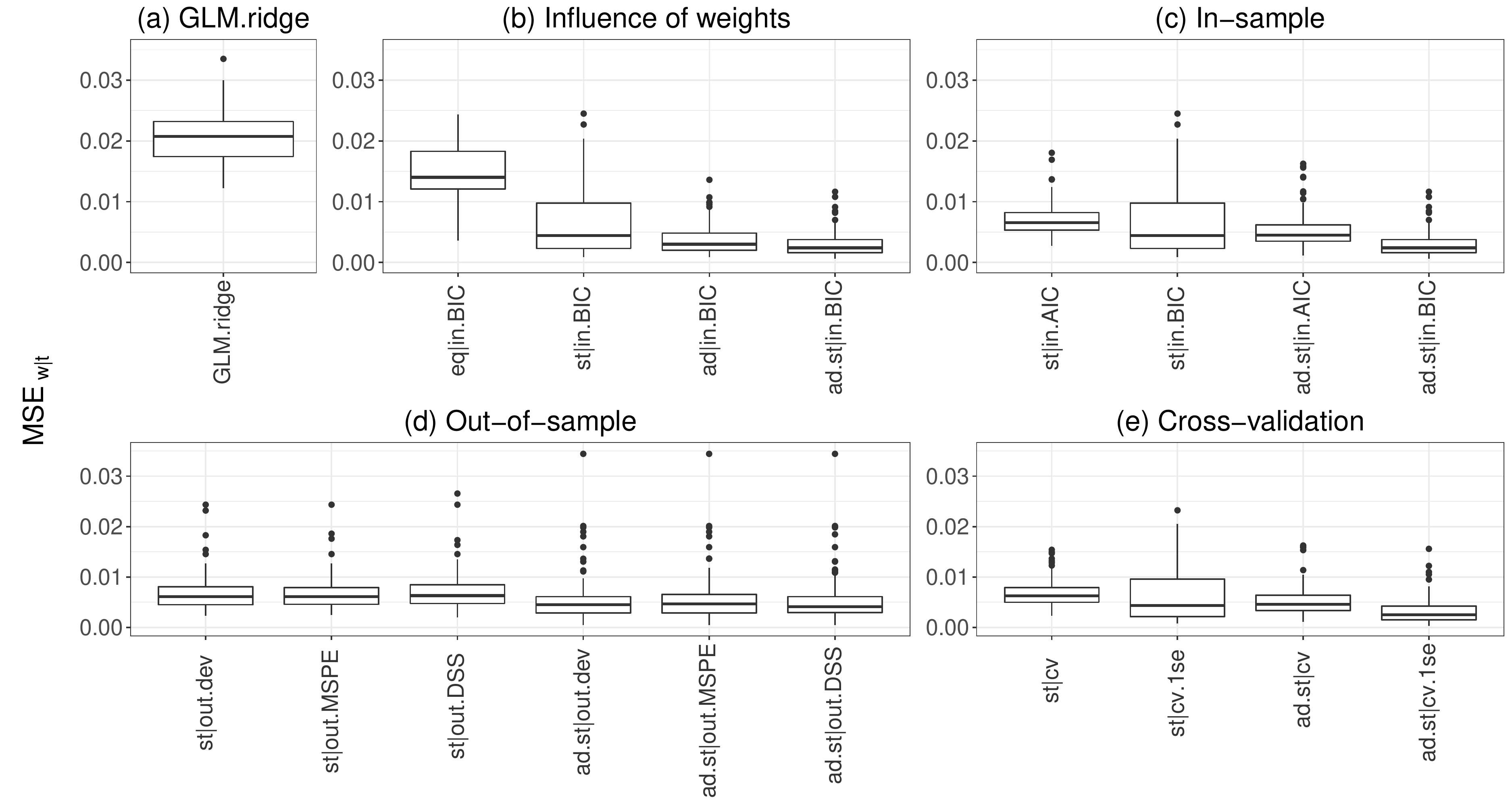}
\phantomsubcaption\label{fig:simmsea}
\phantomsubcaption\label{fig:simmseb}
\phantomsubcaption\label{fig:simmsec}
\phantomsubcaption\label{fig:simmsed}
\phantomsubcaption\label{fig:simmsee}
\end{subfigure}
\caption{Boxplot of $\text{MSE}_{\texttt{w|t}}$ (a) for a standard binomial GLM with a small ridge penalty, (b)-(e) for different settings of SMuRF. Calculations based on 100 simulated data sets.}\label{fig:simmse}
\end{figure}

\paragraph{Coefficient estimation error.}\hspace{-0.2cm}For each setting \texttt{w|t} and each simulated data set $l$, we obtain the parameters $\hat{\bm{\beta}}^{[l]}_{\texttt{w|t}}$ after re-estimation. Firstly, we calculate the mean squared error (MSE) of the re-estimated coefficients $\hat{\bm{\beta}}^{[l]}_{\texttt{w|t}}$ with respect to the true parameter vector $\bm{\beta}$:
\begin{equation}
\text{MSE}^{[l]}_{\texttt{w|t}} = \frac{1}{225}||\bm{\beta} - \hat{\bm{\beta}}^{[l]}_{\texttt{w|t}}||_2^2.
\end{equation}
Figure~\ref{fig:simmse} shows boxplots of the MSE over all simulated data sets for a selection of the settings investigated in Section~\ref{sec:simsettings}. We give the full series of boxplots for all combinations of penalty weight and tuning settings in Appendix~C of the supplementary material. The best settings will have the lowest median MSE with a small box around the median. The \texttt{GLM.ridge} baseline in Figure~\ref{fig:simmsea} performs worse than all the regularized settings. Figure~\ref{fig:simmseb} illustrates the influence of the different penalty weights on the performance of our algorithm. The combined penalty weights perform best with the adaptive weights as a close second. The use of standardization weights is a great improvement over the equal weights setting, but less so than using adaptive or combined weights. This conclusion is independent from the applied tuning method. When selecting $\lambda$ with the in-sample approach, the BIC bests the AIC criterion as shown in Figure~\ref{fig:simmsec}, indicating that for in-sample tuning, the AIC is too lenient towards the model degrees of freedom. However, the larger inter-quartile and whisker range for \texttt{st|in.BIC} indicate that the BIC can be too strict when using the standardized penalty weights, removing or fusing too many coefficients. When using out-of-sample tuning (Figure~\ref{fig:simmsed}), the three criteria perform similarly well. Figure~\ref{fig:simmsee} illustrates that in our simulation study, cross-validation performs best with the one standard error rule, for the different penalty weight settings.

\begin{figure}[!ht]
\centering
\begin{subfigure}[t]{\textwidth}
		\centering
\includegraphics[width = \textwidth]{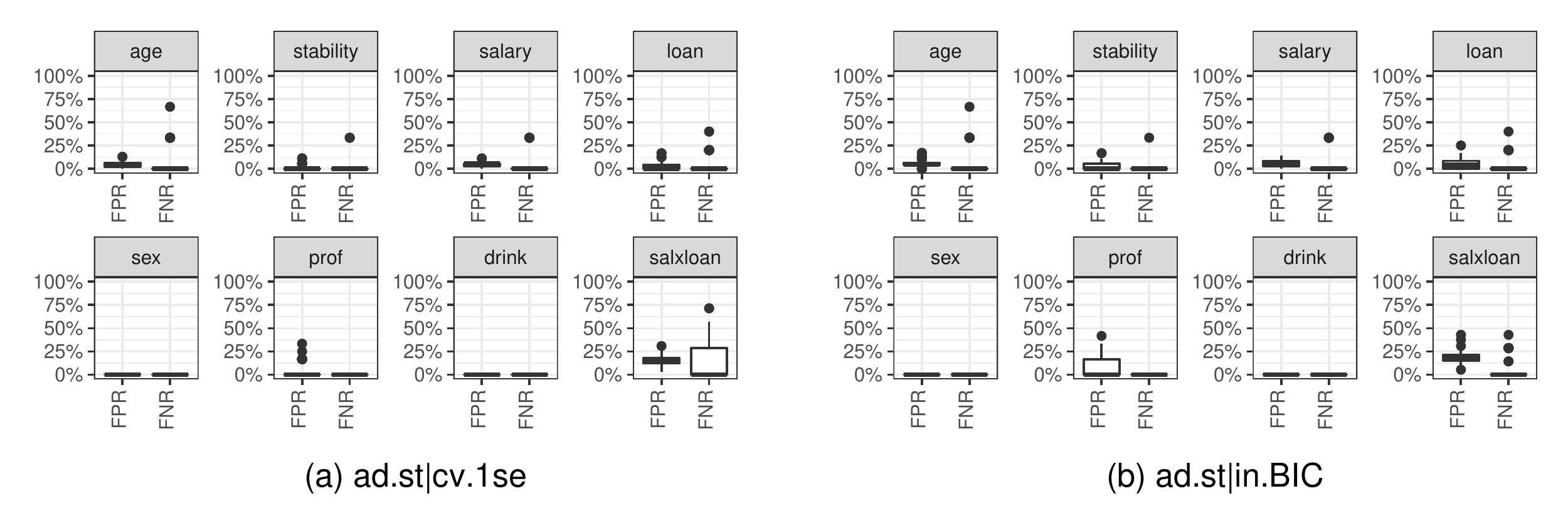}
\phantomsubcaption\label{fig:simfpra}
\phantomsubcaption\label{fig:simfprb}
\end{subfigure}
\caption{FPR and FNR per predictor for different algorithm settings: (a) \texttt{ad.st|cv.1se} and (b) \texttt{ad.st|in.BIC}.}
\label{fig:simfpr}
\end{figure}

\paragraph{Selection and fusion error.}\hspace{-0.2cm}Secondly, we evaluate the selection and fusion properties of our algorithm. We calculate the False Positive Rate (FPR) and False Negative Rate (FNR) of the estimated versus the true coefficient vector, similar to \cite{gertheiss2010}. For predictors \texttt{age}, \texttt{stability}, \texttt{salary}, \texttt{loan}, \texttt{prof} and \texttt{salxloan}, a false positive means that a truly zero regularized coefficient difference is estimated to be non-zero and vice versa for a false negative. The FPR is then the ratio between the number of false positives and the total number of truly zero coefficient differences and the FNR is defined analogously. The FPR for \texttt{sex} is always zero as the true parameter value is non-zero while the FNR is one when the parameter is estimated to be zero. Likewise, for \texttt{drink} the FNR is always zero and the FPR is one when all coefficients have non-zero estimates. Because a GLM estimates all 225 coefficients and their relevant differences to be non-zero, the FPR and FNR is always one and zero respectively, except for \texttt{sex} where the FPR is also zero. Figure~\ref{fig:simfpr} shows boxplots of the FPR and FNR per predictor over all simulations and a selection of the settings discussed in Section~\ref{sec:simsettings}. We display \texttt{ad.st|cv.1se} and \texttt{ad.st|in.BIC} which have similar MSE scores in Figure~\ref{fig:simmse}. Appendix~C from the supplementary material provides similar plots for the other settings. In general, the FPR and FNR are small for all predictors, indicating that the algorithm is able to correctly predict the clusters present in the true parameter vector $\bm\beta$. The clusters for the interaction effect \texttt{salxloan} are slightly more difficult to estimate due to the many parameters involved. Compared to \texttt{ad.st|in.BIC}, \texttt{ad.st|cv.1se} has a better FPR for \texttt{stability}, \texttt{loan}, \texttt{prof} and \texttt{salxloan} and comparable for the other predictors. In contrast \texttt{ad.st|in.BIC} only has a better FPR for \texttt{salxloan}. 

\begin{figure}[!ht]
\centering
\begin{subfigure}[t]{\textwidth}
		\centering
\includegraphics[width = \textwidth]{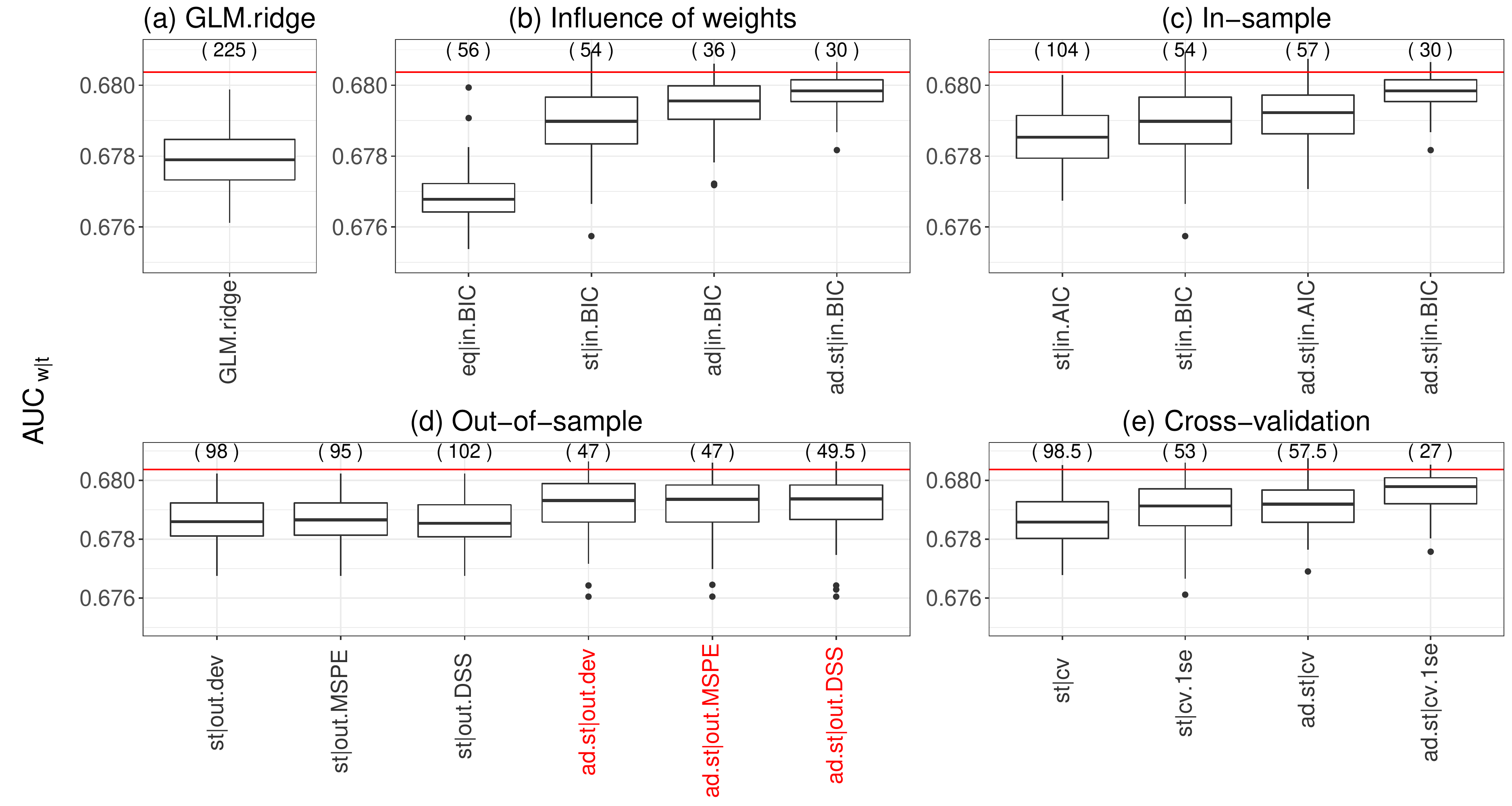}
\phantomsubcaption\label{fig:simauca}
\phantomsubcaption\label{fig:simaucb}
\phantomsubcaption\label{fig:simaucc}
\phantomsubcaption\label{fig:simaucd}
\phantomsubcaption\label{fig:simauce}
\end{subfigure}
\caption{Boxplot of $\text{AUC}_{\texttt{w|t}}$ (a) for a standard binomial GLM approach, (b)-(e) for different settings of SMuRF. Calculations based on 100 simulated data sets. The red lines denote the optimal AUC, reached by a binomial GLM with true coefficients $\bm\beta$. The numbers in brackets denote the median number of unique coefficients for each setting, over all simulated data sets. For plot (d) one outlier is not visible for the red settings.}
\label{fig:simauc}
\end{figure}

\paragraph{Prediction error.}\hspace{-0.2cm}Thirdly, we evaluate the predictive performance of the calibrated models on the hold-out data set of 20,000 observations. We use the re-estimated coefficients to obtain predictions of the credit worthiness on the hold-out observations, for varying cut-off points. We construct the Receiver Operating Characteristic (ROC) and calculate the Area Under this Curve (AUC) for every setting \texttt{w|t} and every simulated data set $l$. The ROC and AUC are standard tools to evaluate the performance of binary classification models, see \cite{roc}, with the AUC reaching 1 for a perfect classifier. Figure~\ref{fig:simauc} shows the performance in terms of AUC as obtained with a selection of different \texttt{w|t} settings. Graphs showing boxplots of the AUC for all possible settings are in Appendix~C of the supplementary material. The median number of unique, non-zero estimated coefficients is between brackets and should be compared to the 17 degrees of freedom from the true model. The ridge GLM always uses the maximum of 225 different coefficients. The red lines in Figure~\ref{fig:simauc} correspond to the AUC calculated from the true parameter vector $\bm\beta$. Again, the \texttt{GLM.ridge} baseline in Figure~\ref{fig:simauca} performs worse compared to most regularization settings, although the differences in terms of AUC are smaller than those observed for MSE. Additionally, the regularized settings use a substantially lower degrees of freedom compared to \texttt{GLM.ridge}. The performance of SMuRF is again improved by using standardization or adaptive penalty weights compared to the equal weights setting, as Figure \ref{fig:simaucb} illustrates. The combined weights result in both the lowest number of unique estimated parameters as well as the best AUC measurements. Figures~\ref{fig:simaucc} and \ref{fig:simaucd} show that the BIC has the best AUC for in-sample selection while for out-of-sample tuning, all settings work equally well. For the cross-validation in Figure \ref{fig:simauce}, the one standard error rule with combined weights again performs best.
\paragraph{Comparison with PIRLS.}In addition to the above simulation study, we performed a comparison between the performance of the PIRLS algorithm proposed by \cite{gertheiss2017}, implemented in the R package \texttt{gvcm.cat}, and the SMuRF algorithm implemented through the \texttt{smurf} package. The detailed results of this comparison can be found in Appendix~C.3. We conclude from this comparison that SMuRF outperforms the PIRLS approach via its efficient implementation strategy. SMuRF achieves a requested level of accuracy faster than the PIRLS implemented in gcvm.cat does. This is particularly advantageous with big data sets.
\\

We conclude from this simulation study that the multi-type regularization strategy greatly improves parameter estimation accuracy and prediction performance compared to a standard GLM approach. The combined adaptive and standardization penalty weights perform best overall. We do not recommend the in-sample criteria to tune $\lambda$, since calibrating and tuning a model on the same data set leads to biased results (see for example \cite{dmbook}). The stratified $K$-fold cross-validation with the one standard error rule performs best across all investigated criteria and at the same time requires the lowest degrees of freedom, compared to all other settings investigated. Hence, we consider \texttt{ad.st|cv.1se} as the superior setting based on this simulation study.

\section{Case study: sparse modeling of count data}\label{sec:MTPLcase}

We consider a data set on Belgian motor third party liability claims, previously analyzed in \cite{denuit2004}, \cite{klein2014} and \cite{henckaerts}. We refer to this as the \texttt{MTPL} data. Our goal is to predict the number of claims a policyholder will file to the insurer, proportional to the length of the insured period. Since the number of claims is an integer response, we opt for a Poisson GLM where the mean is $\text{E}[\texttt{nclaims}] = \bm\mu$ and the link function is the natural logarithm:
\begin{equation}
\log(\bm\mu) = \bm{X}\bm\beta + \log(\texttt{expo}).
\end{equation}

\subsection{Data description}\label{sec:MTPLdescription}

\begin{figure}[!ht]
\includegraphics[width = \textwidth]{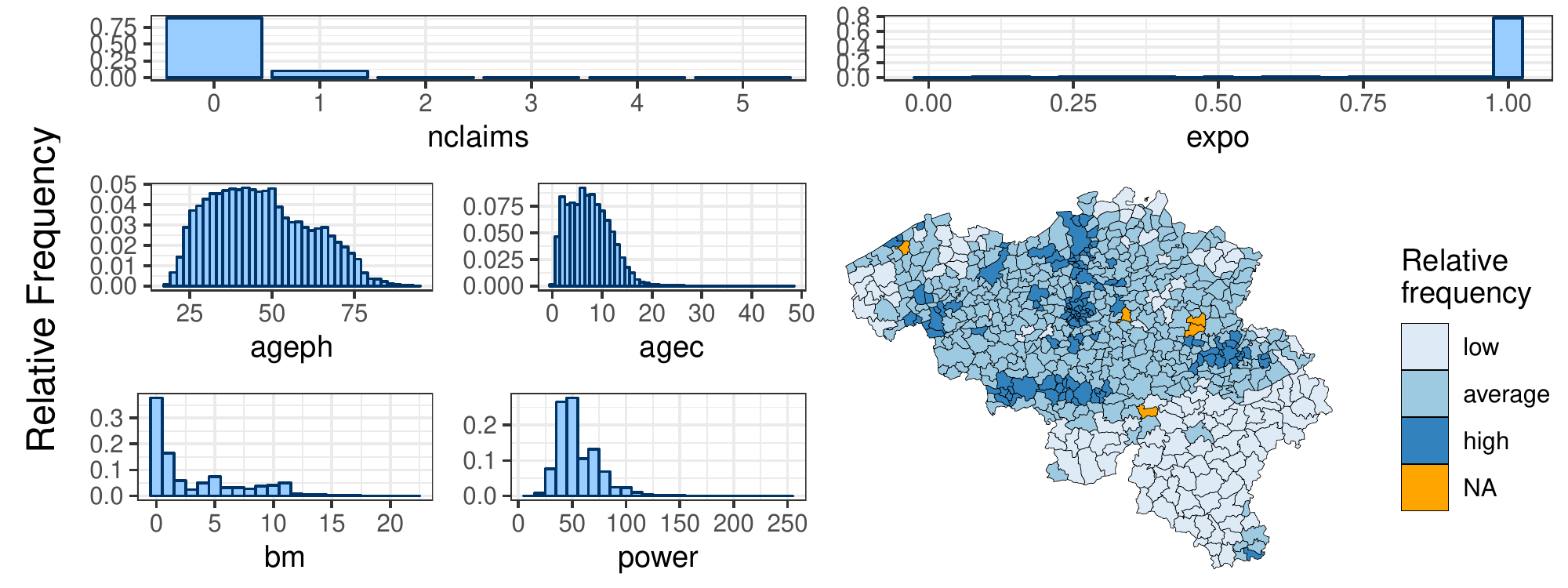}
\caption{Histograms and barplots of \texttt{nclaims}, \texttt{expo},  the ordinal and the spatial predictors in the \texttt{MTPL} data set. The relative frequencies for the spatial information is respectively set to low/high for the municipalities with the 20\% lowest/highest exposure and average otherwise.}
\label{fig:MTPL.rel.freq1}
\end{figure}

\begin{figure}[!ht]
\includegraphics[width = \textwidth]{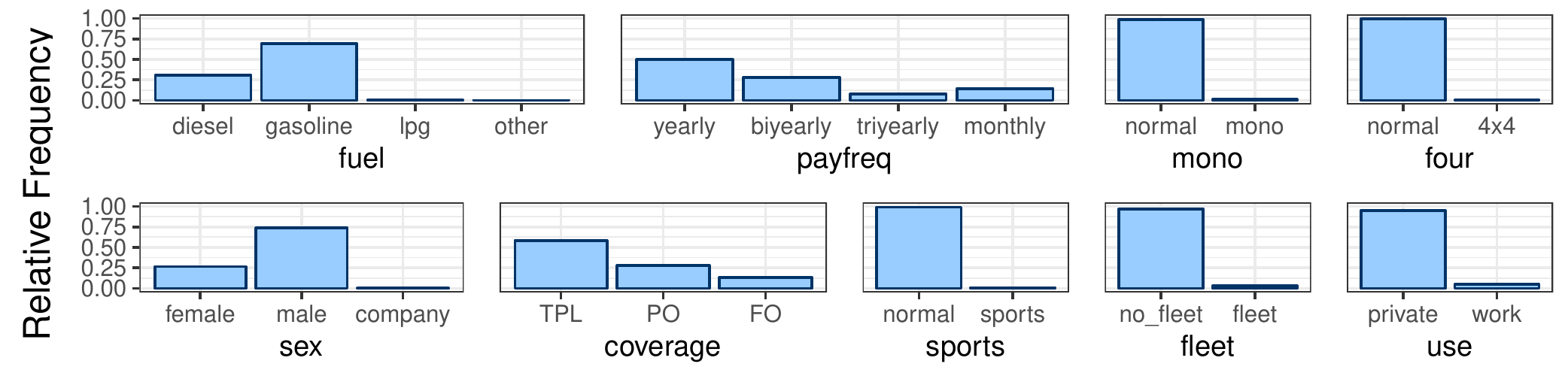}
\caption{Barplots of the binary and nominal predictors in the \texttt{MTPL} data set.}
\label{fig:MTPL.rel.freq2}
\end{figure}

The \texttt{MTPL} data set contains information on 163,660 policyholders from a Belgian insurer in 1997. Each policyholder is observed during an insured period ranging from one day to one year, further denoted as the exposure variable \texttt{expo}. During this period the policyholder is exposed to the risk of being involved in an accident and able to file a claim to the insurer. Policyholders are further distinguished through a set of personal as well as vehicle characteristics displayed in Table~\ref{tab:variablesMTPL}. The aforementioned papers remove some predictors from this data set a priori, such as \texttt{mono}, \texttt{four}, \texttt{sports} and \texttt{payfreq}. We keep these in our analysis and use the data-driven SMuRF algorithm to determine their predictive power.
\begin{table}[H]
    \begin{center}
    \begin{tabular}{p{2cm} p{1.6cm} p{9cm}}
    \arrayrulecolor{black}\toprule
    Type & Name & Description\\
	\arrayrulecolor{black}\midrule
	Response & \texttt{nclaims} & Observed number of claims for the policyholder, used as response: 0-5.\\
	Exposure & \texttt{expo} & Fraction of the year that the policy was active: 0-1.\\
	\arrayrulecolor{lightgray}\midrule
    Ordinal & \texttt{ageph} & Age of the policyholder in whole years: 17-95.\\
    & \texttt{agec} & Age of the insured vehicle in whole years: 0-48.\\
    & \texttt{bm} & Bonus malus level of the policyholder: 0-22, a higher level indicates a worse claim history.\\
    & \texttt{power} & Power of the car in kW: 10-243.\\
    \arrayrulecolor{lightgray}\midrule
    Spatial & \texttt{muni} & Municipality of the policyholder's residence: 589 levels. \\
   \arrayrulecolor{lightgray}\midrule
    Binary & \texttt{use} & Use of the car: \texttt{private} or \texttt{work}.\\
    & \texttt{fleet} & The insured vehicle is part of a fleet: \texttt{no\_fleet} or \texttt{fleet}.\\
    & \texttt{mono} & The insured vehicle is a monovolume: \texttt{normal} or \texttt{mono}.\\
    & \texttt{four} & The insured vehicle has four-wheel drive: \texttt{normal} or \texttt{4x4}.\\
    & \texttt{sports} & The insured vehicle is a sports car: \texttt{normal} or \texttt{sports}.\\
    \arrayrulecolor{lightgray}\midrule
    Nominal & \texttt{coverage} & Coverage type provided by the insurance company: \texttt{TPL}, \texttt{PO} or \texttt{FO}. \newline
    \texttt{TPL}: only third party liability \newline
    \texttt{PO}: partial omnium = \texttt{TPL} + partial material damage \newline
    \texttt{FO}: full omnium = \texttt{TPL} + comprehensive material damage\\
    & \texttt{payfreq} & Payment frequency of the premium: \texttt{yearly}, \texttt{biyearly}, \texttt{triyearly} or \texttt{monthly}\\
    & \texttt{sex} & Sex of the policyholder: \texttt{female}, \texttt{male} or \texttt{company}.\\
    & \texttt{fuel} & Fuel type: \texttt{diesel}, \texttt{gasoline}, \texttt{lpg} or \texttt{other}.\\
    \arrayrulecolor{black}\bottomrule
    \end{tabular}
    \caption{\label{tab:variablesMTPL}Overview of the response, the exposure and the predictors in the \texttt{MTPL} data set.}
    \end{center}
\end{table}
Figure~\ref{fig:MTPL.rel.freq1} shows the histograms and barplots of the response, the exposure, the spatial and the ordinal predictors in the data set. The response \texttt{nclaims} denotes the number of claims filed to the insurer during the exposure period. Figure~\ref{fig:MTPL.rel.freq2} displays the bar plots of the binary and nominal predictors. For more detailed information on the predictors and a more thorough exploratory data analysis, we refer to \cite{henckaerts}.

\subsection{Data cleaning and preparation}

We remove observations which have zero exposure. For the nominal predictors, we remove observations for which \texttt{fuel} is registered as \texttt{lpg} or \texttt{other} as well as observations having \texttt{company} as \texttt{sex}, since these levels contain very few observations. Both of these predictors are thus effectively reduced to binary predictors. Additionally, some predictor levels contain few observations, such as very young or old policyholder ages, or municipalities in the more rural south of Belgium. We require the aggregated exposure in each level of a predictor to exceed 250. This enhances the reliability of the initial GLM fit, and therefore of the adaptive penalty weights resulting from this fit. We bin ordinal predictors by fusing levels without enough exposure with their neighboring level that has the smallest aggregated exposure. The resulting levels of the ordinal predictors are given in Table~\ref{tab:mtpllevels}.
\begin{table}[ht!]
    \begin{center}
    \begin{adjustbox}{max width=\textwidth}
    \begin{tabular}{p{2cm} p{14cm} }
    \arrayrulecolor{black}\toprule
     Predictor & Levels \\
	\arrayrulecolor{black}\midrule
	 \texttt{ageph} & $\{\leq20,21,22,\ldots,81,82-83,\geq84\}$ \\
\addlinespace[1em]
	\texttt{agec} & $\{0-1,2,3,\ldots,19,20-21,\geq22\}$\\
\addlinespace[1em]
	\texttt{bm} &  $\{0,1,\ldots,15,\geq16\}$\\
	 \addlinespace[1em]
\texttt{power} &  $\{\leq22,24-26,28,30,\ldots,104,106-108,110,112-122,124-140,\geq140\}$\\
    \arrayrulecolor{black}\bottomrule
    \end{tabular}
    \end{adjustbox}
    \caption{\label{tab:mtpllevels}List of levels of the ordinal predictors of the \texttt{MTPL} data set after data cleaning.}
    \end{center}
\end{table}
For the spatial predictor \texttt{muni}, we represent the municipalities as polygons on a two-dimensional map of Belgium. If the exposure within one municipality is below 250, we fuse that municipality with the neighbor municipality containing the smallest aggregated exposure. The polygon of this neighbor municipality must have at least one common edge with the polygon of the municipality under consideration. This procedure results in 266 levels, illustrated in Figure~\ref{fig:mtplspatial}. This data cleaning operation particularly fuses municipalities in the more rural south of Belgium. We split the \texttt{MTPL} data into a training data set of $n = 130,587$ observations and a hold-out test set with $n_{\text{test}} = 32,647$ observations.

\subsection{Model settings}

The scaled negative Poisson log-likelihood is used as the loss function $f$:
\begin{equation}
f(\bm\beta; \bm{X}, \bm{y}) = -\frac1n \sum_{i=1}^n \Big( y_i \left(\bm{x}_i \bm\beta + \log(\texttt{expo}_i)\right) - \exp \left(\bm{x}_i \bm\beta + \log(\texttt{expo}_i)\right) - \log(y_i!) \Big),
\end{equation}
where $y_i$ is the observed number of claims during the insured period $\texttt{expo}_i$. We apply a standard Lasso penalty to the binary predictors \texttt{use}, \texttt{fleet}, \texttt{mono}, \texttt{four}, \texttt{sports}, \texttt{sex} and \texttt{fuel}. The Fused Lasso penalty is used for the ordinal predictors. We also choose the Fused Lasso penalty for the \texttt{coverage} and \texttt{payfreq} predictors due to their inherent ordering. For \texttt{payfreq}, each next level corresponds to a more frequent payment of the premiums while for the \texttt{coverage} predictor, the levels are ranked as follows: \texttt{TPL} $<$ \texttt{PO} $<$ \texttt{FO}, in terms of the amount of the protection these coverages offer. For the spatial predictor \texttt{muni}, we employ a Generalized Fused Lasso penalty where all differences between neighboring regions are regularized. The first level of each predictor is taken as the reference level resulting in a parameter vector $\bm\beta$ with 422 coefficients. The full objective function $\mathcal{O}$ is then:
\begin{align*}
\mathcal{O}(\bm\beta; \bm{X}, \bm{y}) =& \ f(\bm\beta; \bm{X}, \bm{y})\\
&+ \lambda \left( \sum_{j \in \texttt{bin}} |w_j\beta_j| + \sum_{j \in \texttt{ord}} ||\bm{D}(\bm{w}_j)\bm\beta_j||_1 + ||\bm{G}(\bm{w}_{\texttt{muni}})\bm{\beta}_{\texttt{muni}}||_1 \right),
\end{align*}
with \texttt{bin} and \texttt{ord} the set of binary and ordinal predictors respectively. We use the combined adaptive and standardization penalty weights: $\bm{w}_j = \bm{w}_j^\text{ad}\cdot\bm{w}_j^\text{st}$ and we tune $\lambda$ with 10-fold stratified cross-validation where the deviance is used as error measure and the one-standard-error rule is applied. We apply the SMuRF algorithm, a GLM and a Generalized Additive Model (GAM, see \cite{gam}) to the training data and compare the predictive accuracy of the methods on the hold-out data. The GLM specification is highly overparameterized since we use the predictor levels of the initial SMuRF fit, as sketched in Section~\ref{sec:MTPLdescription}. The GAM incorporates one-dimensional flexible effects for \texttt{ageph}, \texttt{power}, \texttt{bm}, \texttt{agec} and a two-dimensional effect for \texttt{muni}, based on the longitude and latitude of the center of the municipalities, see \cite{henckaerts}. We fit the GAM in \texttt{R} through the \texttt{mgcv} package of \cite{mgcv}.

\subsection{Results and discussion}

\begin{figure}[!ht]
\centering
\begin{subfigure}[t]{\textwidth}
		\centering
\includegraphics[width = \textwidth]{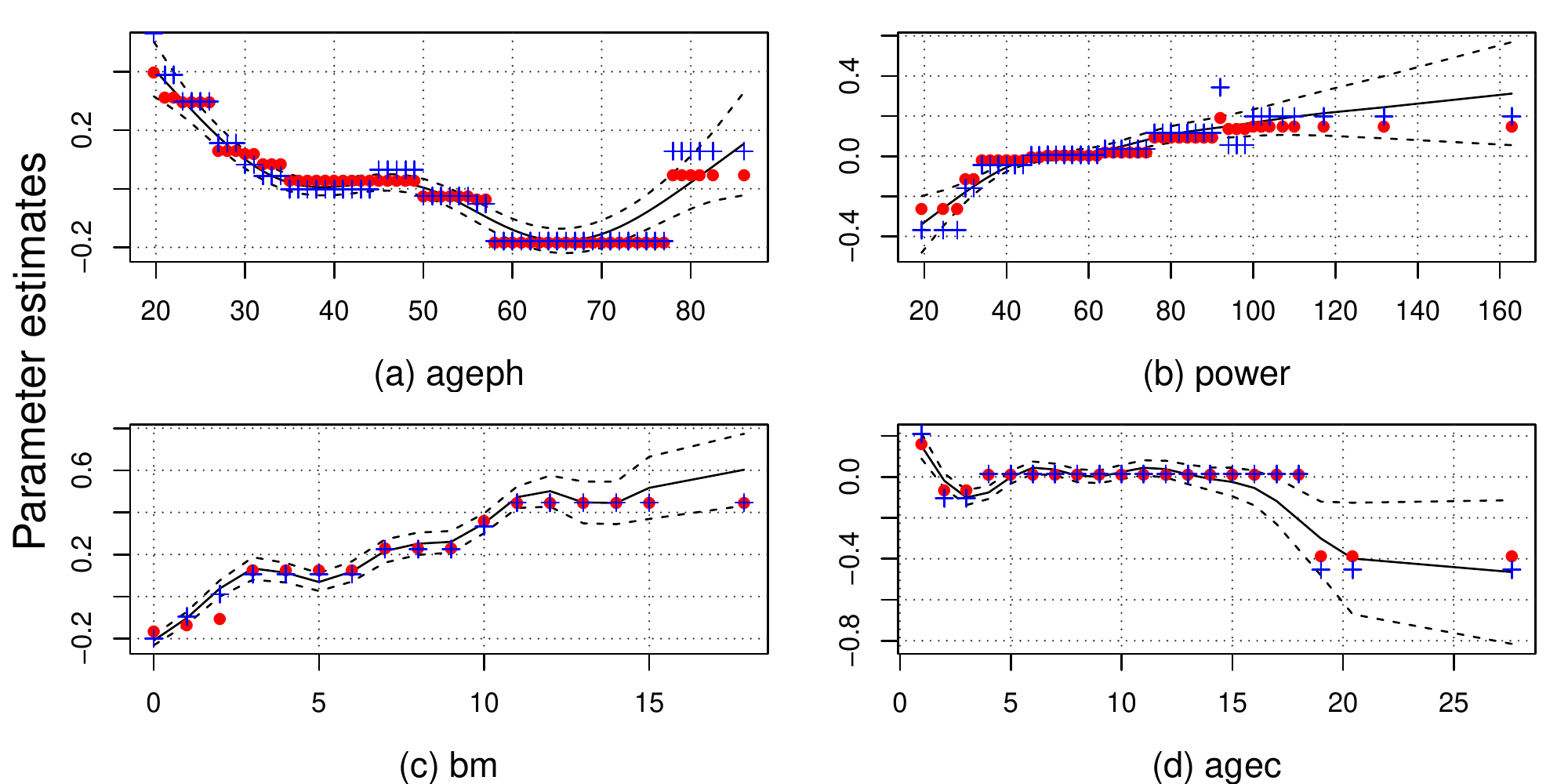}
\phantomsubcaption\label{fig:mtplordinala}
\phantomsubcaption\label{fig:mtplordinalb}
\phantomsubcaption\label{fig:mtplordinalc}
\phantomsubcaption\label{fig:mtplordinald}
\end{subfigure}
\caption{Parameter estimates for the \texttt{ageph}, \texttt{power}, \texttt{bm} and \texttt{agec} predictors of the \texttt{MTPL} data, centered around 0. The dots and crosses denote the parameter estimates from the SMuRF algorithm before and after the re-estimation respectively. The full black line corresponds to the GAM fit and the dotted lines to the fit $\pm$ 2 standard deviations.}
\label{fig:mtplordinal}
\end{figure}

\begin{figure}[!ht]
\centering
\includegraphics[width = \textwidth]{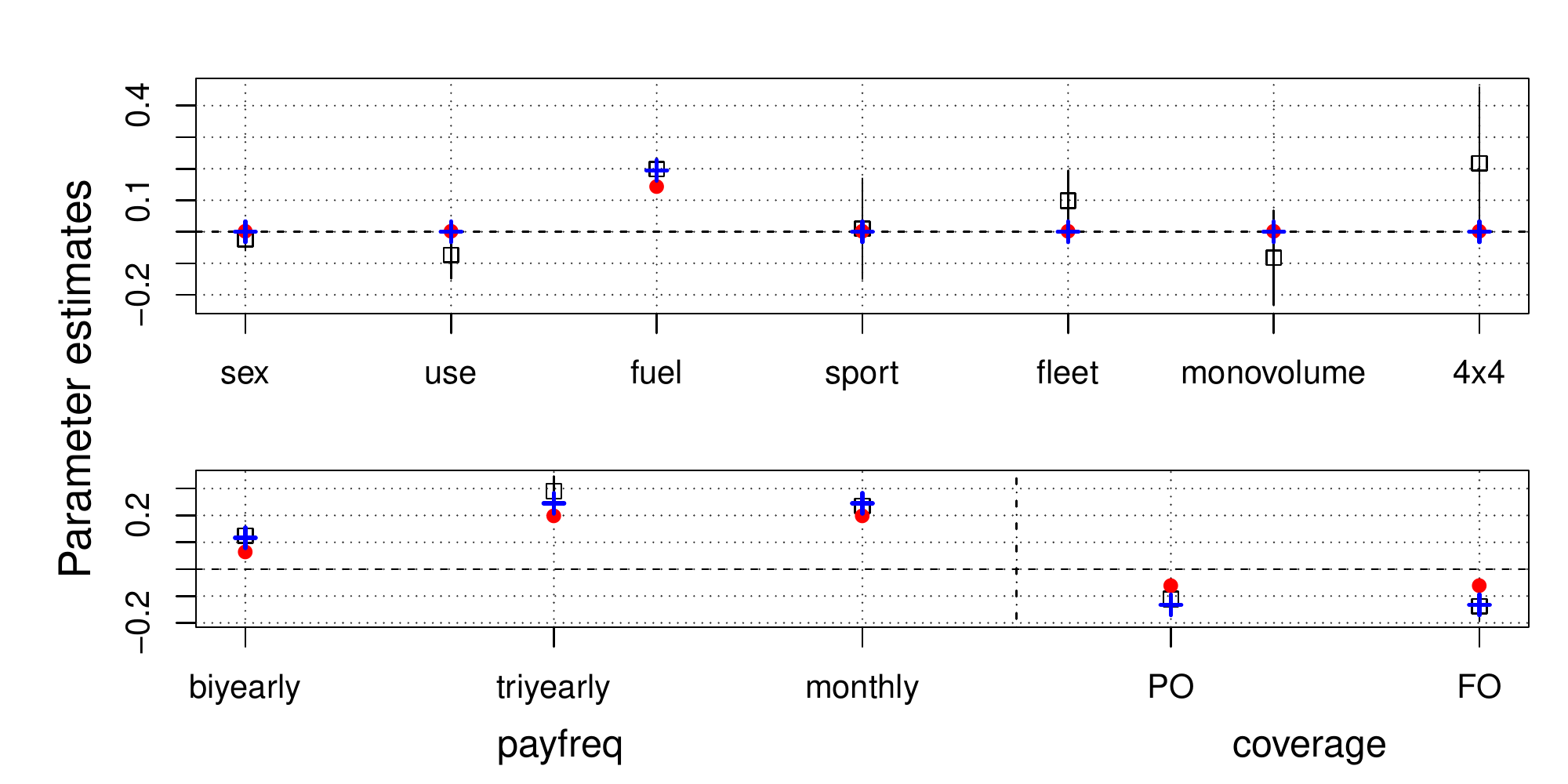}
\caption{Parameter estimates of the binary, \texttt{payfreq} and \texttt{coverage} predictors of the \texttt{MTPL} data. The dots and crosses denote the estimates of the SMuRF algorithm before and after the re-estimation respectively. The black squares correspond to the parameter estimates obtained with the GAM fit and the vertical black lines to the GAM fit $\pm$ 2 standard deviations.}
\label{fig:mtplnominal}
\end{figure}

\begin{figure}[!ht]
\centering
\includegraphics[width = 1\textwidth]{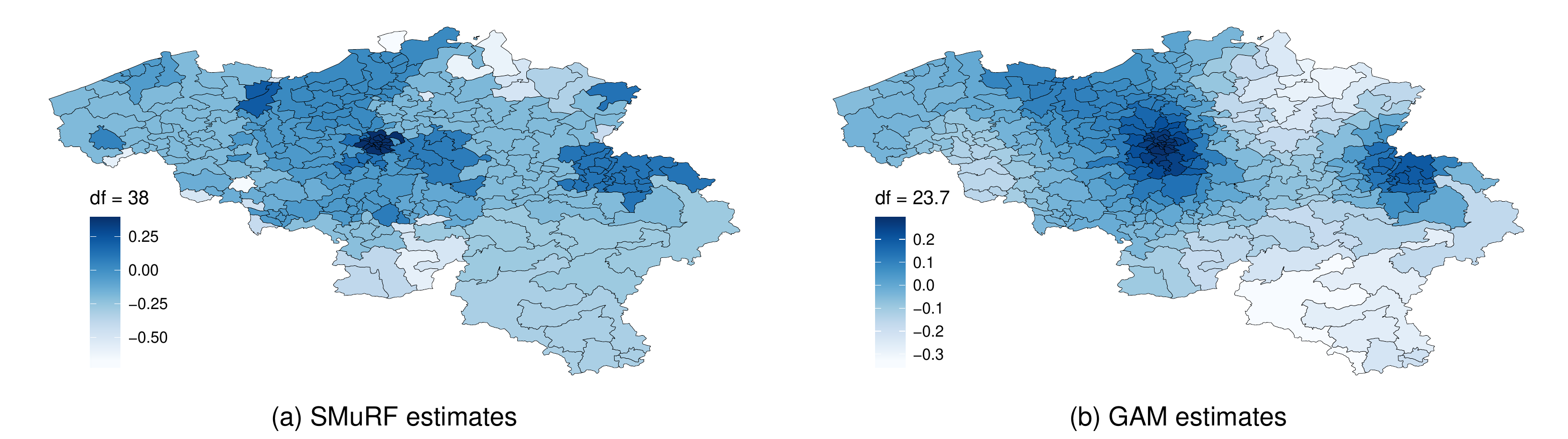}
\caption{Parameter estimates for the spatial \texttt{muni} predictor in the \texttt{MTPL} data obtained with (a) the SMuRF algorithm, and (b) a GAM. Each of the 266 polygons represents a different level in \texttt{muni}. Both color scales are the same.}
\label{fig:mtplspatial}
\end{figure}

We compare the estimated effects as obtained with SMuRF on the one hand and GAM on the other hand in Figures~\ref{fig:mtplordinal}-\ref{fig:mtplspatial}. The dots and crosses in Figures~\ref{fig:mtplordinal} and \ref{fig:mtplnominal} show the parameter estimates as obtained with SMuRF before and after re-estimation respectively. The black lines represent the GAM estimates for ordinal predictors in Figure~\ref{fig:mtplordinal} while the black squares give the GAM estimates for the binary, \texttt{payfreq} and \texttt{coverage} predictors in Figure~\ref{fig:mtplnominal}. Confidence intervals are given as dashed lines or segments respectively. Similar to \cite{henckaerts} we centered the SMuRF parameter estimates to ease the comparison with the GAM estimates. SMuRF leads to $\hat{\bm\beta}$ containing 71 unique coefficients while the GAM calculates 64 degrees of freedom, indicating a comparable model complexity.

Figure~\ref{fig:mtplordinala} illustrates that young, inexperienced drivers report more claims on average and thus represent a higher risk for the insurance company. The riskiness then declines steadily and increases again at older ages. Powerful cars (Figure~\ref{fig:mtplordinalb}) also exhibit increased risk over less powerful cars. Similarly, the model predicts a higher expected claim frequency for policyholders in a high bonus malus scale (Figure~\ref{fig:mtplordinalc}) due to their claims history. In \cite{henckaerts}, \texttt{agec} is not considered in the analysis but SMuRF recognizes it to have some predictive power. Especially for older cars, such as old timers, the expected claim frequency is lower. The parameter estimates obtained for the fused levels of the ordinal predictors in Figure~\ref{fig:mtplordinal} follow nicely the behavior of the GAM fit while greatly (and automatically) reducing the dimensionality compared to a standard GLM. Most parameters are estimated close or within the confidence interval of the GAM fit. The coefficients before and after re-estimation are close to each other for wider bins and are relatively farther apart for smaller bins.

Figure~\ref{fig:mtplnominal} shows the parameter estimates for the binary predictors, and the predictors \texttt{payfreq} and \texttt{coverage}. From the set of binary predictors, only \texttt{fuel} is selected while the others are put to 0, effectively removing them from the model. The parameter estimates obtained with the GAM fit confirm this behaviour, as 0 is within the confidence interval for all removed predictors except \texttt{4x4}. As the levels of \texttt{4x4} are highly imbalanced (see Figure~\ref{fig:MTPL.rel.freq2}), the influence of this predictor on the negative log-likelihood is minor and the regularized estimates remove the predictor from the model. The expected number of claims rises as \texttt{payfreq} increases up to monthly or triyearly payments, which are fused in the final model obtained with SMuRF. When a policyholder buys a partial or full omnium, the expected claim frequency decreases compared to the standard third party liability option. The SMuRF estimates fuse the levels for partial and full omnium.

Figure~\ref{fig:mtplspatial} illustrates the estimated parameters for the spatial effect, captured by \texttt{muni}. The SMuRF algorithm estimates 38 unique coefficients for the initial 266 different levels whereas the GAM calculates 23.7 degrees of freedom. For both the SMuRF as the GAM estimates, we see a higher expected claim frequency for people living in and around the larger cities in Belgium, though this distinction is less clear for the GAM estimates. In contrast, the models predict less claims for people living in the rural parts to the south, northeast and west of Belgium. Similar to the ordinal predictors, the GAM estimates are smoother than the SMuRF estimates and need less degrees of freedom to represent the data. However, the range of the parameter estimates of the SMuRF algorithm is wider than for the GAM, allowing for larger differences in expected claim frequency.

We also compare the out-of-sample predictive performance of SMuRF, the GLM and the GAM on the hold-out dataset. For each model, we sort the predictions such that $\hat{y}_{(1)}$ refers to the observation with the highest predicted claims frequency. We define the proportion of the sorted population, $\text{prop}_i$ and the proportion of observed claims, $\text{obs}_i$, by:
\begin{equation*}
\text{prop}_i = \frac{i}{n_{\text{out}}}, \qquad\qquad \text{obs}_i = \frac{\sum_{j=1}^{i} y_{(j)}}{\sum_{j=1}^{n_{\text{out}}} y_{(j)}},
\end{equation*}
with $n_{\text{out}}$ the number of observations in the hold-out data set and $y_{(j)}$ the observed number claims for observation $(j)$. We construct a cumulative capture rate curve by plotting $\text{prop}_i$ versus $\text{obs}_i$ for all $i\in\{1,\ldots, n_{\text{out}}\}$ and calculate the area under this curve (AUCC) for each model. The model with the highest AUCC score is best at ranking the individual risk of each policyholder. This technique is based on the Lorenz curves for insurance ratemaking, proposed by \cite{Frees2014}. Table~\ref{tab:mtploos} gives the out-of-sample results for the log-likelihood, the DSS and the AUCC score. For all computed measures, the relative difference between the SMuRF and GAM approach is very small with GAM performing slightly better for the log-likelihood and DSS score whereas SMuRF has a better AUCC score. In contrast, the GLM performs worst in all measures.
\begin{table}[ht!]
    \begin{center}
    \begin{tabular}{p{1cm} p{0.5cm} p{2.3cm} p{2.3cm} p{2.3cm}}
    \arrayrulecolor{black}\toprule
    model & df & log-likelihood & DSS & AUCC\\
	\arrayrulecolor{black}\midrule

	GLM    & 422 & -12475.8 & -35801.8 & 0.61418\\
	GAM    & 64 & \textbf{-12456.4} & \textbf{-36259.9} & 0.61652\\
	SMuRF & 71 & -12457.1 & -36176.8 & \textbf{0.61712}\\
	
	\arrayrulecolor{black}\bottomrule
    \end{tabular}
    \caption{\label{tab:mtploos}Out-of-sample results for the SMuRF, GLM and GAM approach on the \texttt{MTPL} hold-out data.}
    \end{center}
\end{table}

In conclusion, we see that SMuRF is competitive with the well established GAM approach, both in model complexity as in predictive accuracy, while additionally performing automatic predictor selection, fusion of levels and improving the interpretability.


\section{Discussion}\label{sec:conclusion}

We studied the general problem of convex optimization with a differentiable loss function and multi-type penalty terms. This setting is highly relevant when the level structure of different predictor types needs to be taken into account in the regularization. Our contribution is twofold. First, we developed the SMuRF algorithm that accurately and effectively solves this general optimization problem. The algorithm extends other proximal gradient algorithms found in the literature for convex, regularized optimization.

Secondly, we showed in full detail how this algorithm creates sparse models, using varying combinations of Lasso-type penalties, and investigating and documenting all possible model choices. The choice of penalty weights and the tuning strategy have a substantial influence on the performance of the estimated model. In our simulation study, the cross-validation approach with one standard error rule and combined adaptive and standardization penalty weights provided the best results. Additionally, the re-estimation of coefficients provided good results in this paper, where we used relatively large data sets. However, when dealing with smaller data sets, it might be preferable to work with the original regularized estimates as the performance gain due to the bias reduction might be canceled out by the increase in variance. Our implementation of the algorithm is available on CRAN (\url{https://cran.r-project.org/web/packages/smurf/}). We propose several paths for future research.

\paragraph{Extending SMuRF to other loss functions and penalties.} In the current implementation of SMuRF, only Lasso, Group Lasso and (Generalized) Fused Lasso are available. This can be straightforwardly extended to the Ridge and Elastic-net (\cite{elnet}) penalties. Another extension is the Generalized Lasso penalty, which replaces the graph structured matrix $\bm{G}(\bm{w}_j)$ of the Generalized Fused Lasso by an arbitrary matrix $\bm{M}$. By construction, the associated proximal operator can be solved with the same ADMM algorithm used for the Generalized Fused Lasso. This allows for more elaborate modeling options such as piece-wise polynomial regression or wavelet smoothing. Additionally, the current implementation of our algorithm can handle the superposition of the Lasso or Group Lasso with other penalties, all acting on the same subvector $\beta_j$. Examples of these in the literature are the Sparse Group or the Sparse Generalized Fused Lasso. However, these joint penalties need extra tuning parameters, making the model training more difficult. Further theoretical work needs to be done to find efficient ways of choosing or tuning these extra parameters. Additionally, SMuRF can be extended to handle other optimization problems such as Cox regression, Generalized Estimating Equations or M-estimators.

\paragraph{Stochastic version of SMuRF.} In current machine learning literature, stochastic versions of gradient descent algorithms exist where only part of the data is used every iteration. This speeds up the calculation per iteration while requiring more iterations for convergence. Additionally, due to its stochastic nature, stochastic optimization methods are less prone to get stuck in a local optimum, a useful property in the context of non-convex optimization. Therefore it is interesting to adapt SMuRF into a stochastic variant. This opens up the use of the algorithm with non-convex penalties such as the $L_0$ norm.


\section*{Acknowledgement}
Sander Devriendt, Katrien Antonio, Tom Reynkens and Roel Verbelen are grateful for the financial support of \textit{Ageas Continental Europe} and the support from KU Leuven through the C2 COMPACT research project. We also thank professor Jed Frees from the University of Wisconsin-Madison, several referees, the editor and the managing editor for their helpful comments, feedback and reviews on this work.

\bibliographystyle{apalike}
\bibliography{References.Sparsity}

\end{document}


\maketitle

\appendix

Section~\ref{apdsec:algorithm} gives a detailed overview of the implementation of the SMuRF algorithm. The calculation of the proximal operator (PO) for the Generalized Fused Lasso is explained in Section~\ref{apdsec:pos} including details on the convergence of the SMuRF algorithm, and Section~\ref{apdsec:simulation} expands on the simulation results, including a comparison with a Penalized Iteratively Reweighted Least Squares algorithm implementation.

\section{Algorithm overview}\label{apdsec:algorithm}

The naive version of the SMuRF algorithm can be found in Algorithm~1 in the paper. In this section, we discuss several improvements that are used in the implemented SMuRF algorithm.
Pseudo code for the full algorithm is given in Algorithm~\ref{alg:mtpPGA2} and numeric values for the algorithm parameters are given in Table~\ref{tab:apdalgo}.

\subsection{Stopping criterion}
The algorithm stops after $m$ steps if the following relative stopping criterion is met:
\[\frac{|\mathcal{O}(\bm\beta^{(m)}) - \mathcal{O}(\bm\beta^{(m-1)})|}{\mathcal{O}(\bm\beta^{(m-1)})} \leq \varepsilon,\] 
where $\varepsilon$ is a numerical tolerance value,
or if $m$ is equal to the maximum number of iterations.

\subsection{Backtracking of step size}\label{apdsec:backtracking}

The step size $s$ is a crucial parameter in the convergence of the algorithm. In many applications it is not possible to determine analytically an optimal value for this parameter. Backtracking is a popular solution to obtain a step size guaranteeing the convergence of the algorithm, see e.g.\ \citet{fista}.
The idea is to start from a large initial guess for the step size and to reduce it while the inequality in step~\ref{alg:step_crit} holds. 
The inequality is motivated by the convergence analysis of (accelerated) proximal gradient methods, see e.g.\ \citet{fista} and Section~9.2 in \citet{convexopt}.
Note that the backtracking parameter $\tau$ in step~\ref{alg:step1} needs to be strictly smaller than 1 in order to reduce the step size $s$.
In case the step size drops below $10^{-14}$, backtracking of the step size (steps~\ref{alg:step1} and \ref{alg:step2}) is no longer performed.

\subsection{Accelerated gradient descent}
As explained in Section~3.1, we use a standard procedure from the gradient descent method \citep{nesterov} to approximate the objective function.
Instead of using a standard gradient update as in Algorithm~1, \citet{nesterov} suggests to use acceleration.
Here, a new point $\bm{\theta}^{(k)}$ is found by moving along the line determined by $\bm{\beta}^{(k-1)}$ and $\bm{\beta}^{(k-2)}$, $\bm{\theta}^{(k)} = \bm{\beta}^{(k-1)} + \frac{\alpha^{(k-1)} - 1}{\alpha^{(k)}} \left(\bm{\beta}^{(k-1)} - \bm{\beta}^{(k-2)}\right)$ (see step~\ref{alg:gradient} in Algorithm~\ref{alg:mtpPGA2}), and then performing a gradient update on $\bm{\theta}^{(k)}$. We start with acceleration weight $\alpha^{(1)}=0$, and then $\alpha^{(k)}$ is found iteratively using the formula in step~\ref{alg:acc1}. The use of the extra point $\bm{\theta}^{(k)}$ provides optimal convergence for first-order algorithms used to minimize smooth convex functions \citep{nesterov}, while only requiring an easy additional calculation.

\subsection{Adaptive restarts}

Accelerated methods are often interpreted as momentum methods since the acceleration step size (or momentum) $\frac{\alpha^{(k)} - 1}{\alpha^{(k+1)}}$ depends on the previous iteration and gets larger in every iteration, i.e.\ bigger leaps are taken. 
They can lead to faster convergence, however they do not necessarily monotonically decrease the objective function unlike (standard) gradient descent methods.
\citet{adrestart} indicate that this non-monotone behavior occurs when the momentum exceeds its optimal value \citep{nesterov_book}.
Therefore, they propose to restart the momentum after a fixed number of iterations or if the objective function increases.
We choose to use the latter restart scheme which is adaptive and easy to implement since the values for the objective function are readily available.
\citet{adrestart} show that using adaptive restarts leads to a convergence rate close to the optimal convergence which is obtained with the problem specific, and hard to calculate, optimal series for $\alpha^{(k)}$.

\par If a restart is performed in step $k$, the momentum is reset and the previous estimate for $\bm{\beta}$ is kept, see steps \ref{alg:restart1} and \ref{alg:restart2} in Algorithm~\ref{alg:mtpPGA2}. Note that the acceleration updates then become $\alpha^{(k+1)} = 1$ and $\bm{\theta}^{(k+1)}=\bm{\beta}^{(k)}=\bm{\beta}^{(k-1)}$.

\par In the algorithm we use a slightly different criterion to perform a restart. If we would use $\mathcal{O}({\bm\beta}^{(k)})> \mathcal{O}(\bm\beta^{(k-1)})$ as the restart criterion as in \citet{adrestart}, a restart can take place even though  
\[\frac{\mathcal{O}(\bm\beta^{(k)}) - \mathcal{O}(\bm\beta^{(k-1)})}{\mathcal{O}(\bm\beta^{(k-1)})} \leq \varepsilon.\] 
In that case we actually want the algorithm to end. Therefore, we only restart if 
\[\mathcal{O}(\bm\beta^{(k)})> \mathcal{O}(\bm\beta^{(k-1)}) (1+\varepsilon)\]
with $\varepsilon$ the numerical tolerance. This avoids that a restart is performed when the stopping criterion is met.

\begin{algorithm}
	\caption{Full SMuRF algorithm}\label{alg:mtpPGA2}
	{\fontsize{10}{20}\selectfont
		\begin{algorithmic}[1]
			\State \textbf{Input} $\bm{\beta}^{(0)}, \bm{\theta}^{(1)} = \bm{\beta}^{(0)}, \alpha^{(1)} = 1, \bm{X}, \bm{y}, s, \lambda$
			\For{$k=1$ to $m$}
			\State\label{alg:gradient} $\tilde{\bm\beta} \gets \bm{\theta}^{(k)} - s\nabla f(\bm{\theta}^{(k)})$ \Comment gradient update
			\State $\left(\tilde{\beta}_0, \tilde{\bm\beta}_1, \ldots, \tilde{\bm\beta}_J\right) \gets \tilde{\bm\beta}$ \Comment partition full vector in components per feature
			\State $\bm{\beta}^{(k)}_j \gets \text{prox}_{s\lambda g_j}\left(\tilde{\bm\beta}_j\right)$\label{alg:proxupdate} \Comment calculate PO for all $j$ in $\{0,\ldots,J\}$
			\State $\bm\beta^{(k)} \gets \left(\beta^{(k)}_0, \bm{\beta}^{(k)}_1, \ldots, \bm{\beta}^{(k)}_J\right)$ \Comment recombine to full vector
			\While{$\mathcal{O}(\bm\beta^{(k)}) >  f(\bm{\theta}^{(k)}) + (\bm{\beta}^{(k)} - \bm{\theta}^{(k)})^T \nabla f(\bm{\theta}^{(k)})  + \frac{\|\bm{\beta}^{(k)} - \bm{\theta}^{(k)}\|^2_2}{2s}  + g(\bm{\beta}^{(k)} )$} \label{alg:step_crit}
			\State\label{alg:step1} $s \gets s\times \tau$ \Comment backtrack step size
			\State\label{alg:step2} Perform steps 3 to 6.
			\EndWhile
			\If{$\mathcal{O}(\bm\beta^{(k)}) > \mathcal{O}(\bm\beta^{(k-1)}) (1+\varepsilon)$} \Comment adaptive restart
			\State\label{alg:restart1} $\bm\beta^{(k)} \gets \bm\beta^{(k-1)}$ \Comment use old estimates
			\State\label{alg:restart2} $\alpha^{(k)} \gets 0$ \Comment reset momentum
			\EndIf 
			\State\label{alg:acc1} $\alpha^{(k+1)}  \gets \frac{1 + \sqrt{1 + 4 \left(\alpha^{(k)}\right)^2}}2$ \Comment acceleration (part~1)
			\State\label{alg:acc2} $\bm{\theta}^{(k+1)}  \gets \bm{\beta}^{(k)} + \frac{\alpha^{(k)} - 1}{\alpha^{(k+1)}} \left(\bm{\beta}^{(k)} - \bm{\beta}^{(k-1)}\right)$ \Comment acceleration (part~2)
			\EndFor
			\State \textbf{return} $\bm{\beta^{(m)}}$
		\end{algorithmic}
	}
\end{algorithm}

\subsection{Convergence of the SMuRF algorithm}

The SMuRF algorithm is a gradient descent algorithm using the acceleration method of \cite{nesterov} equivalent to the acceleration in the Fast Iterative Soft-Thresholding Algorithm for the Lasso problem in \cite{fista}. Therefore, the SMuRF algorithm inherits the convergence and complexity properties of these algorithms. The resulting convergence rate for SMuRF is $\mathcal{O}(1/k^2)$, corresponding to the theoretical optimal convergence rate for a gradient descent method (see \cite{nesterov} and \cite{parikh}).

 \section{Proximal operators}\label{apdsec:pos}
For Lasso and Group Lasso, the proximal operators can be computed analytically using the soft-thresholding and group soft-thresholding operators, respectively.
To compute the proximal operators for the (Generalized) Fused Lasso,
\[\text{prox}_{s\lambda g_j}\left(\tilde{\bm\beta}_j\right)=\underset{\mathbf{x}}{\mathrm{argmin}}\ \frac12 \| \tilde{\bm\beta}_j-\mathbf{x}\|_2^2 + s\lambda  \|\bm{G}(\bm w_j)\mathbf{x}\|_1,\]
no analytical solutions are available and we hence rely on numerical methods.
We can rewrite the problem as \[\underset{\mathbf{x}}{\mathrm{argmin}}\  \frac12 \|\tilde{\bm\beta}_j-\mathbf{x}\|_2^2 + s\lambda \|\mathbf{z}\|_1 \quad \text{subject to } \bm{G}(\bm w_j)\mathbf{x}-\bm{I}_d\mathbf{z}=\mathbf{0},\]
where $\mathbf{z}$ is the dual variable and $\bm{I}_d$ the identity matrix of dimension $d$.
This reformulated problem can be solved using the Alternating Direction Method of Multipliers (ADMM - \cite{admmoriginal, admmoriginal2}) algorithm. This iterative method has the following equations in iteration $l$ (see Section~6.4.1 in \citet{admm}):
\begin{align*}
\mathbf{x}^{(l)} &= \left(\bm{I}_d + \rho^{(l-1)} \bm{G}(\bm w_j)^T \bm{G}(\bm w_j)\right)^{-1} \left(\tilde{\bm\beta}_j + \rho^{(l-1)} \bm{G}(\bm w_j)^T (\mathbf{z}^{(l-1)}-\mathbf{u}^{(l-1)})\right) \\
\mathbf{z}^{(l)} &= \mathcal{S}\left(\xi \bm{G}(\bm w_j) \mathbf{x}^{(l)} + (1-\xi)\mathbf{z}^{(l-1)} + \mathbf{u}^{(l-1)}; (s\lambda)/\rho^{(l-1)}\right) \\
\mathbf{u}^{(l)} &= \mathbf{u}^{(l-1)} + \left(\xi \bm{G}(\bm w_j) \mathbf{x}^{(l)} + (1-\xi)\mathbf{z}^{(l-1)}\right) - \mathbf{z}^{(l)},
\end{align*}
where $\rho^{(l-1)}>0$ is the augmented Lagrangian parameter, $\mathbf{u}$ the scaled dual variable and $\xi \in (1,2)$ the relaxation parameter. 
Starting values $\mathbf{z}^{(0)}$ and $\mathbf{u}^{(0)}$ need to be given. We use $\mathbf{z}^{(0)}=\bm{G}(\bm w_j)\mathbf{x}^{(0)}$ where $\mathbf{x}^{(0)}=\bm\beta_j^{(k-1)}$ is the estimate from the previous iteration in the SMuRF algorithm, and $\mathbf{u}^{(0)}=\mathbf{0}$. 
The updates for $\mathbf{x}$ and $\mathbf{z}$ happen in an alternating way, in contrast to the method of multipliers where the updates happen simultaneously.
\par The stopping criterion is based on the primal and dual residuals $\mathbf{r}^{(l)}$ and $\mathbf{s}^{(l)}$:
\begin{align*}
\mathbf{r}^{(l)} &= \bm{G}(\bm w_j) \mathbf{x}^{(l)} - \mathbf{z}^{(l)}\\
\mathbf{s}^{(l)} & = -\rho^{(l-1)} \bm{G}(\bm w_j)^T (\mathbf{z}^{(l)}-\mathbf{z}^{(l-1)}).
\end{align*}
Both residuals converge to the zero vector as ADMM proceeds. The stopping criterion is then
\[\|\mathbf{r}^{(l)}\|_2 \leq \varepsilon_{\text{pri}} \qquad \text{and} \qquad\|\mathbf{s}^{(l)}\|_2 \leq \varepsilon_{\text{dual}} \]
with $\varepsilon^{\text{pri}}$ and $\varepsilon^{\text{dual}}$ the primal and dual tolerance defined as
\begin{align*}
\varepsilon_{\text{pri}} &= \sqrt{m}\, \varepsilon_{\text{abs}} +  \varepsilon_{\text{rel}} \max\{\|\bm{G}(\bm w_j)\mathbf{x}^{(l)}\|_2,\|-\mathbf{z}^{(l)}\|_2\}\\
\varepsilon_{\text{dual}} &= \sqrt{d}\, \varepsilon_{\text{rel}} +  \varepsilon_{\text{abs}} \rho^{(l-1)}\|\bm{G}(\bm w_j)^T\mathbf{u}^{(l)}\|_2.
\end{align*}
Here, $\varepsilon_{\text{abs}}$ is the absolute tolerance and $\varepsilon_{\text{rel}}$ is the relative tolerance. 

\par As starting value for $\rho$ we use $\rho^{(0)}=1$. Afterwards, $\rho$ can be updated according to the scheme discussed in \citet{admm} which was further improved in \citet{zhu2017}:
\[\rho^{(l)}=\begin{cases}
\eta_{\rho} \rho^{(l-1)} & \text{if } \|\mathbf{r}^{(l)}\|_2 /\varepsilon_{\text{pri}} \geq \mu_{\rho} \|\mathbf{s}^{(l)}\|_2 /\varepsilon_{\text{dual}}\\
\rho^{(l-1)}/\eta_{\rho}  & \text{if } \|\mathbf{s}^{(l)}\|_2/\varepsilon_{\text{dual}} \geq \mu_{\rho} \|\mathbf{r}^{(l)}\|_2 /\varepsilon_{\text{pri}} \\
\rho^{(l-1)} &  \text{otherwise.}
\end{cases}
\]
\citet{admm} suggest to use $\mu_{\rho}=10$ and $\eta_{\rho}=2$. If $\rho$ is changed, one also needs to change $u^{(l)}$ according to $u^{(l)} = u^{(l)} \frac{\rho^{(l-1)}}{\rho^{(l)}}.$  	This means for example that if $\rho$ is halved, $u$ needs to be doubled.

\par The implementation of the ADMM algorithm was done in \texttt{C++} using the \textit{Armadillo} library \citep{Armadillo} which is called through the \texttt{R} package \textit{RcppArmadillo} \citep{RcppArmadillo}.

\par The matrix inverse in the update for $\mathbf{x}$ only needs to be recomputed when $\rho$ is updated. Instead of using a general function to compute the matrix inverse, we take its special structure into account. 
Since $\bm{G}(\bm w_j)^T \bm{G}(\bm w_j)$ is symmetric, we can compute its eigenvalue decomposition $\bm{G}(\bm w_j)^T \bm{G}(\bm w_j) = \bm{Q} \bm{\Lambda} \bm{Q}^T$ with $\bm{Q}$ an orthogonal matrix with the eigenvectors in the columns and $\bm{\Lambda}$ a diagonal matrix with the eigenvalues $\ell_1,\ldots, \ell_d$ on the diagonal. Application of the Woodbury matrix identity then gives
\begin{align*}
\left(\bm{I}_d + \rho^{(l-1)} \bm{G}(\bm w_j)^T \bm{G}(\bm w_j)\right)^{-1} = \bm{I}_d- \rho^{(l-1)} \bm{Q} \bm{\Lambda}' \bm{Q}^T,
\end{align*}
where $\bm{\Lambda}'$ is a diagonal matrix with main diagonal 
\[\frac1{\frac1{l_1}+\rho^{(l-1)}}, \ldots, \frac1{\frac1{l_d}+\rho^{(l-1)}}.\]
When $\rho^{(l-1)}$ changes, the inverse can easily be recomputed as the eigenvector and eigenvalues are independent of $\rho^{(l-1)}$. Note that the eigenvalue decomposition needs to be computed only once and not at every computation of the proximal operator. Therefore, this approach is faster than using a general function to compute the matrix inverse of $\bm{I}_d + \rho^{(l-1)} \bm{G}(\bm w_j)^T \bm{G}(\bm w_j)$. Table~\ref{tab:apdalgo} provides a list of the implemented numeric values for the ADMM algorithm.

\begin{table}
	\centering
	\begin{tabular}{r l}
	    \arrayrulecolor{black}\toprule
		Parameter & Value \\
		    \arrayrulecolor{black}\midrule

		$\varepsilon$ & $10^{-8}$ \\ 
		maximum number of iterations & $10^4$ \\
		    \arrayrulecolor{lightgray}\midrule

		$\tau$ & 0.5 \\ 
		s & $0.1\times n$ \\
		    \arrayrulecolor{lightgray}\midrule

		$\rho^{(0)}$ & 1	\\
		maximum number of iterations (ADMM) & $10^4$ \\
		$\varepsilon_{rel}$ & $10^{-10}$\\
		$\varepsilon_{abs}$ & $10^{-12}$ \\
		$\xi$ & 1.5 \\
		$\mu_{\rho}$ & 10 \\
		$\eta_{\rho}$ & 2 \\
			    \arrayrulecolor{black}\bottomrule
	\end{tabular}
\caption{Parameter choices for SMuRF, the backtracking and the ADMM algorithm.}\label{tab:apdalgo}
\end{table}

\section{Simulation study}\label{apdsec:simulation}

\subsection{Simulation parameters}

A detailed overview of variables, their 
levels and the individual true parameters used for the simulation study is found in Table~\ref{tab:apdsim}.

\begin{table}[ht!]
    \begin{center}
    \begin{adjustbox}{max width=\textwidth}
    \begin{tabular}{p{2cm} p{1.85cm} p{9cm} p{6cm}}
    \arrayrulecolor{black}\toprule
    Type & Name & Description & True parameter $\bm\beta^{(var)}$\\
	\arrayrulecolor{black}\midrule
	 & \texttt{score} & Credit score, used as response variable: 0 for bad and 1 for good customers. & \\
	\arrayrulecolor{lightgray}\midrule
    Ordinal & \texttt{age} & Age of the customer: 20-70. & $\beta_{i}^{(age)}$ = 0 for $i\in \left[20,25\right]$,\newline
$\beta_{i}^{(age)}$ = 0.25 for $i\in \left[26,40\right]$,\newline
$\beta_{i}^{(age)}$ = 0.5 for $i\in \left[41,60\right]$,\newline
$\beta_{i}^{(age)}$ = 0.75 for $i\in \left[61,70\right]$. \\
\addlinespace[1em]
    & \texttt{stability} & Consecutive time in years spent with current job/employer: 0-20. & $\beta_{i}^{(stab)}$ = 0 for $i\in \left[0,2\right]$,\newline
$\beta_{i}^{(stab)}$ = 0.3 for $i\in \left[3,6\right]$,\newline
$\beta_{i}^{(stab)}$ = 0.5 for $i\in \left[7,20\right]$.\\
\addlinespace[1em]
    & \texttt{salary} & Monthly net income of customer in EUR, rounded to the nearest 100: 1000-5000. & $\beta_{i}^{(sal)}$ = 0 for $i\in \left[1,10\right]$,\newline
$\beta_{i}^{(sal)}$ = 0.4 for $i\in \left[11,20\right]$,\newline
$\beta_{i}^{(sal)}$ = 0.6 for $i\in \left[21,30\right]$,\newline
$\beta_{i}^{(sal)}$ = 1 for $i\in \left[31,41\right]$.\\
\addlinespace[1em]
    & \texttt{loan} & Monthly loan payment, in EUR, rounded to the nearest 100: 100-3000. & $\beta_{i}^{(loan)}$ = 0 for $i\in \left[1,5\right]$,\newline
$\beta_{i}^{(loan)}$ = -0.2 for $i\in \left[6,10\right]$,\newline
$\beta_{i}^{(loan)}$ = -0.4 for $i\in \left[11,15\right]$,\newline
$\beta_{i}^{(loan)}$ = -0.6 for $i\in \left[16,20\right]$,\newline
$\beta_{i}^{(loan)}$ = -0.8 for $i\in \left[21,25\right]$,\newline
$\beta_{i}^{(loan)}$ = -1 for $i\in \left[26,30\right]$.\\
    \arrayrulecolor{lightgray}\midrule
    Binary & \texttt{sex} & Gender of the customer: \texttt{female} or \texttt{male}. & $\beta_{i}^{(sex)}$ = 0 for female and -0.3 for male clients.\\
    \arrayrulecolor{lightgray}\midrule
    Nominal & \texttt{prof} & Profession of the customer, coded in 10 levels. & $\beta_{i}^{(prof)}$ = 0 for $i\in \{1,3,7\}$,\newline
$\beta_{i}^{(prof)}$ = 0.25 for $i\in \{4,5,8,10\}$,\newline
$\beta_{i}^{(prof)}$ = 0.5 for $i\in \{2,6,9\}$. \\
\addlinespace[1em]
& \texttt{drink} & Type of drink had during interview, coded in 5 levels. & $\beta_{i}^{(drink)}$ = 0 for $i\in \{1,\ldots,5\}$. \\
    \arrayrulecolor{lightgray}\midrule
    Interaction & \texttt{salxloan} & Interaction effect between \texttt{salary} and \texttt{loan} variables. & $\beta_{i}^{(sxl)}$ = 0.5 if \texttt{salary} $\geq 3,500$ and \texttt{loan} $\geq 2,000$, otherwise $\beta_{i}^{(sxl)}$ = 0\\
    \arrayrulecolor{black}\bottomrule
    \end{tabular}
    \end{adjustbox}
    \caption{\label{tab:apdsim}Overview of the variables and their levels for the simulated dataset.}
    \end{center}
\end{table}

\subsection{Simulation results}

\begin{figure}[!ht]
\centering
\includegraphics[width = \textwidth]{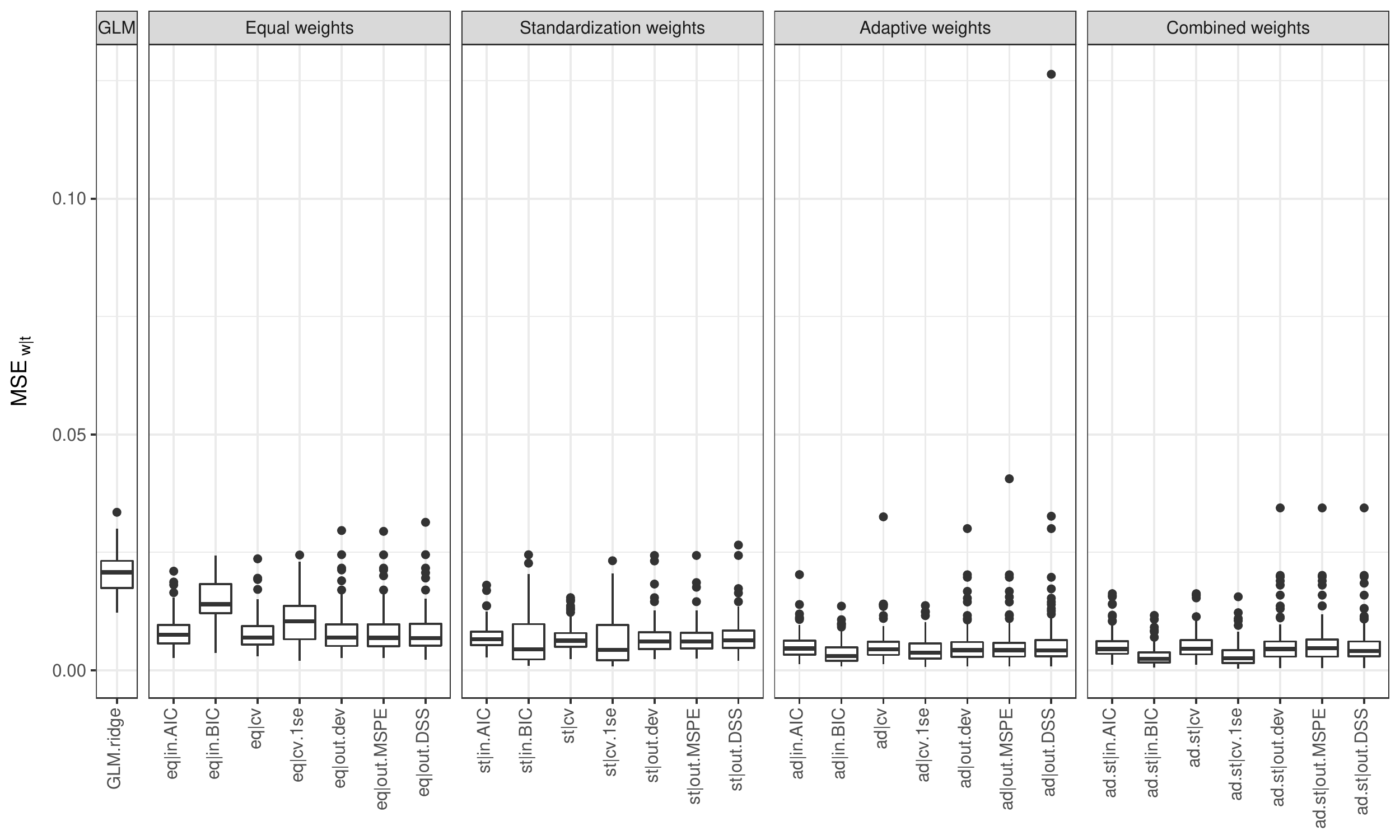}
\caption{Boxplot of parameter MSE for the binomial GLM with a small ridge penalty and for the different settings of the SMuRF algorithm. }
\label{fig:apdbetasim}
\end{figure}

\begin{figure}[!ht]
\centering
\includegraphics[width = \textwidth]{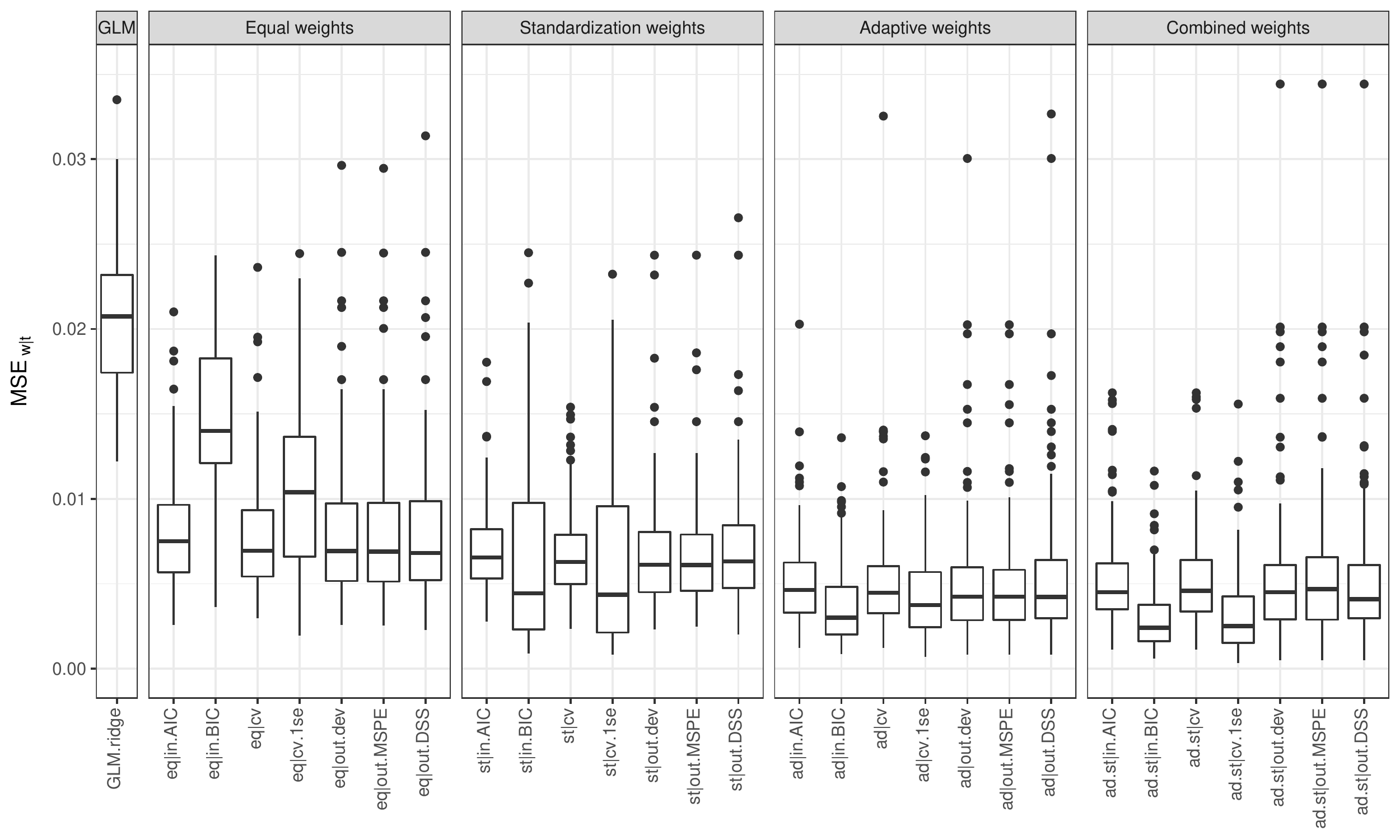}
\caption{Zoomed-in boxplot of parameter MSE for the binomial GLM with a small ridge penalty and for the different settings settings of the SMuRF algorithm. }
\label{fig:apdbetasimzoomed}
\end{figure}

Figure~\ref{fig:apdbetasim} provides boxplots of $\text{MSE}^{[i]}_{\texttt{w|t}}$ for the binomial GLM with a small ridge penalty and for the different settings of the SMuRF algorithm. An additional zoomed-in version of the graph is provided in Figure~\ref{fig:apdbetasimzoomed}.

\begin{figure}[!ht]
\centering
\includegraphics[width = 0.45\textwidth]{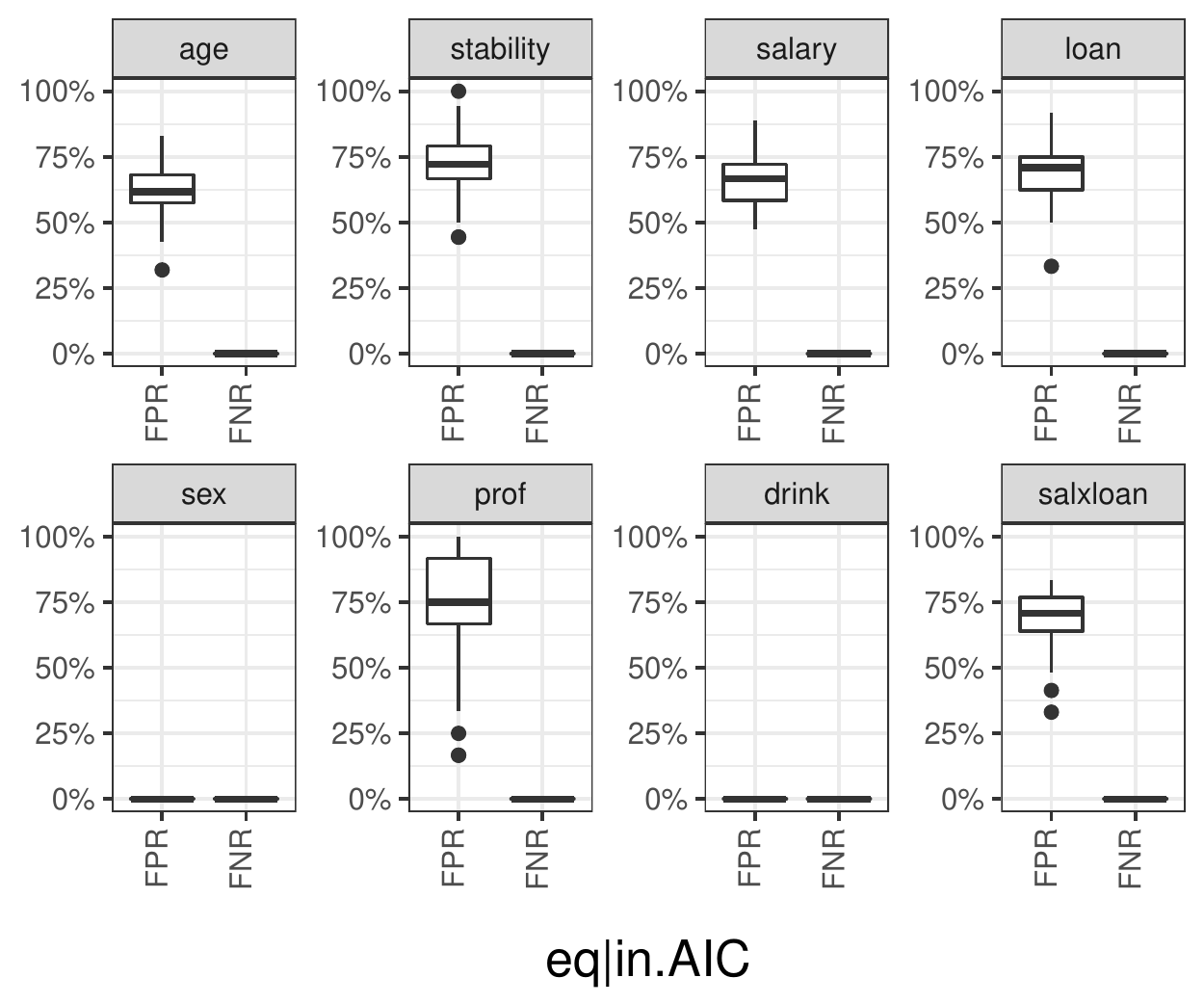}
\hspace{0.2cm}
\includegraphics[width = 0.45\textwidth]{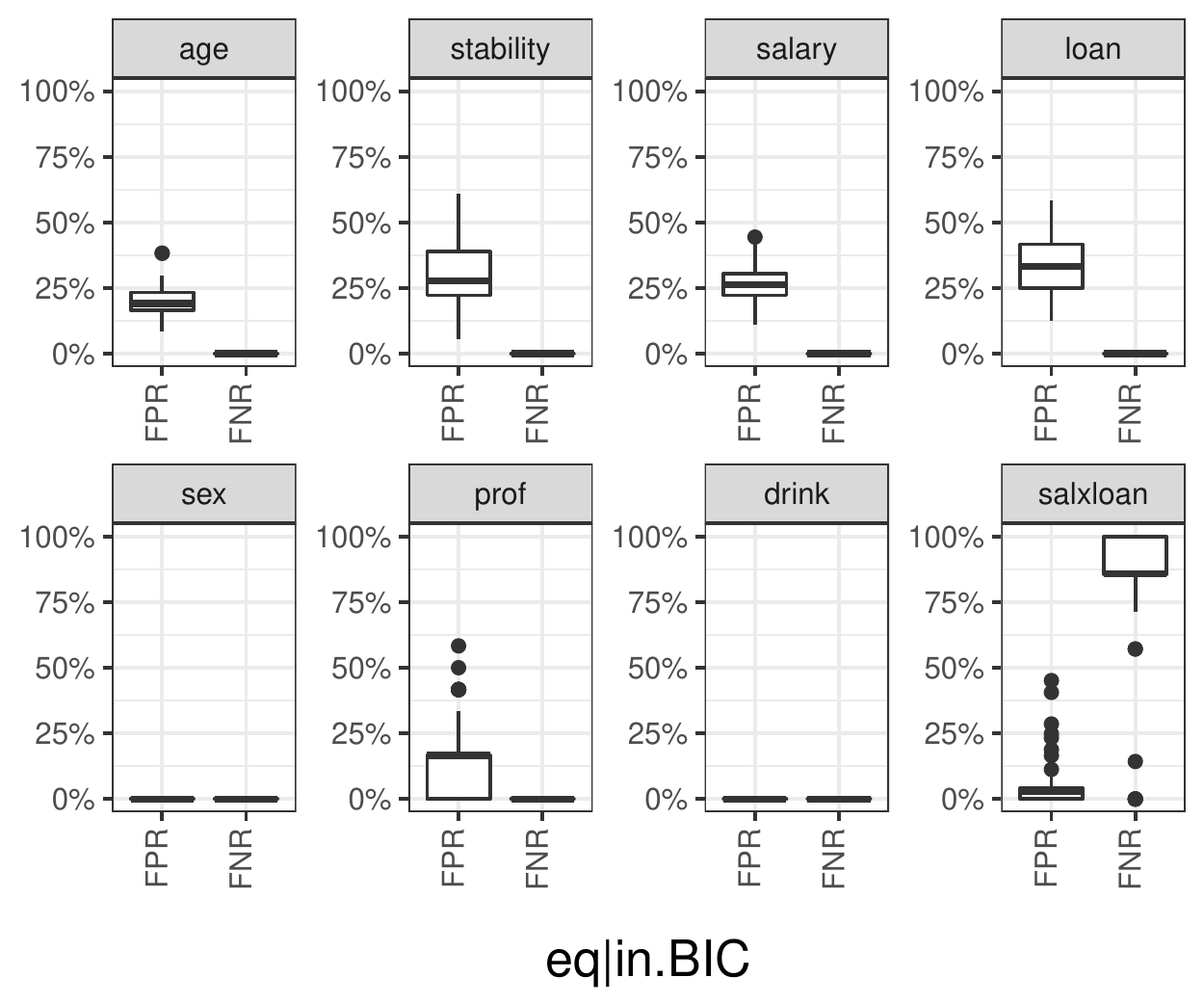}
\caption{Boxplots of the FPR and FNR per variable for settings \texttt{eq|in.AIC} and \texttt{eq|in.BIC} of the SMuRF algorithm.}
\label{fig:apdfpnr1}
\end{figure}

\begin{figure}[!ht]
\centering
\includegraphics[width = 0.45\textwidth]{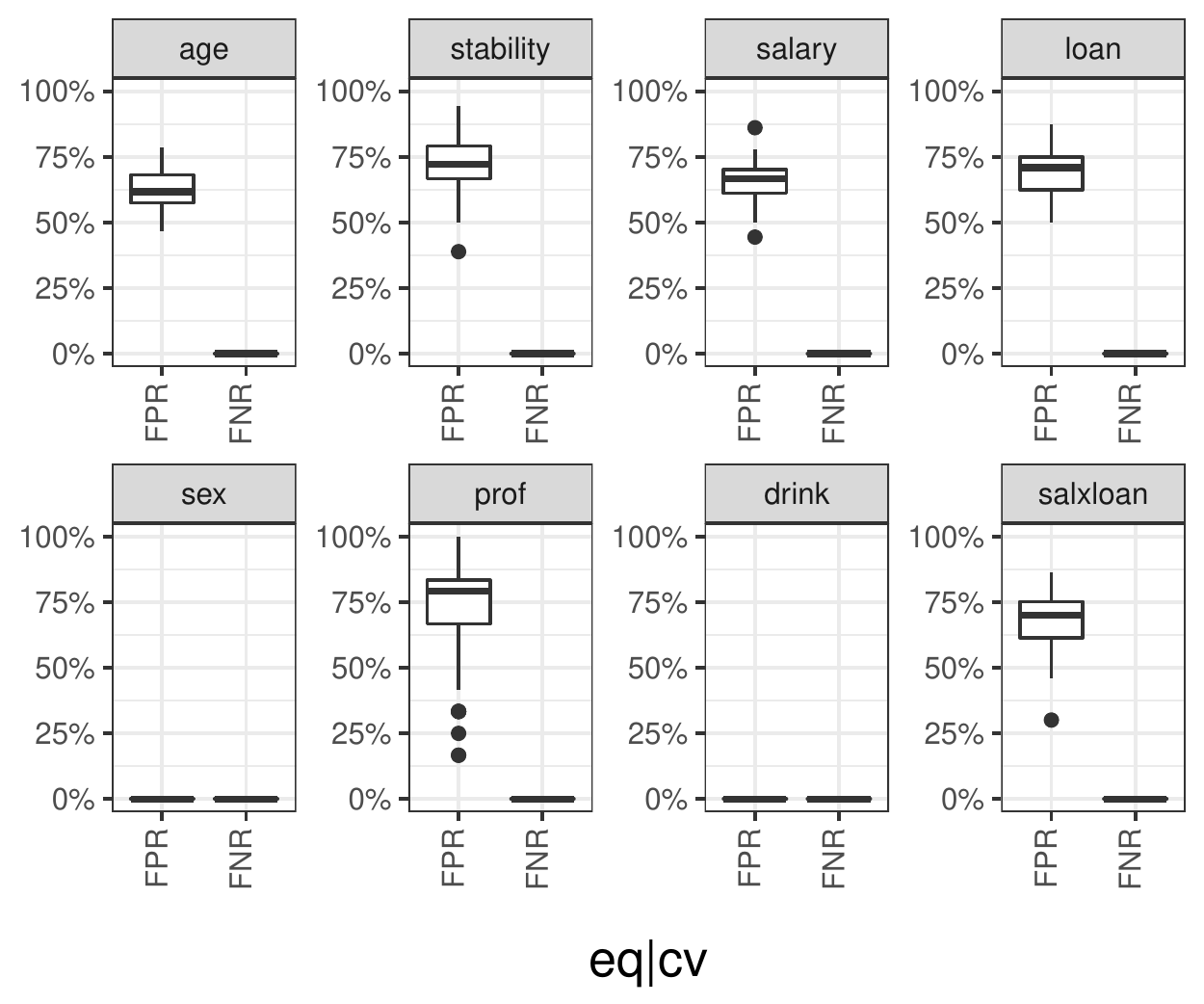}
\hspace{0.2cm}
\includegraphics[width = 0.45\textwidth]{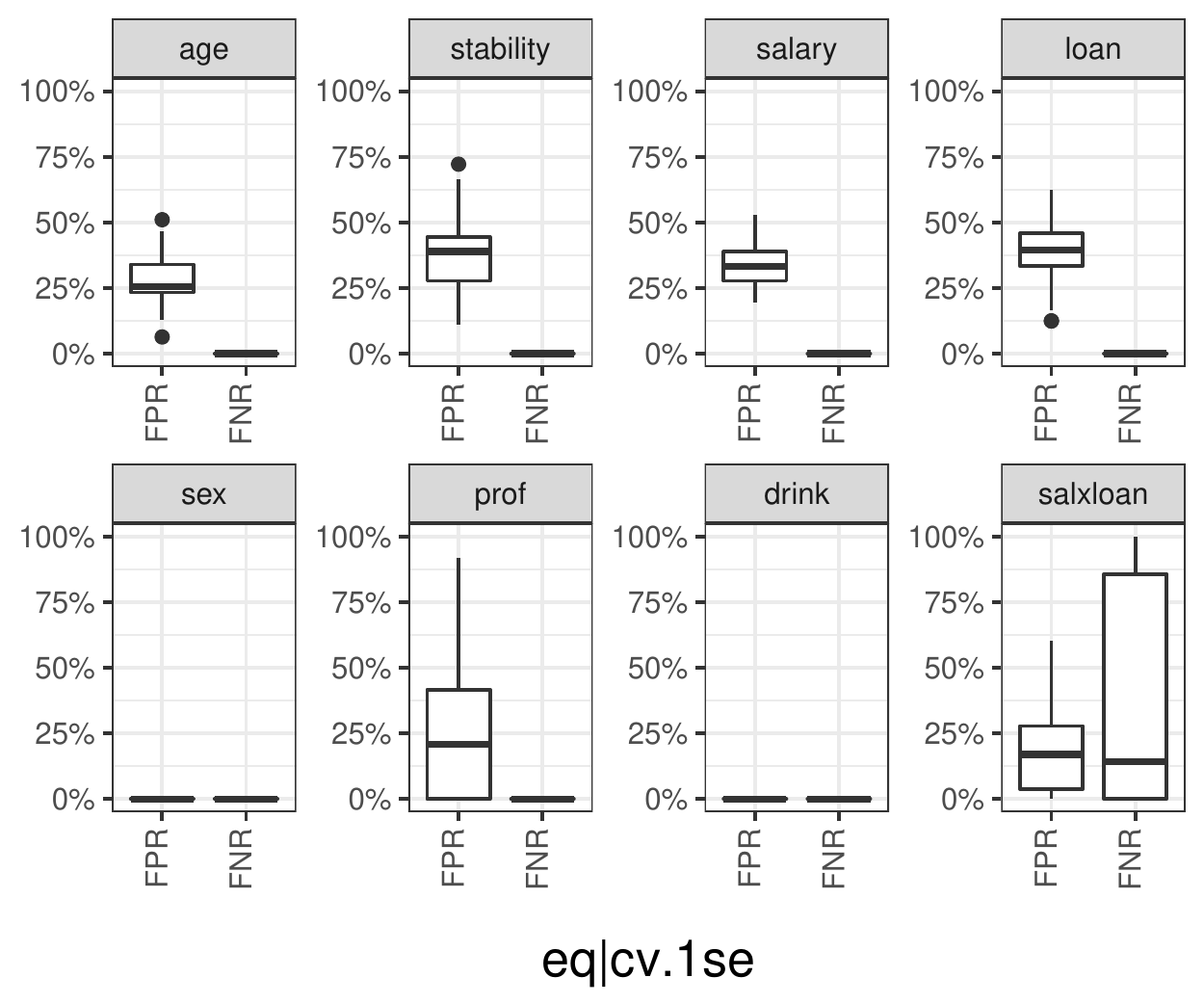}
\caption{Boxplots of the FPR and FNR per variable for settings \texttt{eq|cv} and \texttt{eq|cv.1se} of the SMuRF algorithm.}
\label{fig:apdfpnr2}
\end{figure}

\begin{figure}[!ht]
\centering
\includegraphics[width = 0.45\textwidth]{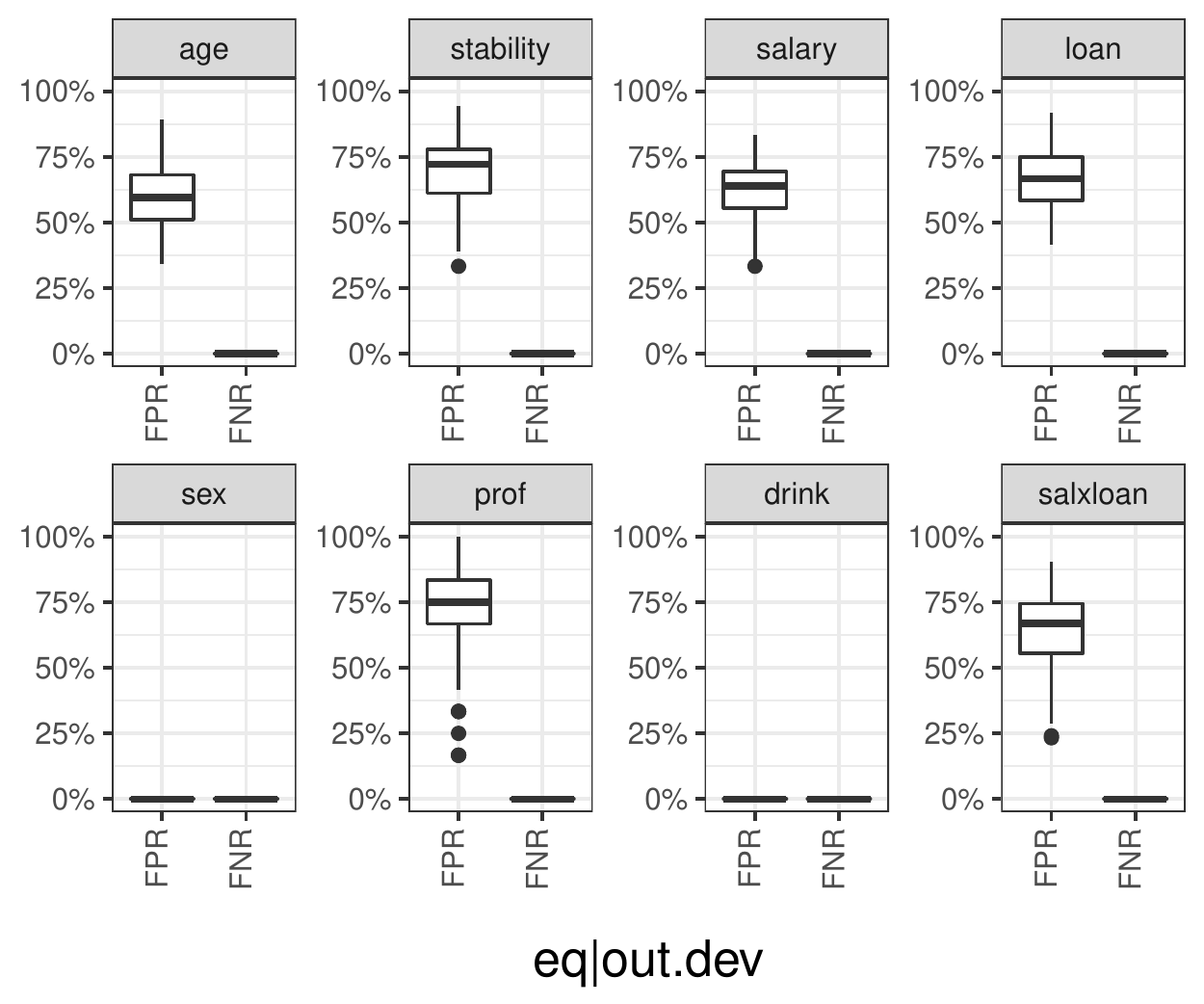}
\caption{Boxplots of the FPR and FNR per variable for settins \texttt{eq|out.dev} of the SMuRF algorithm.}
\label{fig:apdfpnr3}
\end{figure}

\begin{figure}[!ht]
\centering
\includegraphics[width = 0.45\textwidth]{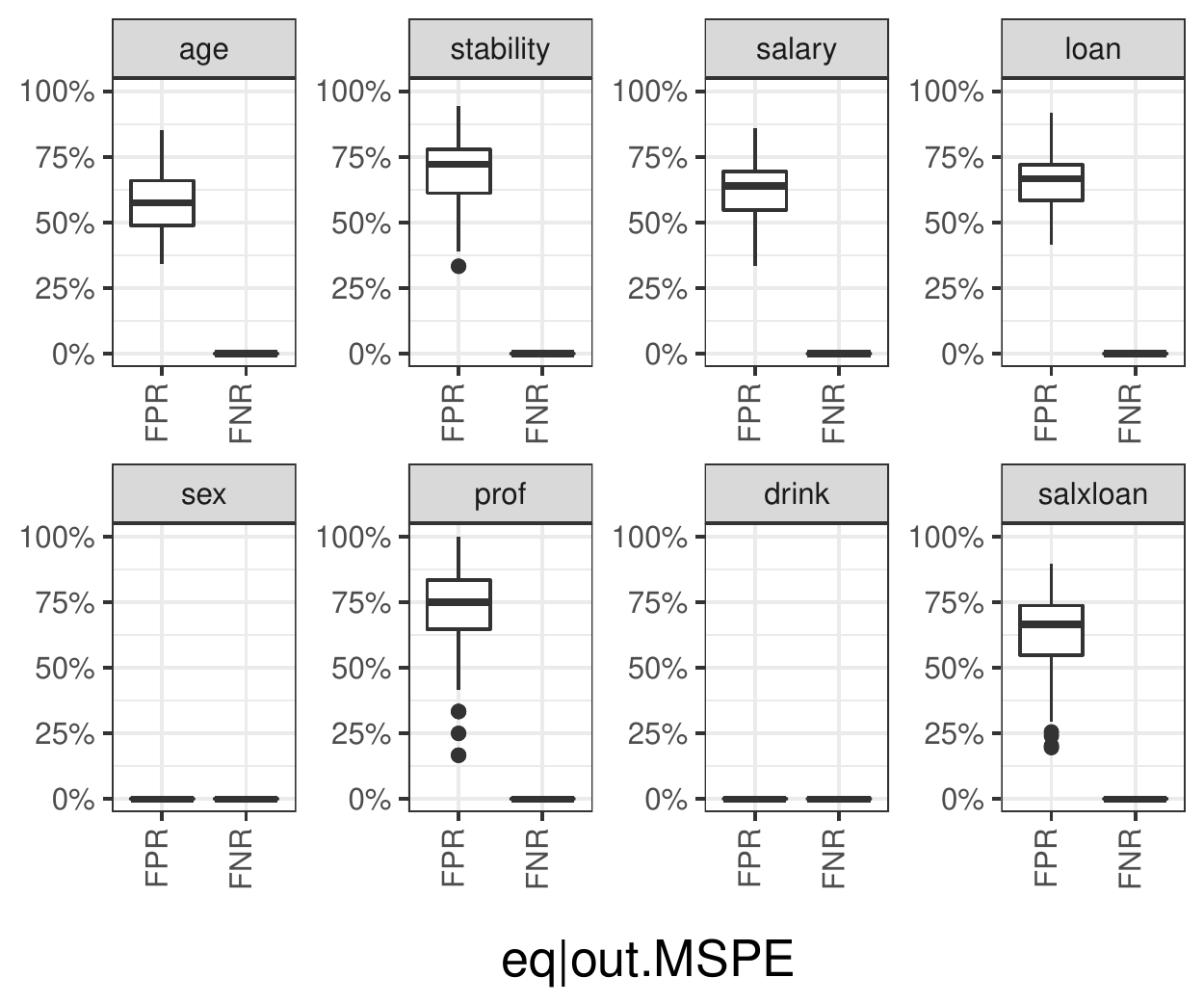}
\hspace{0.2cm}
\includegraphics[width = 0.45\textwidth]{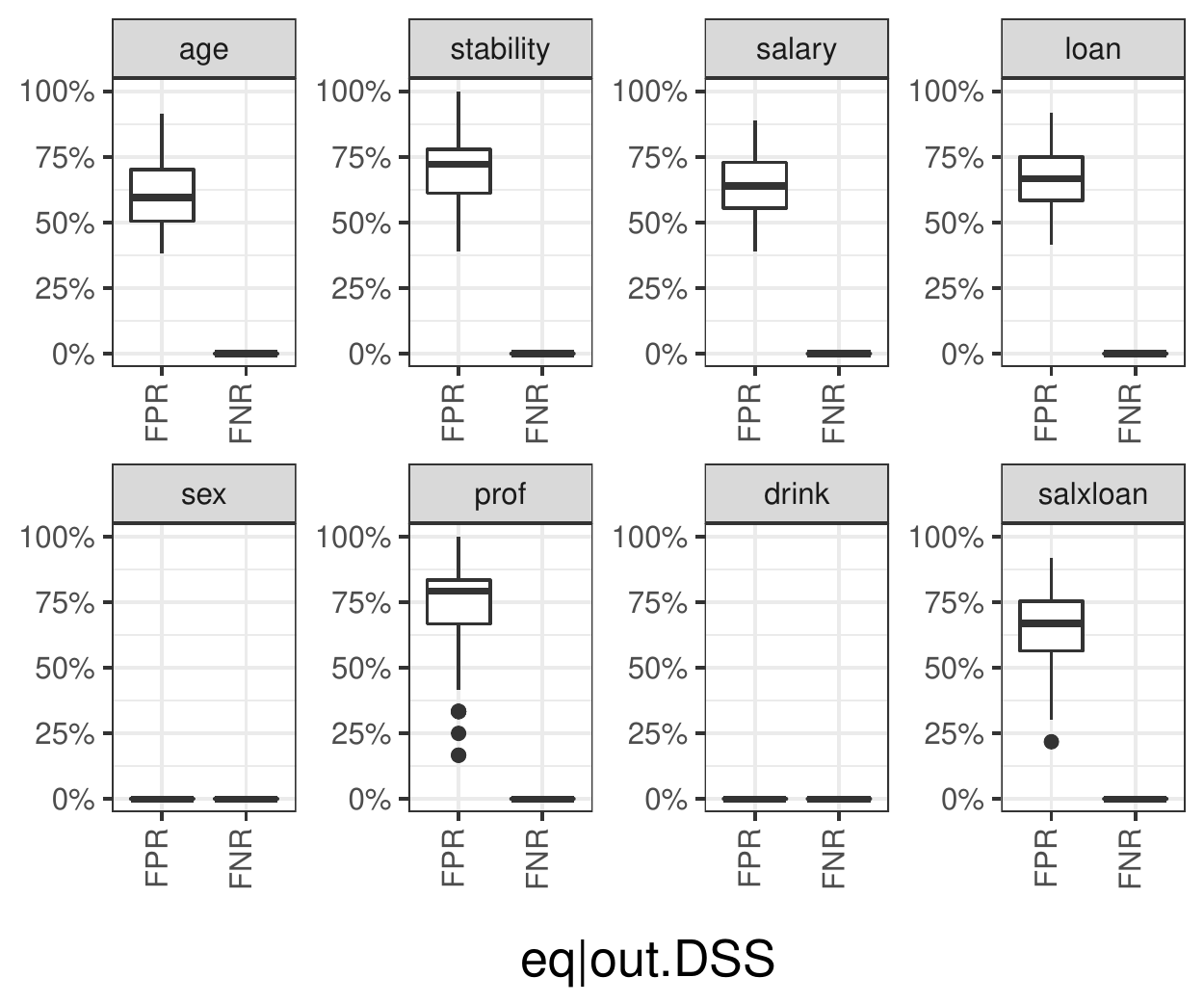}
\caption{Boxplots of the FPR and FNR per variable for settings \texttt{eq|out.MSPE} and \texttt{eq|out.DSS} of the SMuRF algorithm.}
\label{fig:apdfpnr4}
\end{figure}

\begin{figure}[!ht]
\centering
\includegraphics[width = 0.45\textwidth]{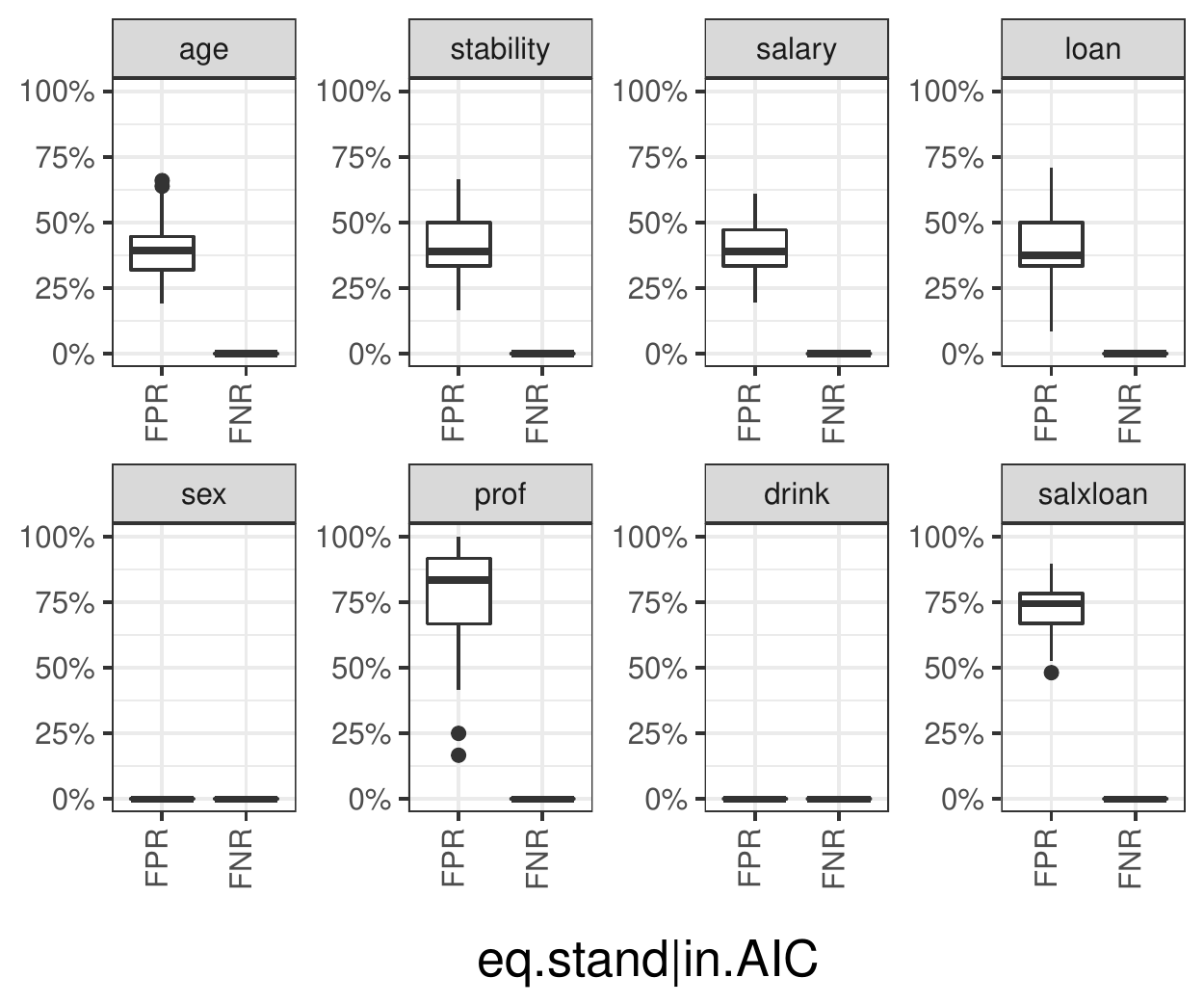}
\hspace{0.2cm}
\includegraphics[width = 0.45\textwidth]{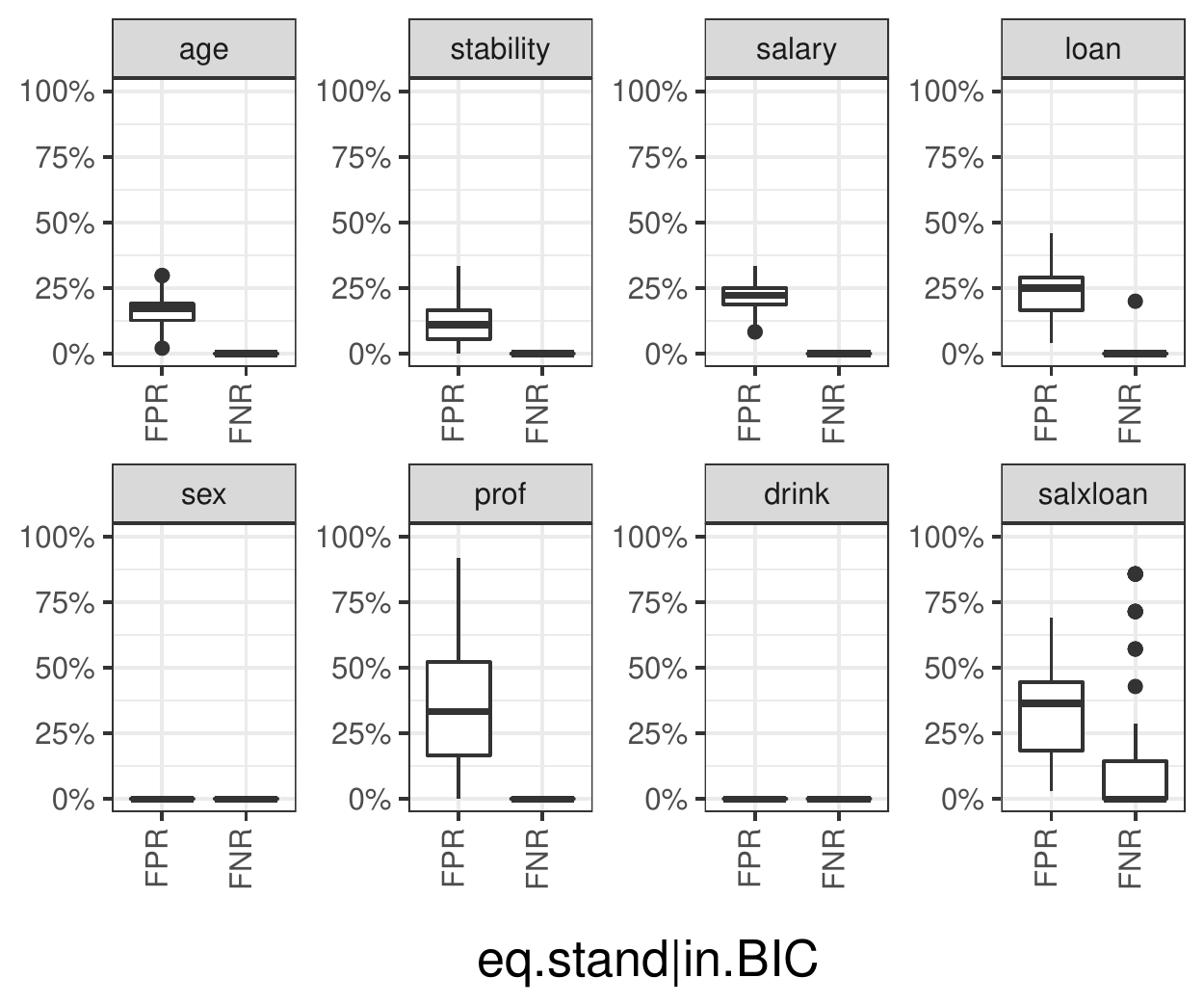}
\caption{Boxplots of the FPR and FNR per variable for settings \texttt{eq.stand|in.AIC} and \texttt{eq.stand|in.BIC} of the SMuRF algorithm.}
\label{fig:apdfpnr5}
\end{figure}

\begin{figure}[!ht]
\centering
\includegraphics[width = 0.45\textwidth]{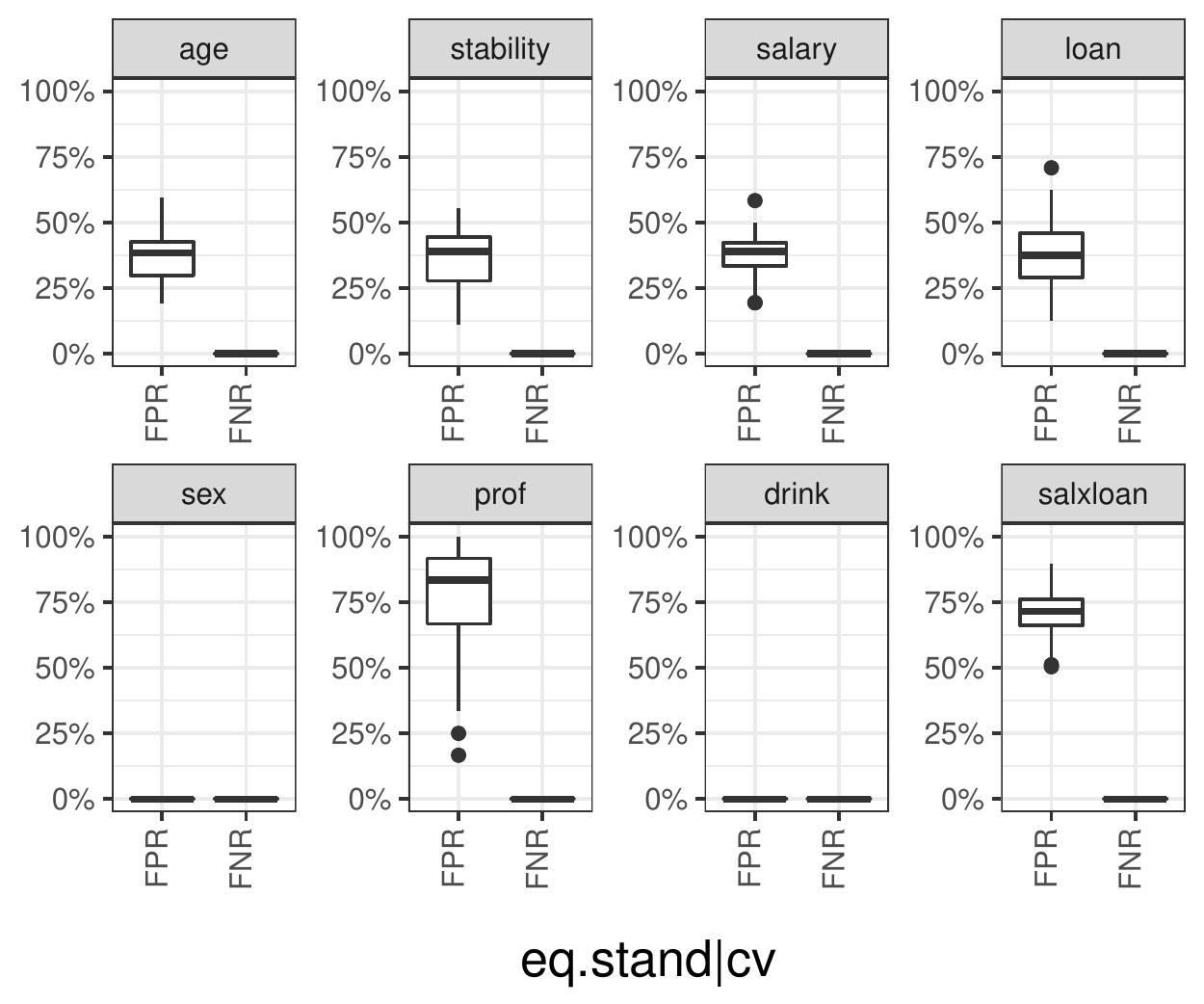}
\hspace{0.2cm}
\includegraphics[width = 0.45\textwidth]{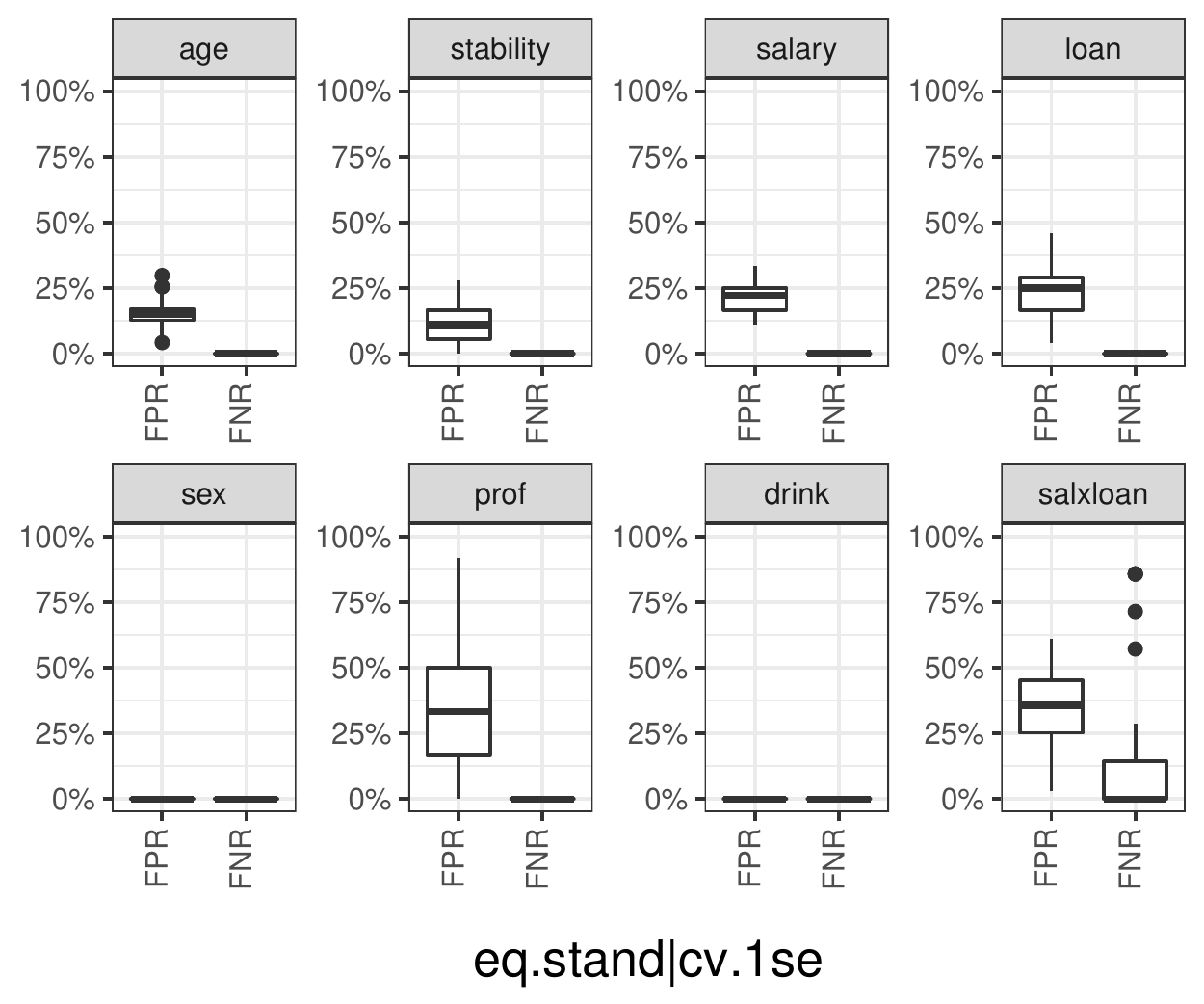}
\caption{Boxplots of the FPR and FNR per variable for settings \texttt{eq.stand|cv} and \texttt{eq.stand|cv.1se} of the SMuRF algorithm.}
\label{fig:apdfpnr6}
\end{figure}

\begin{figure}[!ht]
\centering
\includegraphics[width = 0.45\textwidth]{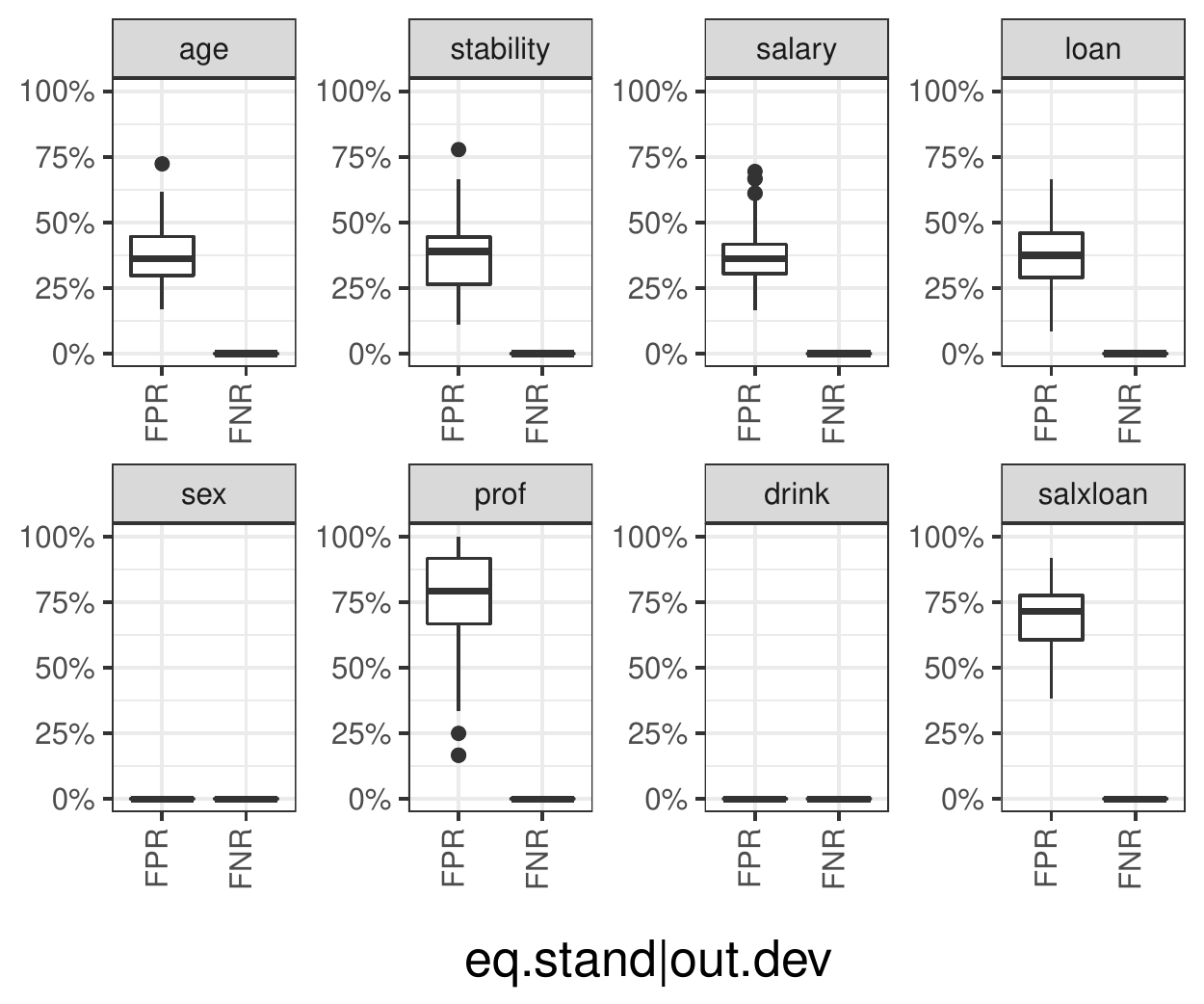}
\caption{Boxplots of the FPR and FNR per variable for setting \texttt{eq.stand|out.dev} of the SMuRF algorithm.}
\label{fig:apdfpnr7}
\end{figure}

\begin{figure}[!ht]
\centering
\includegraphics[width = 0.45\textwidth]{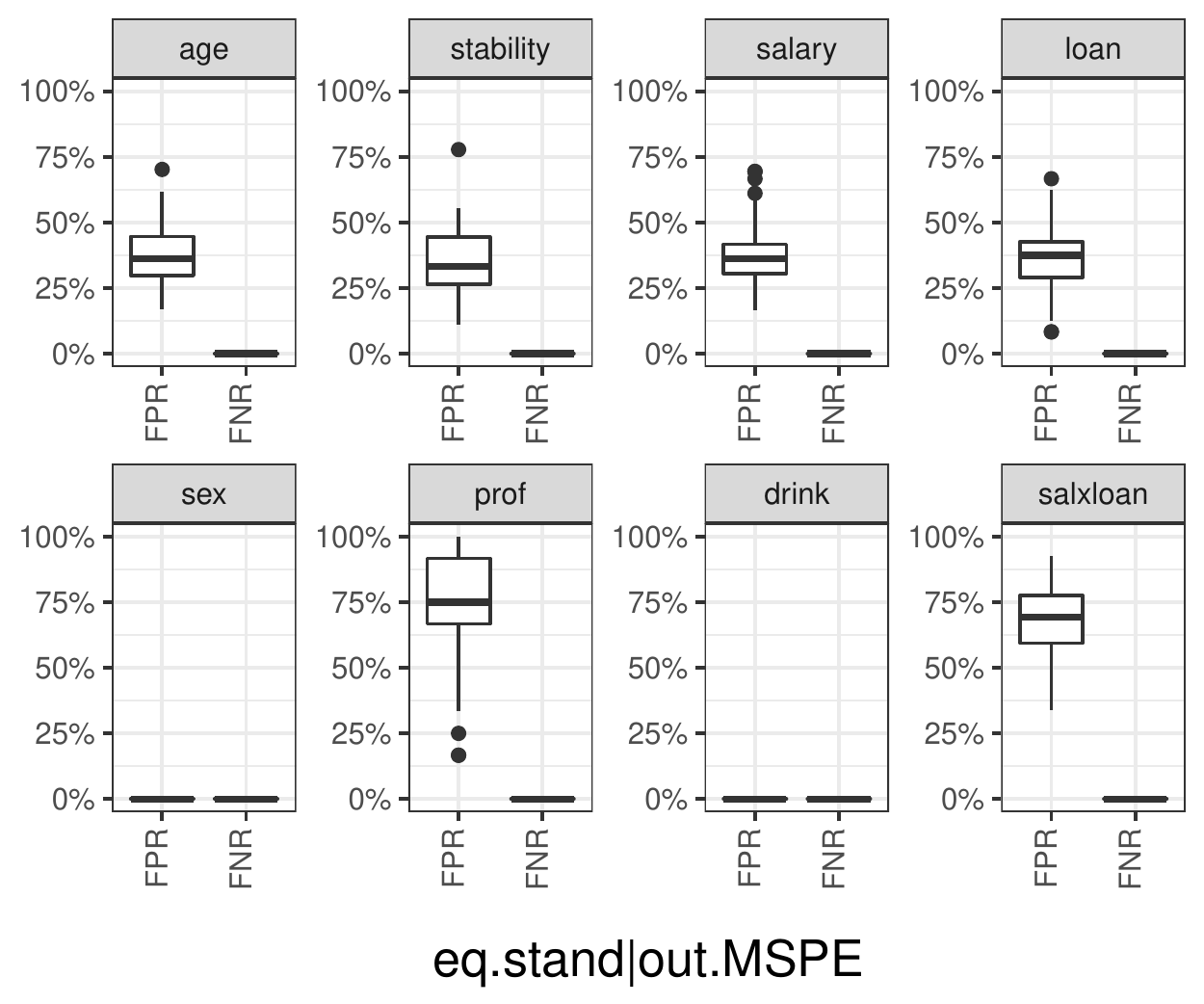}
\hspace{0.2cm}
\includegraphics[width = 0.45\textwidth]{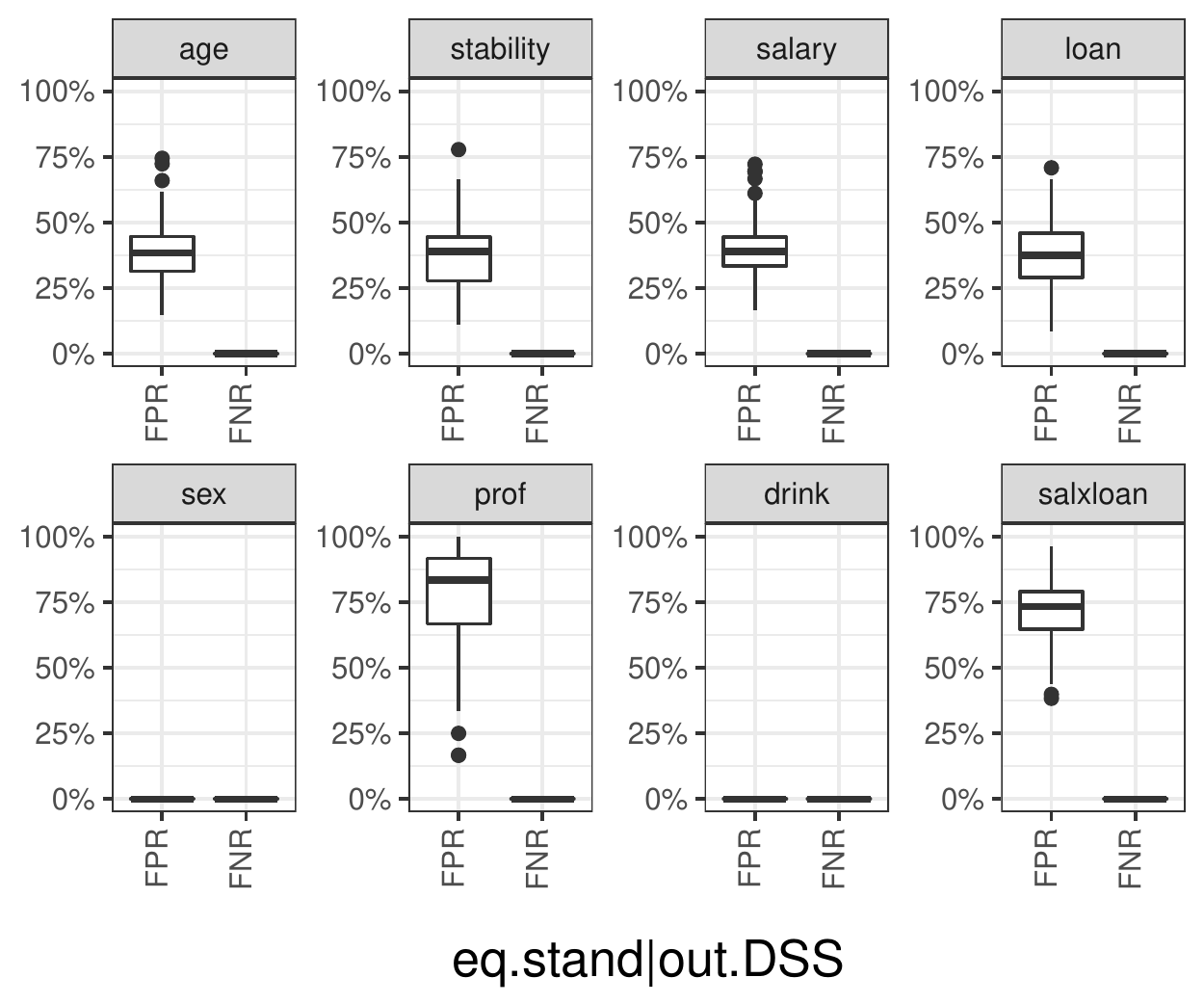}
\caption{Boxplots of the FPR and FNR per variable for settings \texttt{eq.stand|out.MSPE} and \texttt{eq.stand|out.DSS} of the SMuRF algorithm.}
\label{fig:apdfpnr8}
\end{figure}

\begin{figure}[!ht]
\centering
\includegraphics[width = 0.45\textwidth]{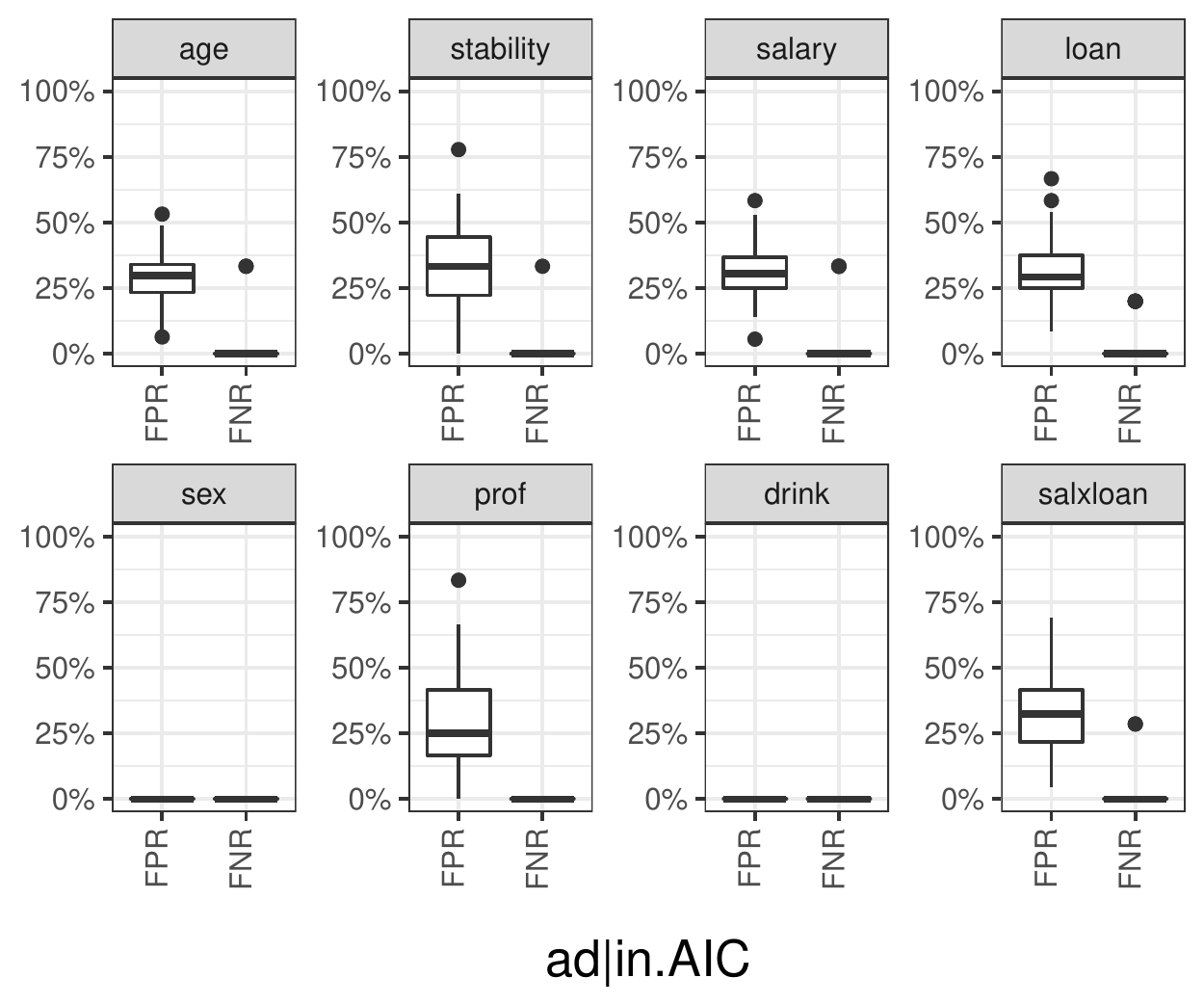}
\hspace{0.2cm}
\includegraphics[width = 0.45\textwidth]{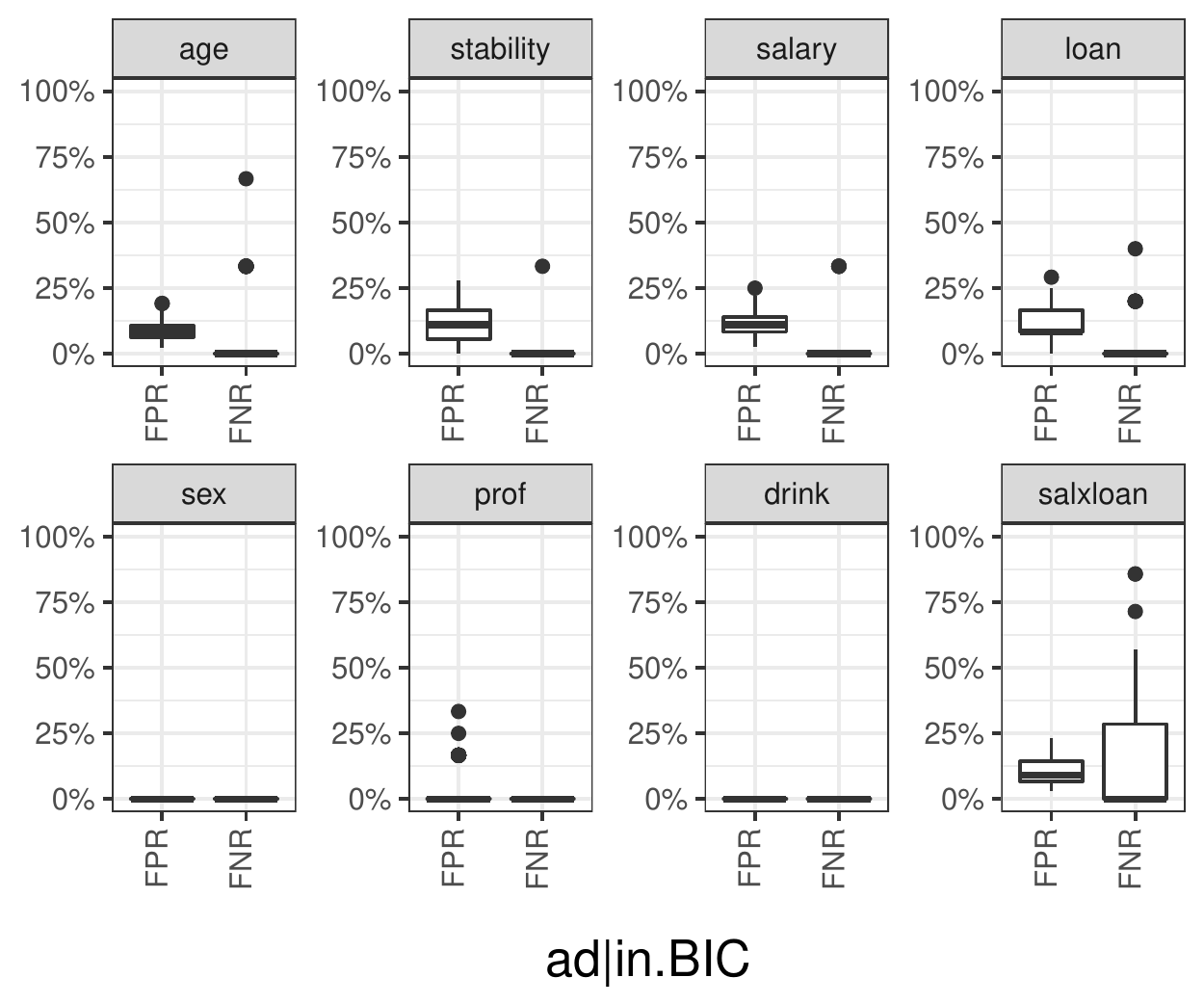}
\caption{Boxplots of the FPR and FNR per variable for settings \texttt{ad|in.AIC} and \texttt{ad|in.BIC} of the SMuRF algorithm.}
\label{fig:apdfpnr9}
\end{figure}

\begin{figure}[!ht]
\centering
\includegraphics[width = 0.45\textwidth]{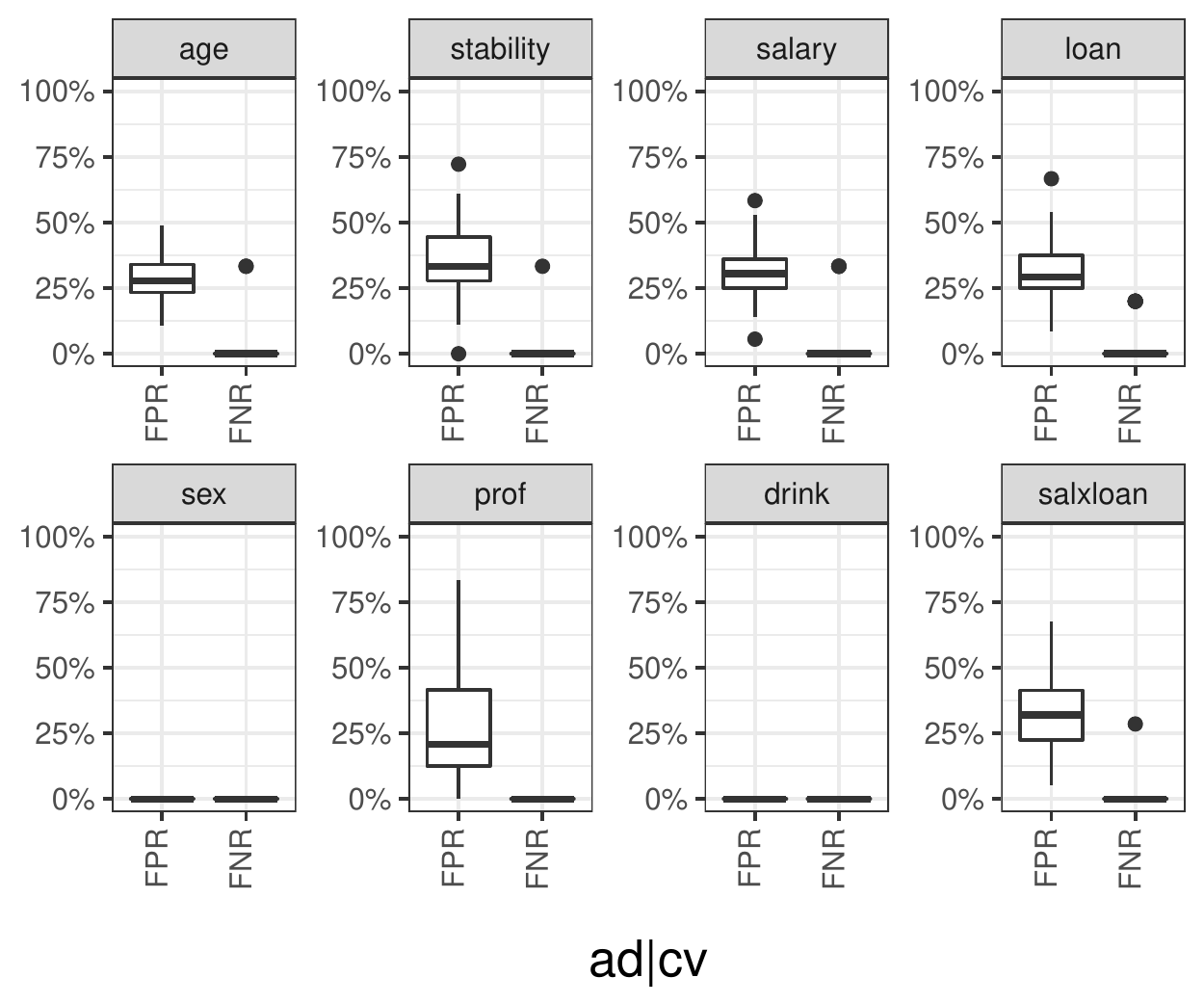}
\hspace{0.2cm}
\includegraphics[width = 0.45\textwidth]{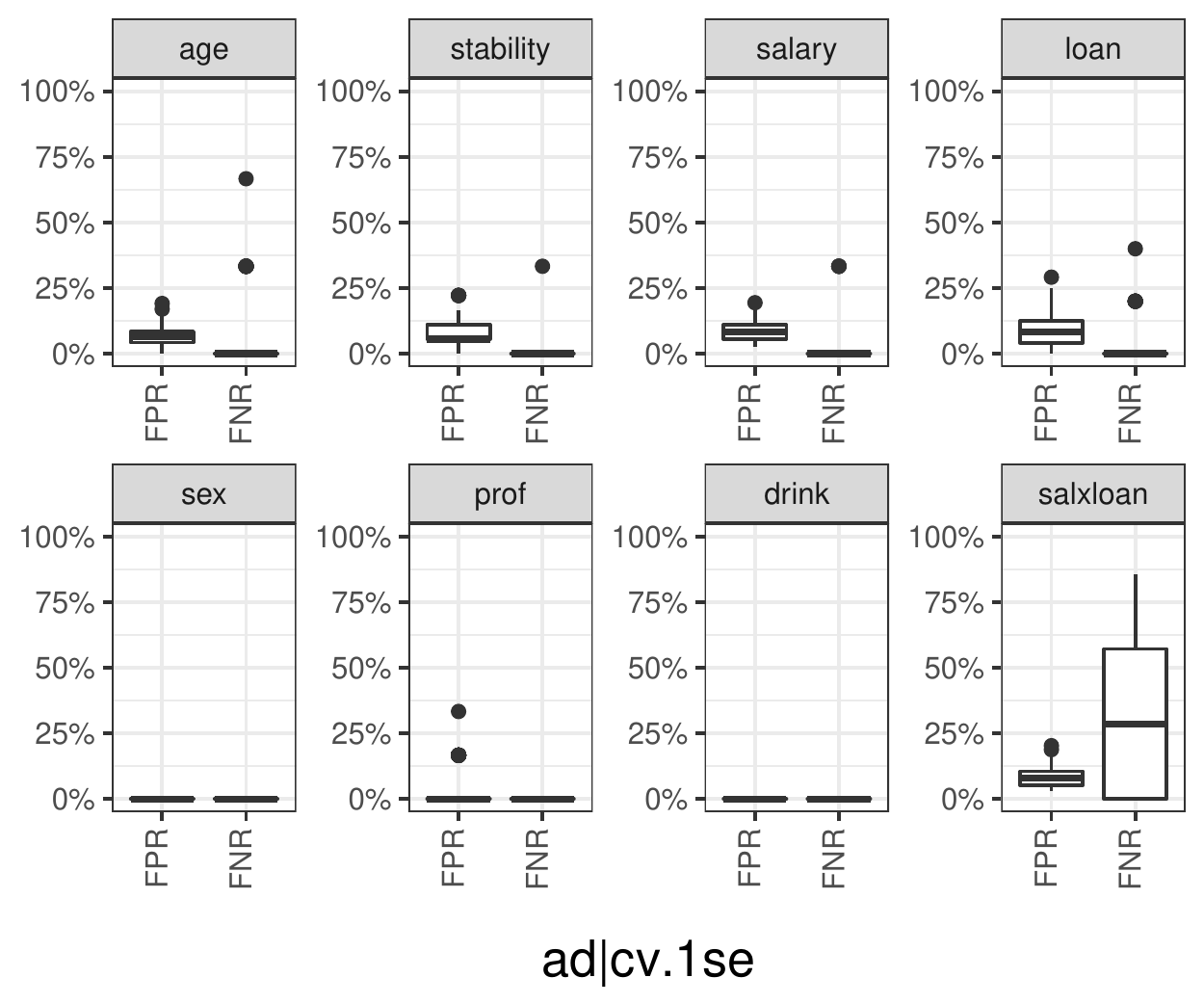}
\caption{Boxplots of the FPR and FNR per variable for settings \texttt{ad|cv} and \texttt{ad|cv.1se} of the SMuRF algorithm.}
\label{fig:apdfpnr10}
\end{figure}

\begin{figure}[!ht]
\centering
\includegraphics[width = 0.45\textwidth]{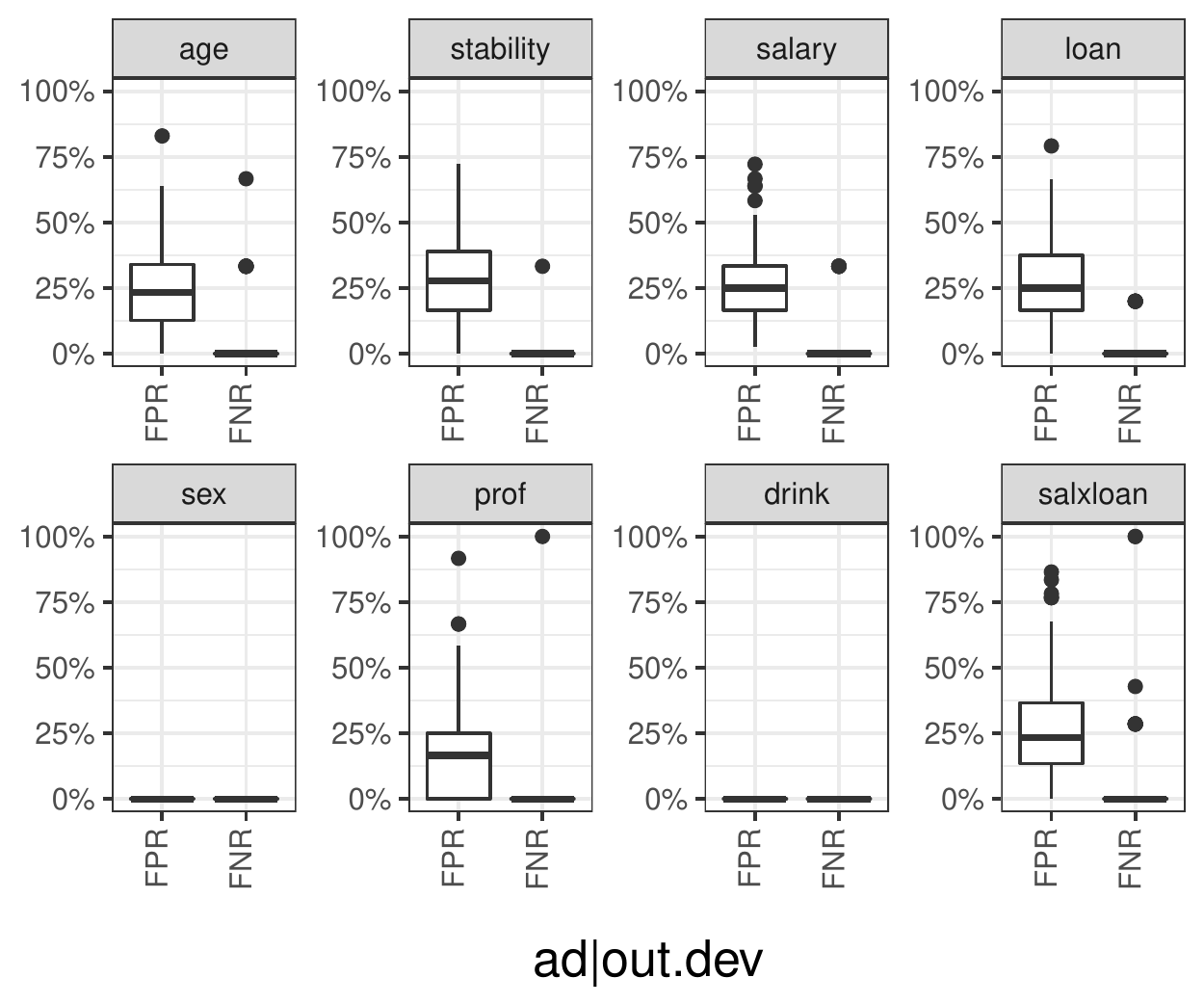}
\caption{Boxplots of the FPR and FNR per variable for setting \texttt{ad|out.dev} of the SMuRF algorithm.}
\label{fig:apdfpnr11}
\end{figure}

\begin{figure}[!ht]
\centering
\includegraphics[width = 0.45\textwidth]{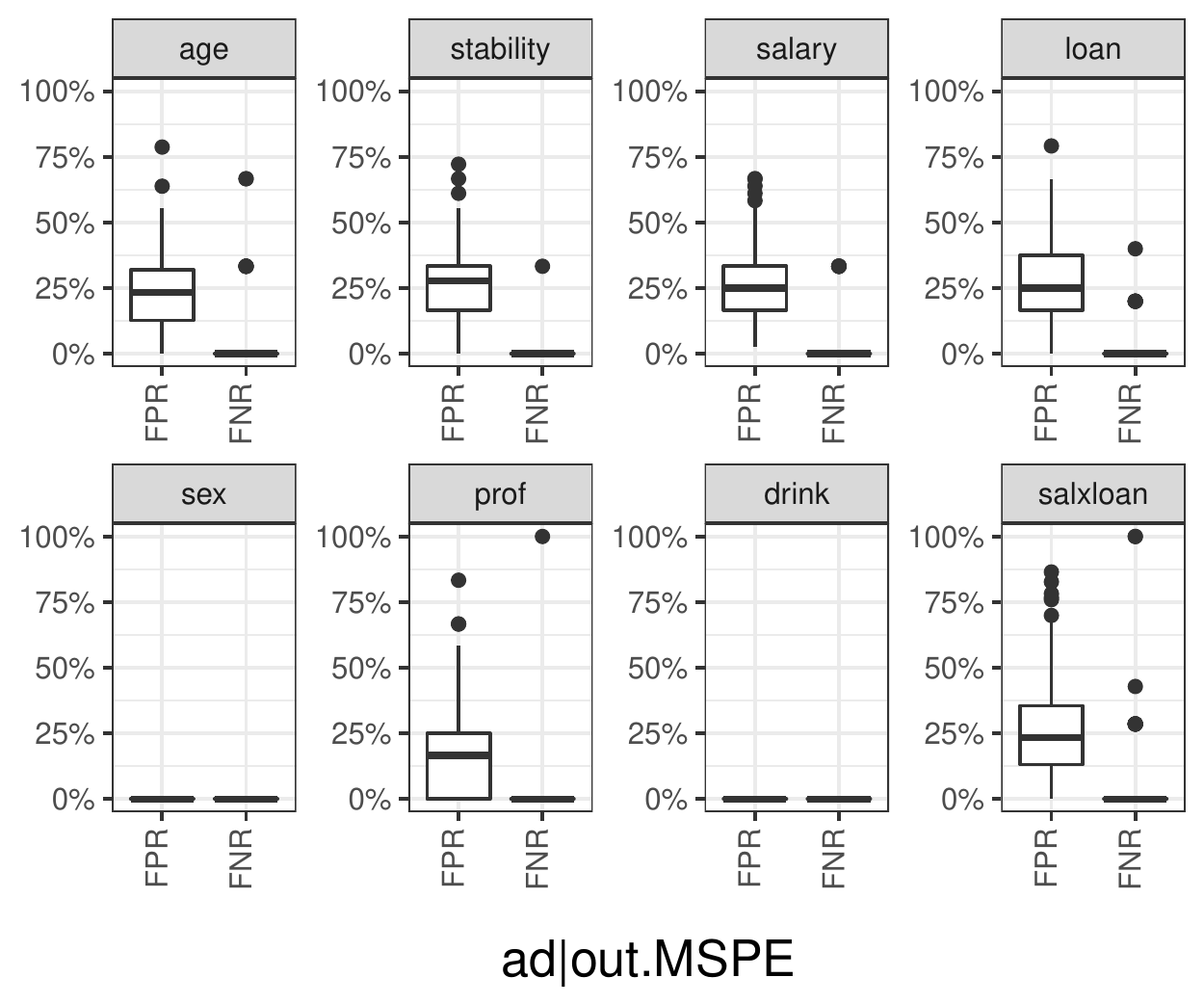}
\hspace{0.2cm}
\includegraphics[width = 0.45\textwidth]{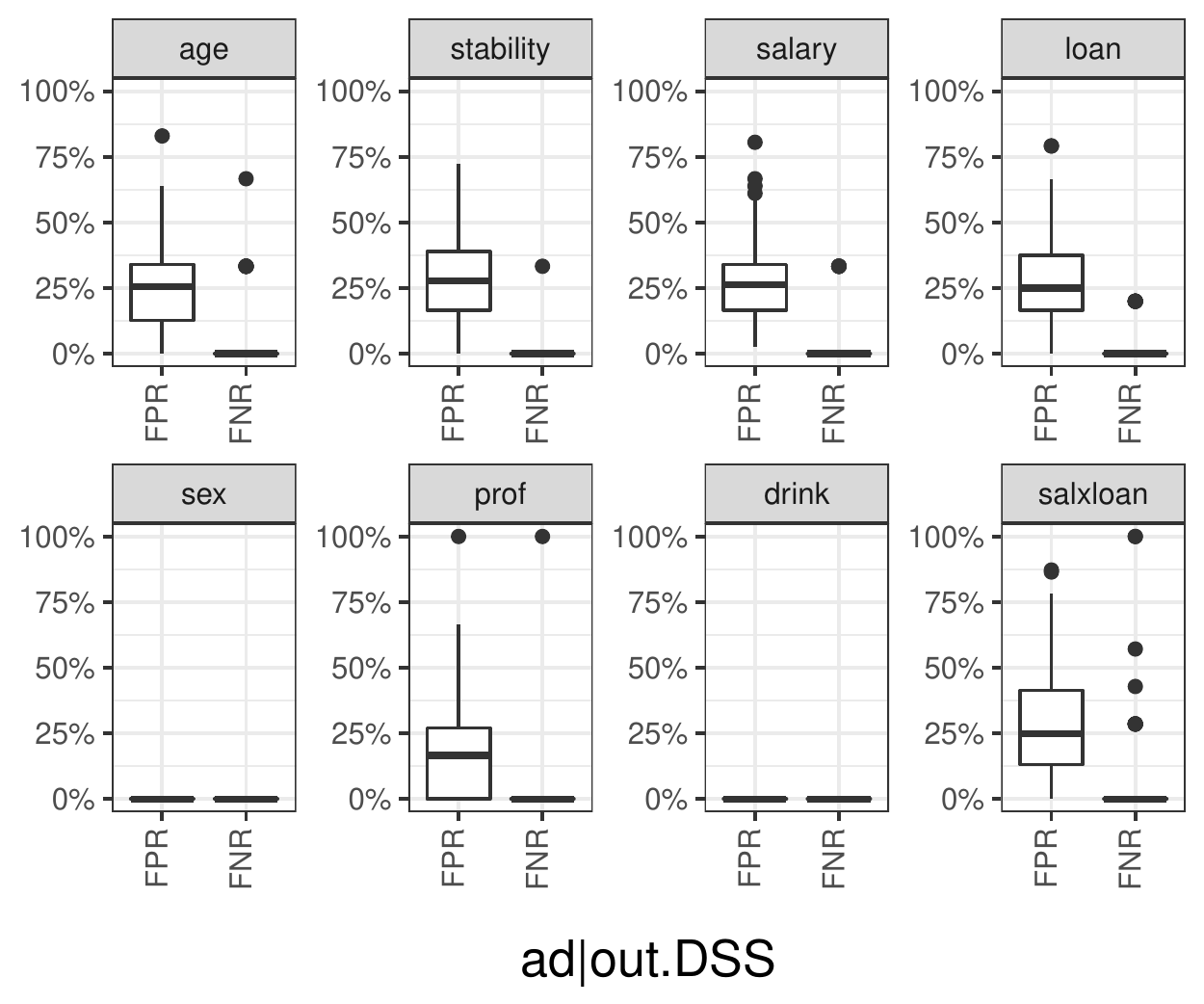}
\caption{Boxplots of the FPR and FNR per variable for settings \texttt{ad|out.MSPE} and \texttt{ad|out.DSS} of the SMuRF algorithm.}
\label{fig:apdfpnr12}
\end{figure}

\begin{figure}[!ht]
\centering
\includegraphics[width = 0.45\textwidth]{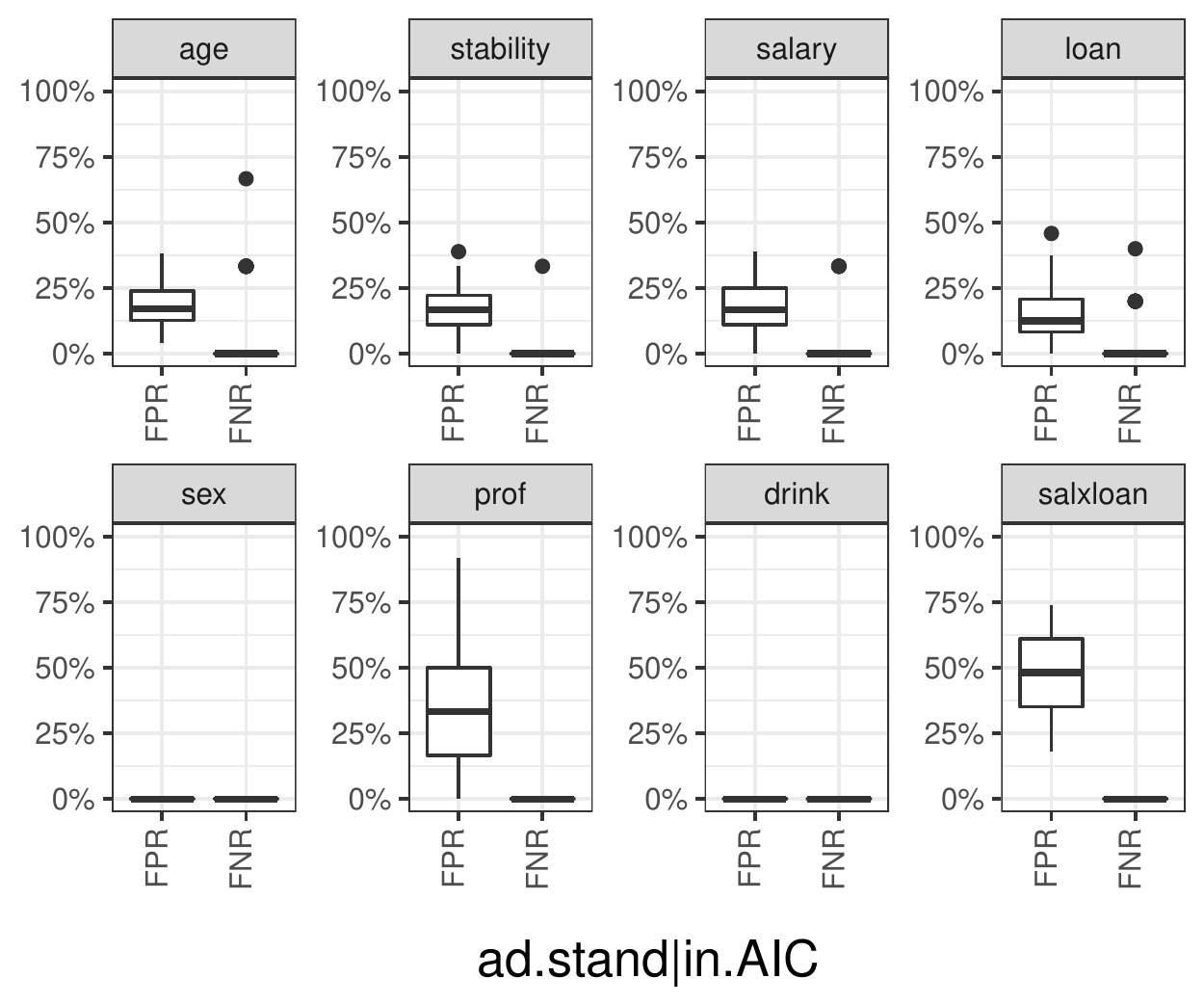}
\hspace{0.2cm}
\includegraphics[width = 0.45\textwidth]{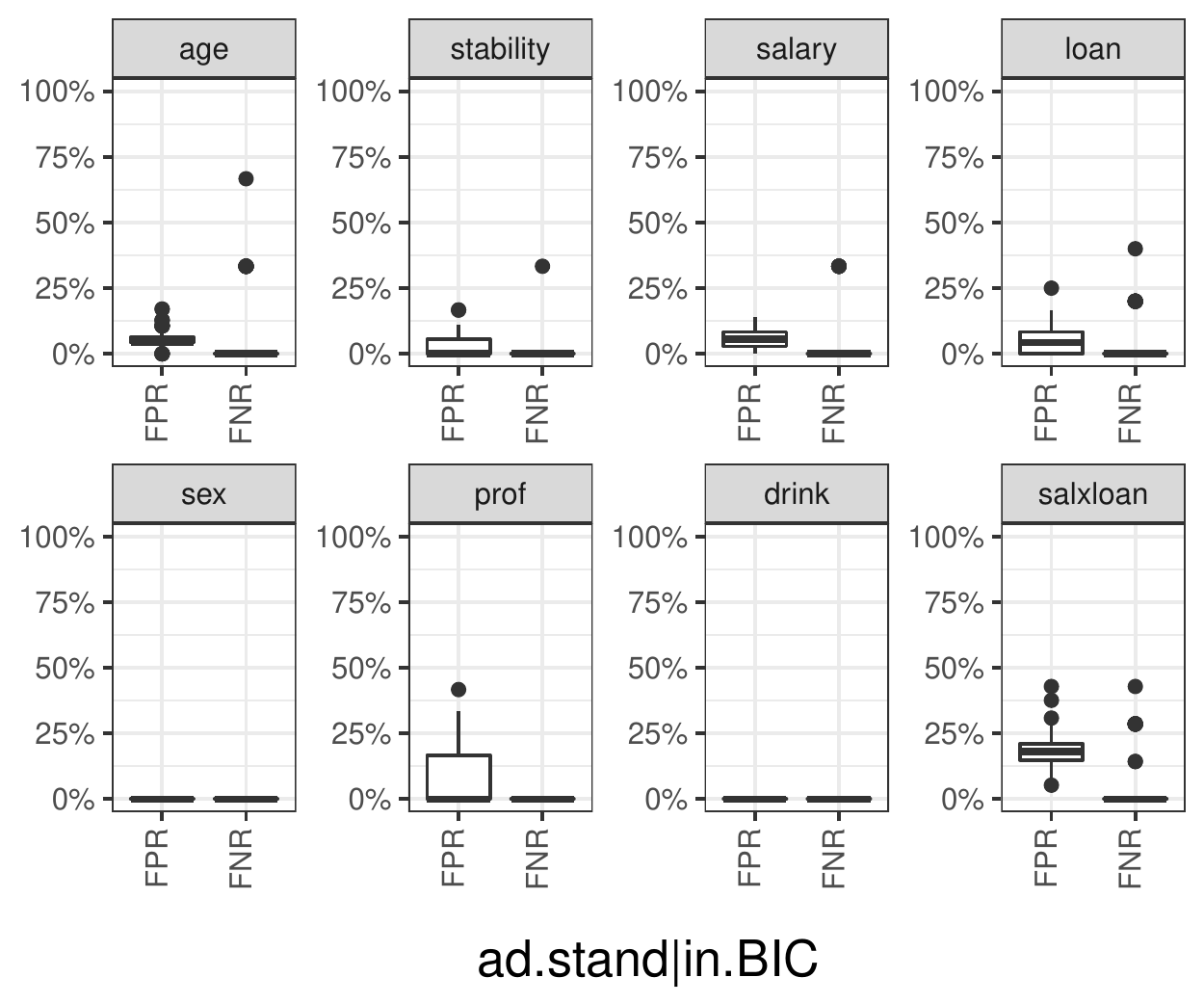}
\caption{Boxplots of the FPR and FNR per variable for settings \texttt{ad.stand|in.AIC} and \texttt{ad.stand|in.BIC} of the SMuRF algorithm.}
\label{fig:apdfpnr13}
\end{figure}

\begin{figure}[!ht]
\centering
\includegraphics[width = 0.45\textwidth]{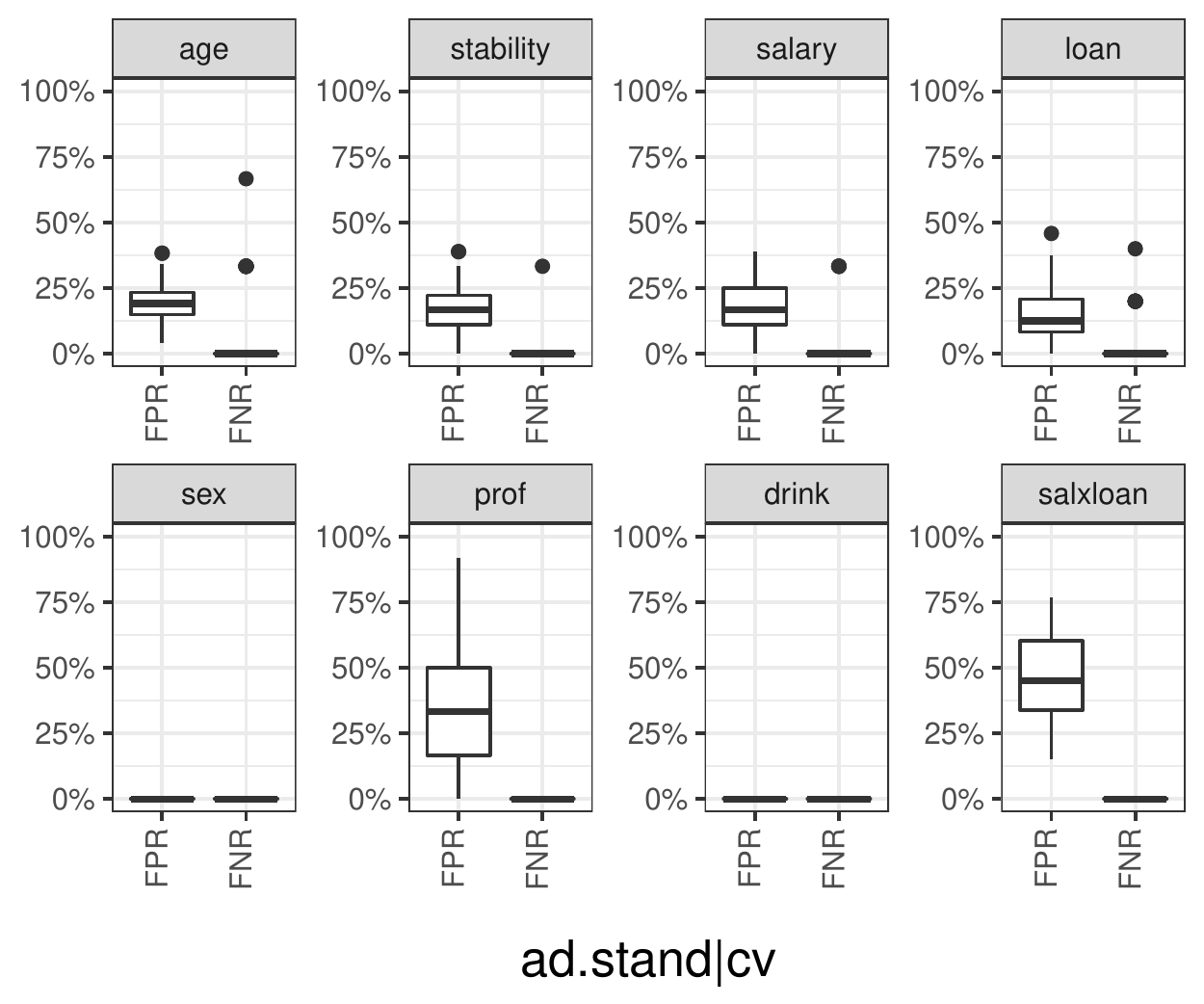}
\hspace{0.2cm}
\includegraphics[width = 0.45\textwidth]{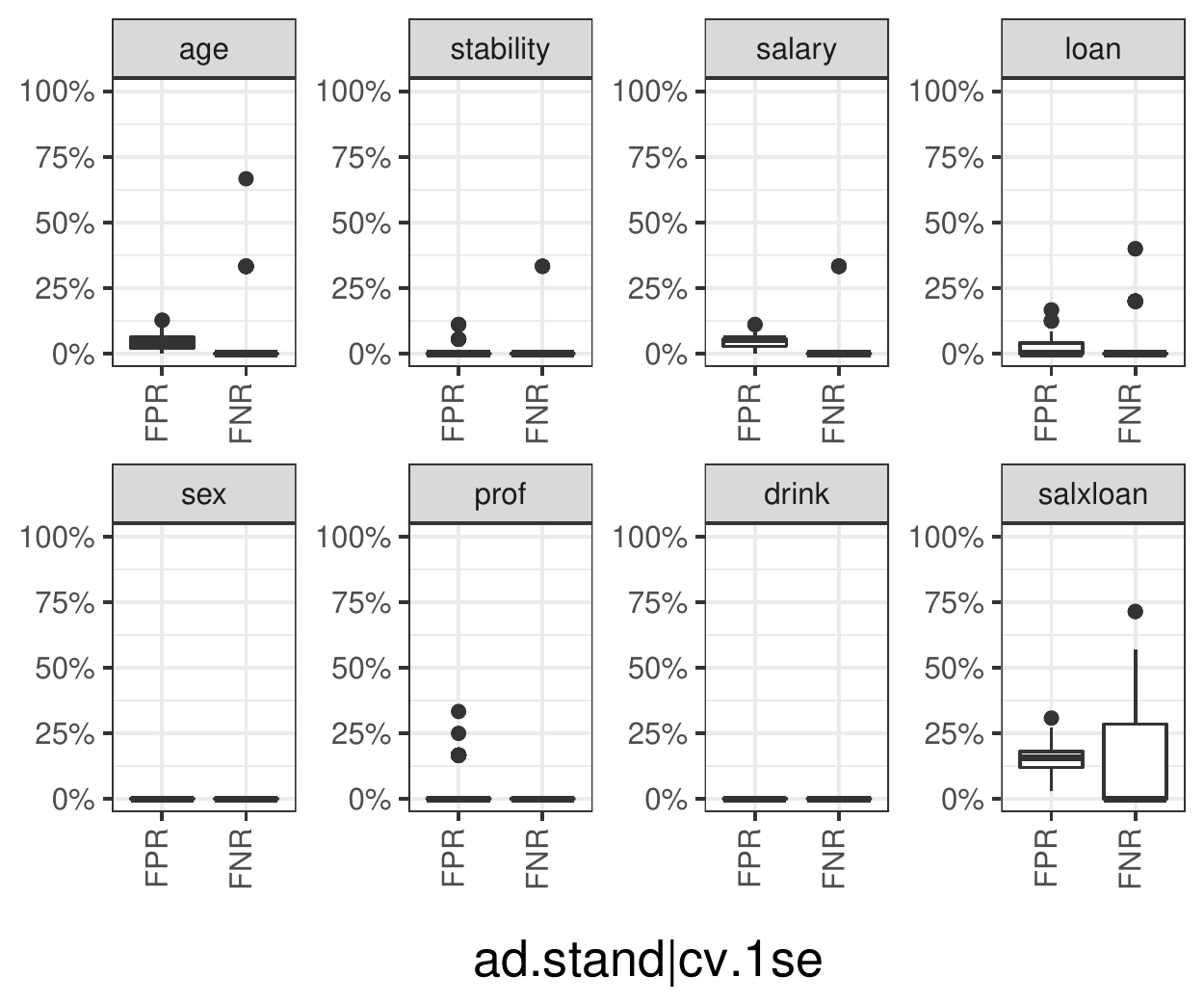}
\caption{Boxplots of the FPR and FNR per variable for settings \texttt{ad.stand|cv} and \texttt{ad.stand|cv.1se} of the SMuRF algorithm.}
\label{fig:apdfpnr14}
\end{figure}

\begin{figure}[!ht]
\centering
\includegraphics[width = 0.45\textwidth]{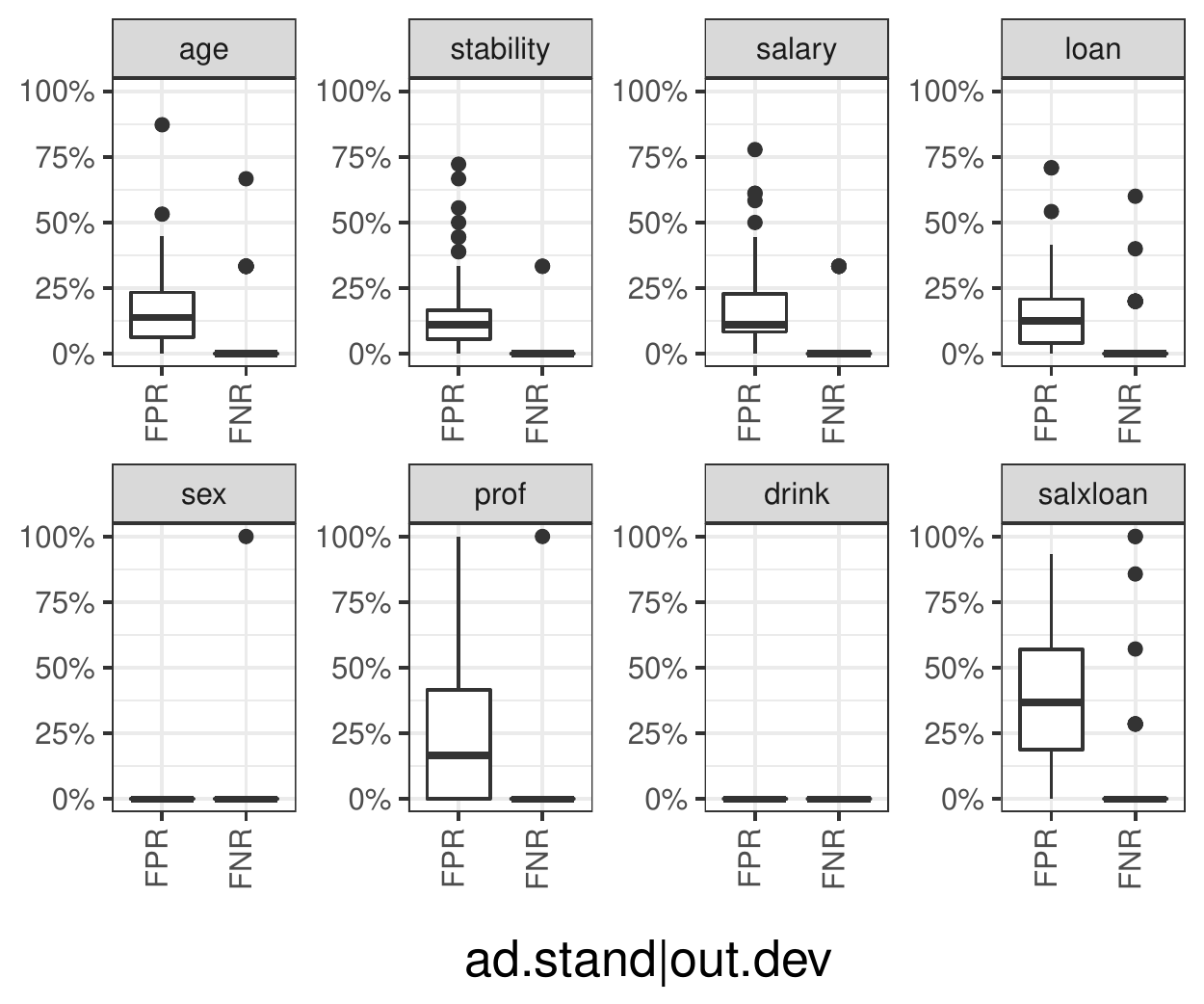}
\caption{Boxplots of the FPR and FNR per variable for setting \texttt{ad.stand|out.dev} of the SMuRF algorithm.}
\label{fig:apdfpnr15}
\end{figure}

\begin{figure}[!ht]
\centering
\includegraphics[width = 0.45\textwidth]{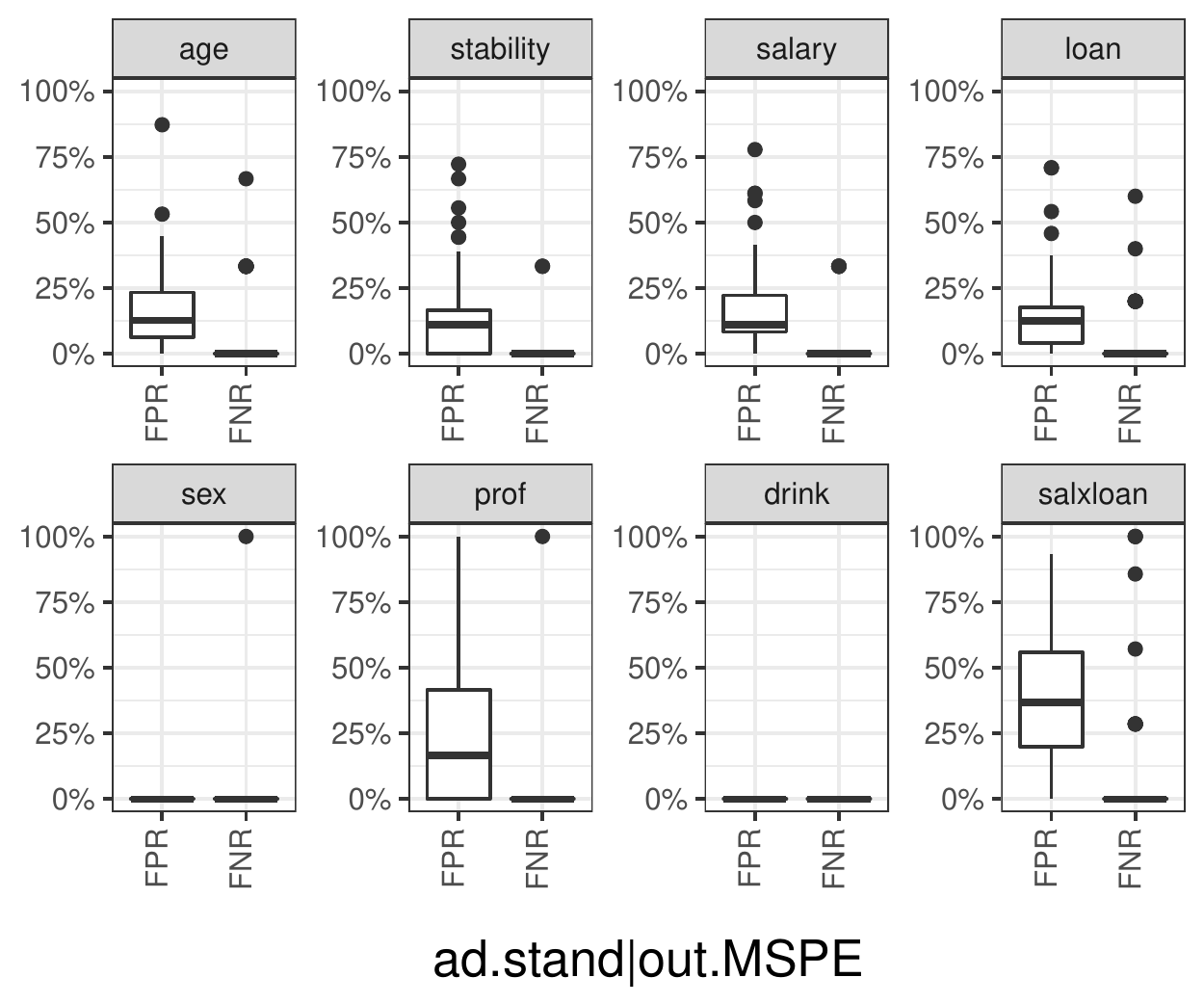}
\hspace{0.2cm}
\includegraphics[width = 0.45\textwidth]{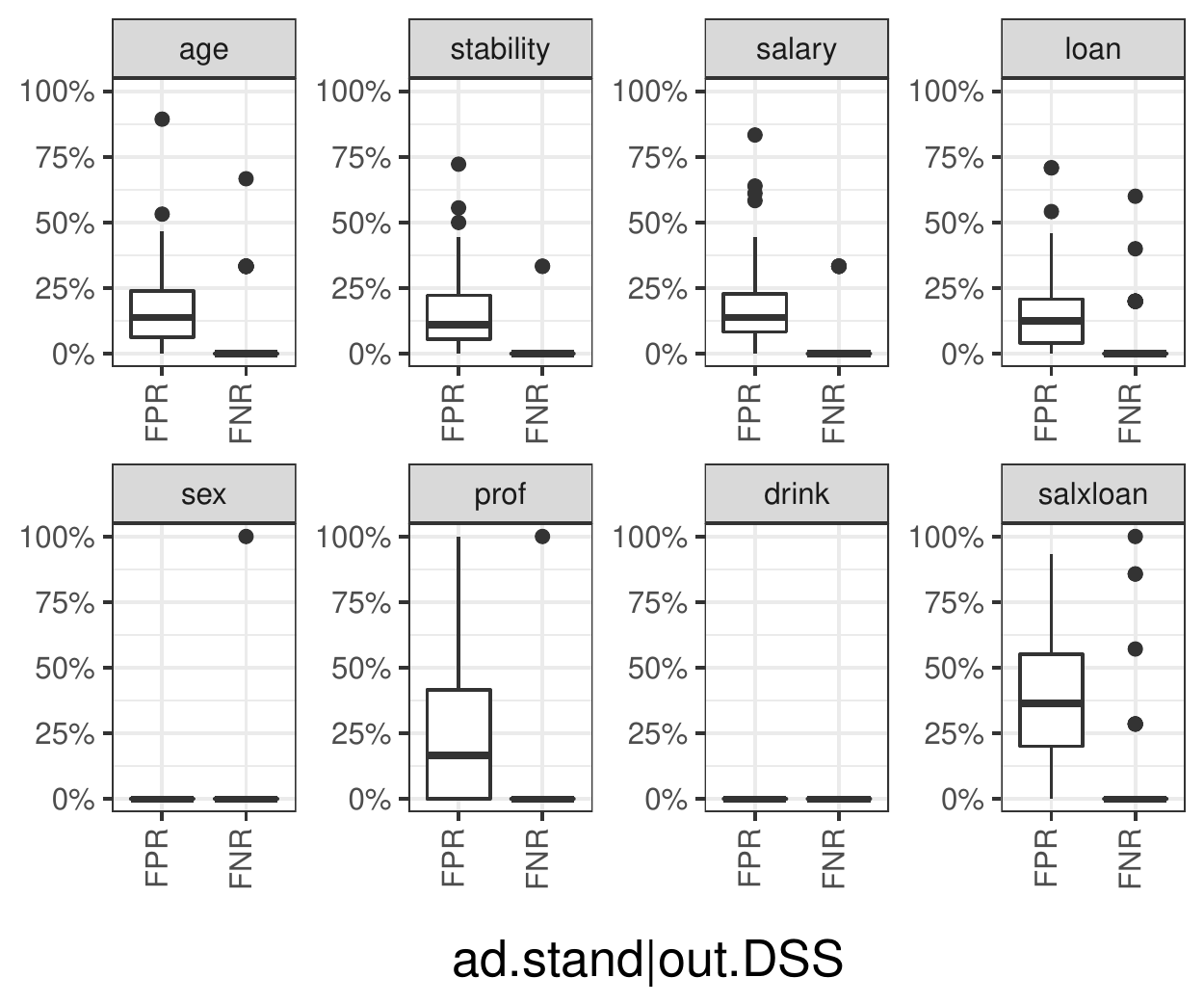}
\caption{Boxplots of the FPR and FNR per variable for settings \texttt{ad.stand|out.MSPE} and \texttt{ad.stand|out.DSS} of the SMuRF algorithm.}
\label{fig:apdfpnr16}
\end{figure}

Figures \ref{fig:apdfpnr1}-\ref{fig:apdfpnr16} show the results of the false positive and negative rate for all settings.

\begin{figure}[!ht]
\centering
\includegraphics[width = \textwidth]{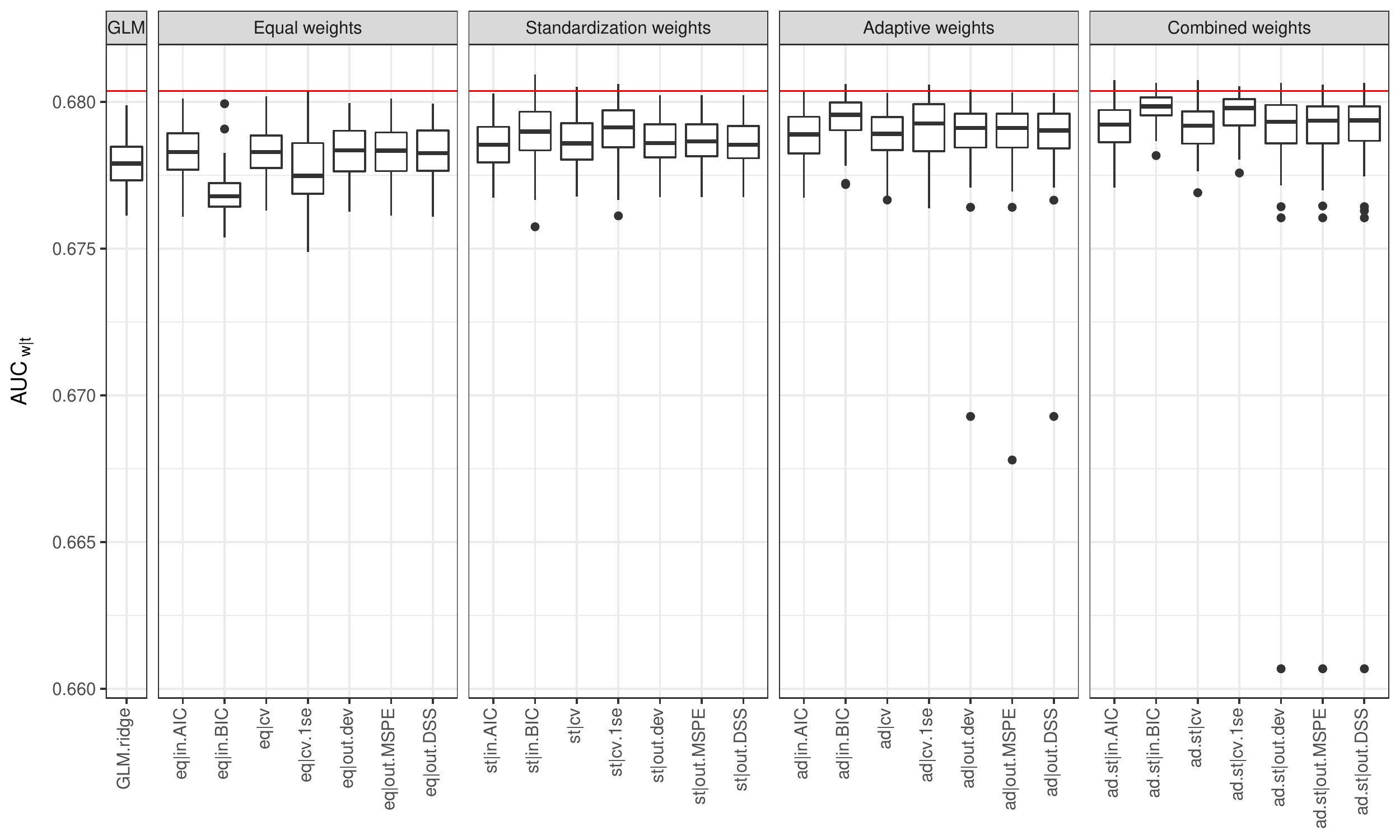}
\caption{Boxplot of the AUC for the binomial GLM with a small ridge penalty and for the different settings settings of the SMuRF algorithm. }
\label{fig:apdaucfull}
\end{figure}

\begin{figure}[!ht]
\centering
\includegraphics[width = \textwidth]{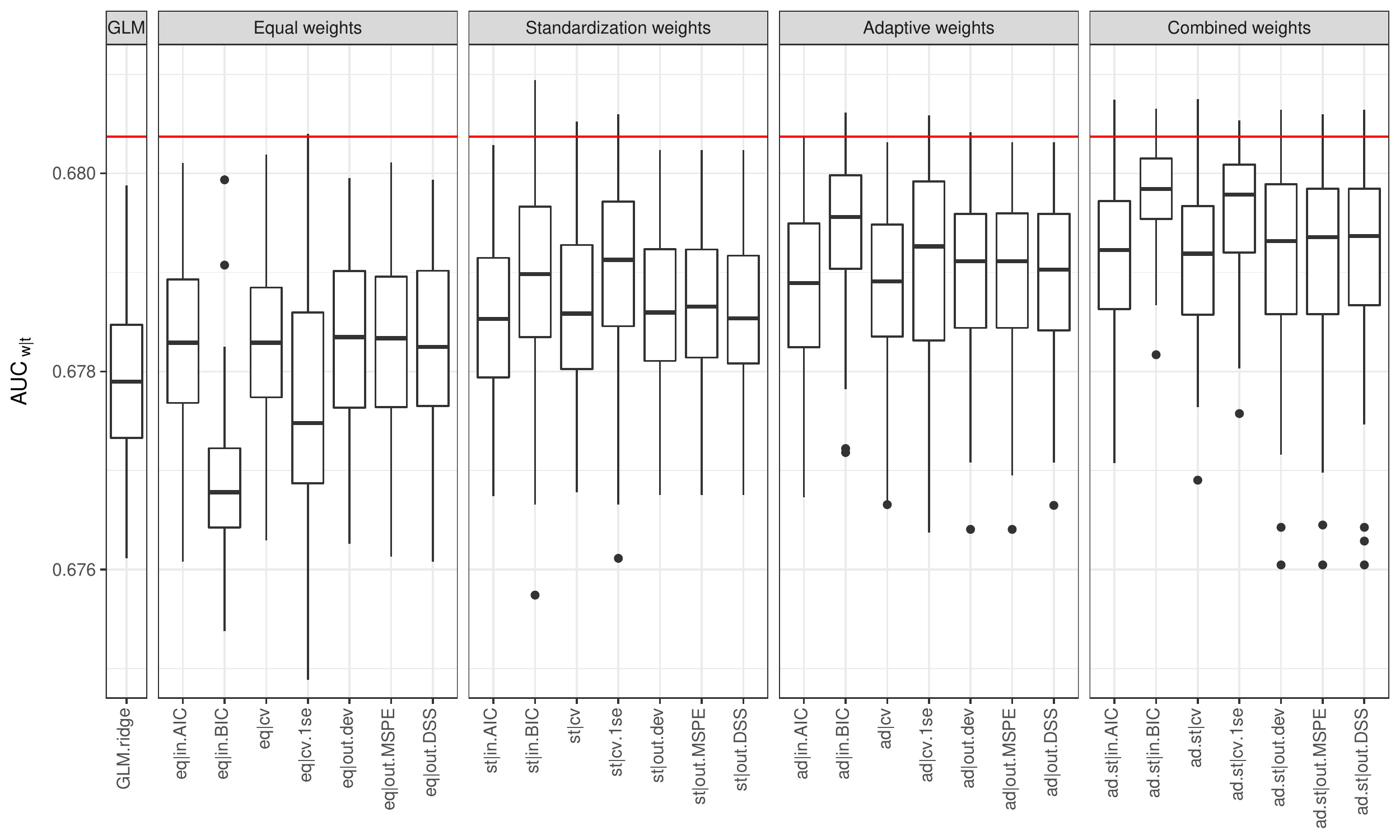}
\caption{Zoomed-in boxplot of the AUC for the binomial GLM with a small ridge penalty and for the different settings settings of the SMuRF algorithm. }
\label{fig:apdauczoomed}
\end{figure}

We provide an overview of the results of the AUC for the different settings in Figure~\ref{fig:apdaucfull} and a zoomed version in Figure~\ref{fig:apdauczoomed}.

\clearpage

\subsection{Comparison with PIRLS implementation in \texttt{gvcm.cat}}\label{apdsec:smurfgvcm}

In this section, we compare the R implementation of the SMuRF algorithm through the \texttt{smurf} package (see \url{https://cran.r-project.org/web/packages/smurf/}) to the PIRLS algorithm proposed by \cite{gertheiss2010} and \cite{gertheiss2017}, implemented in the R package \texttt{gvcm.cat}. The Penalized Iteratively Re-weighted Least Squares (PIRLS) algorithm applies a quadratic approximation in the region of the non-differentiable points of the Lasso-type penalties to make the objective function - at least twice - differentiable everywhere. As such, the IRLS approach can be used to this adjusted objective function. In essence the PIRLS is a second order, Newton-type optimization algorithm which we expect to scale less well with respect to the amount of available data than our first-order approach in SMuRF. All calculations below were performed on a standard business laptop with 8 logical processors of which 5 were used for parallel computations in the \texttt{smurf} package.

\begin{figure}[!ht]
\centering
\includegraphics[width = \textwidth]{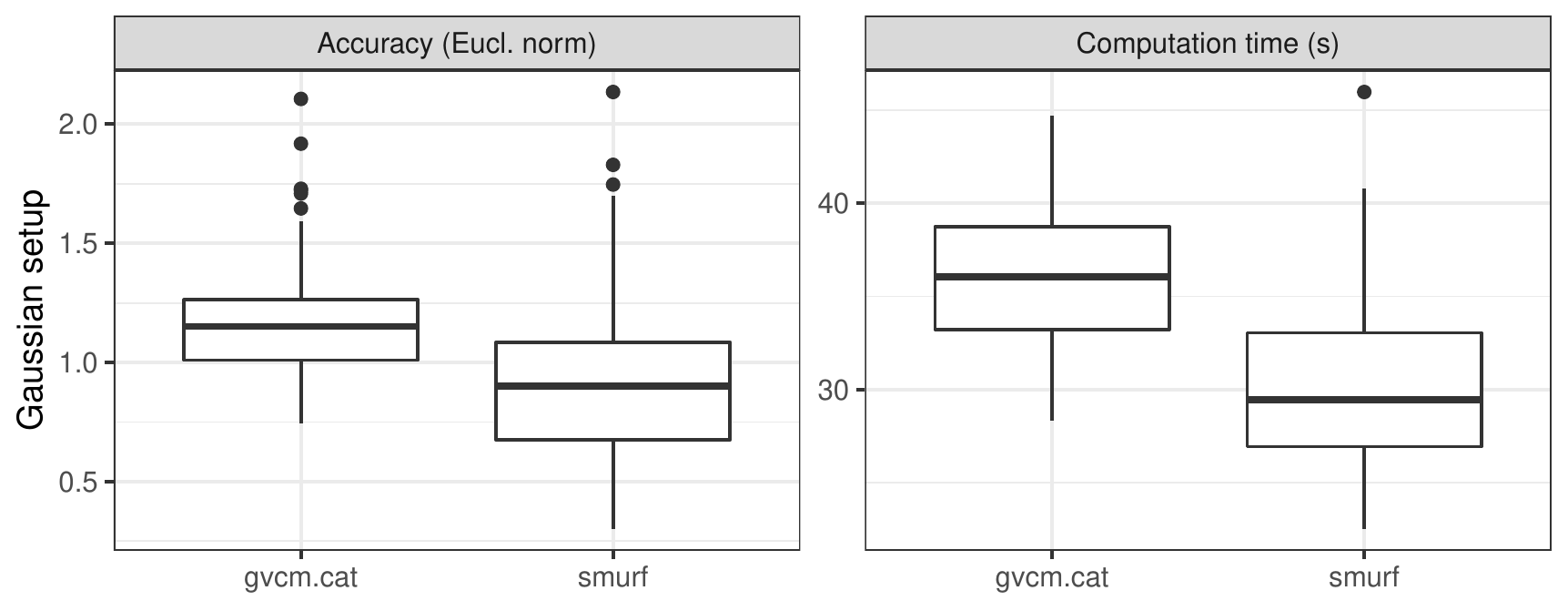}
\caption{Comparison of the parameter accuracy (using the Euclidean norm between true and estimated coefficients) and the computation time (in seconds) of the R implementation \texttt{smurf} versus \texttt{gvcm.cat} for 100 simulations of a Gaussian regression problem with $n=2,500$ data points. The data was simulated incorporating a mix of continuous and ordinal predictors with 11 degrees of freedom but an initial overparametrization of 31 parameters.}
\label{fig:algocompareGaussian}
\end{figure}

In a first comparison, we simulate a Gaussian regression problem based on $n=2,500$ data points. The data exists of a response generated from a Gaussian distribution with the mean depending on 3 continuous predictors and 4 ordinal factors. The true degrees of freedom of the regression problem is 11 while the initial overparametrization starts with 31 coefficients. The Lasso and Fused Lasso are applied to the overparametrized predictor respectively. Both the \texttt{smurf} and \texttt{gvcm.cat} package are programmed to find the optimal solution of this penalized regression setting by tuning $\lambda$ with 5-fold cross-validation using the deviance as performance criterion and adaptive weights based on an initial GLM fit. As both approaches can in theory reach any accuracy with respect to the true underlying coefficients, we compare the computation time when similar accuracy is reached. Therefore, any futher settings (approximation precision and estimation rounding for PIRLS, numerical precision stopping criterion for SMuRF, lambda search vector for both algorithms,...) are taken such that we expect the obtained accuracy after estimation to be comparable. We perform this simulation 100 times to reduce sampling effects. Figure~\ref{fig:algocompareGaussian} shows boxplots of the accuracy (measured by the Euclidean distance with respect to the true underlying parameters) and the computation time (in seconds) of each implementation over the 100 simulations. The \texttt{smurf} implementation performs slightly better but the computation time and accuracy of \texttt{gvcm.cat} is comparable.

\begin{figure}[!ht]
\centering
\includegraphics[width = \textwidth]{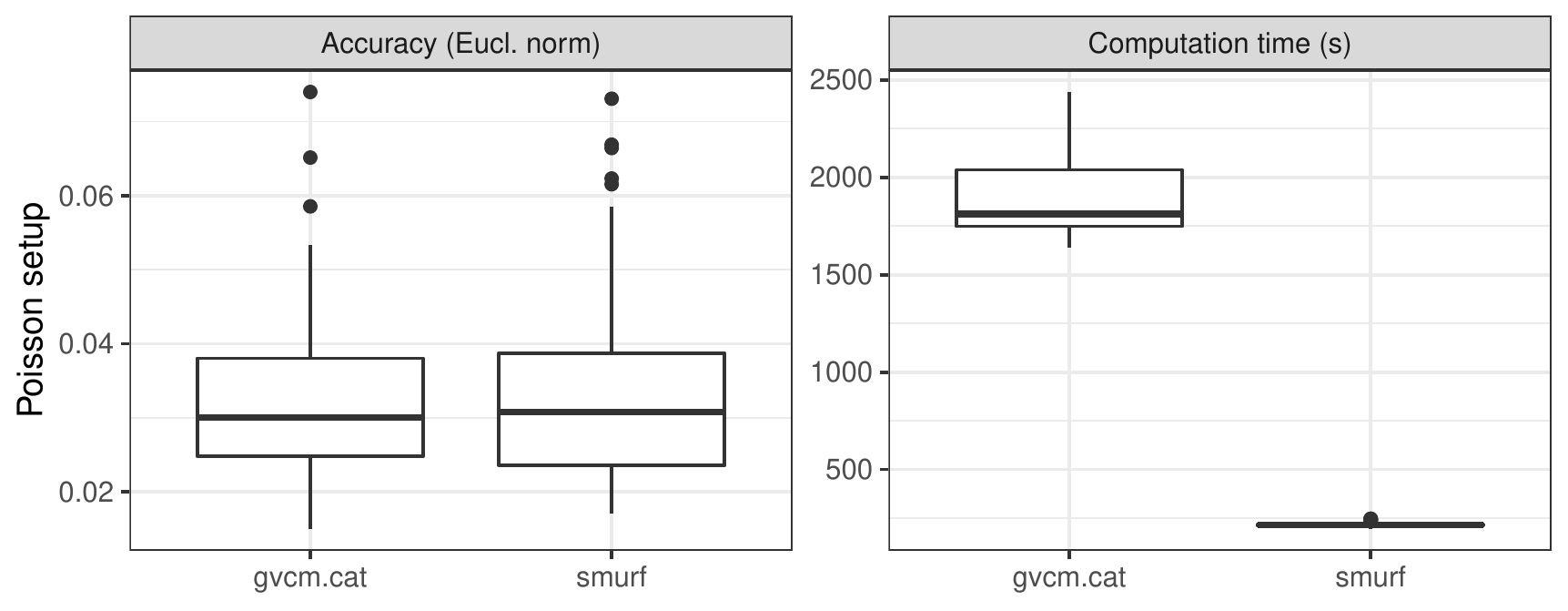}
\caption{Comparison of the parameter accuracy (using the Euclidean norm between true and estimated coefficients) and the computation time (in seconds) of the R implementation \texttt{smurf} versus \texttt{gvcm.cat} for 100 simulations of a Poisson regression problem with $n=50,000$ data points. The data was simulated incorporating a mix of continuous, ordinal and nominal predictors with 17 degrees of freedom but an initial overparametrization of 70 parameters.}
\label{fig:algocomparePoisson}
\end{figure}

Secondly, we perform a similar exercise, but in a Poisson regression framework with $n=50,000$ data points, 3 continuous predictors, 3 ordinal factors and 3 nominal factors resulting in a true model with 17 degrees of freedom with an initial overparametrization of 70 coefficients. This predictive problem thus has more data points and coefficients to penalize. Figure~\ref{fig:algocomparePoisson} shows the accuracy and computation time for both algorithms over 100 instances of the simulated Poisson regression. Both algorithms reach the desired accuracy, driven by our initialization of the algorithm settings. However, the computation time is on average almost 9 times higher for the PIRLS implementation in \texttt{gvcm.cat} compared to our implementation of SMuRF. Due to implementation differences, it is hard to attribute this run-time difference to exact properties of the algorithms. The \texttt{smurf} package uses parallel computation over 5 logical processors for performing the 5-fold cross-validation while \texttt{gvcm.cat} did not. This is an advantage for larger datasets which improves the computation time by a factor 5 theoretically. Additionally, implementation efficiency in matrix multiplications may play a role. However, since the speed-up of our approach is almost a factor 9, it is our belief that the SMuRF algorithm itself, besides possible implementation advantages, is more suited to large predictive problems due to the lack of second order derivatives to be calculated and the approach to solve smaller, easier subproblems instead of one full optimization over all penalties simultaneously.

\clearpage
\bibliographystyle{apalike} 
\bibliography{References.Sparsity}